\documentclass[journal]{IEEEtran}
\usepackage{amsfonts}
\usepackage{color}
\usepackage{amsmath}
\DeclareMathOperator*{\argmax}{arg\,max}
\usepackage{graphicx}
\newtheorem{Proposition}{Proposition}[section]
\newtheorem{Corollary}{Corollary}[section]
\newtheorem{Lemma}{Lemma}[section]
\newtheorem{Remark}{Remark}
\usepackage[linesnumbered,lined,ruled,commentsnumbered]{algorithm2e}%
\usepackage{bm}
\usepackage{multirow}

\usepackage{float}

\usepackage[caption=false,font=normalsize]{subfig}
\usepackage{cite}

\ifCLASSINFOpdf
\else
\fi
\begin{document}
%



\title{On Communication-Efficient Multisensor Track Association via Measurement Transformation (Extended Version)}


%
\author{Haiqi~Liu,
		Jiajie~Sun,
		Xuqi~Zhang,
		Fanqin~Meng,
		Xiaojing~Shen,
		and~Pramod~K.~Varshney
\thanks{This work was supported in part by Sichuan Youth Science and Technology Innovation Team under Grant 2022JDTD0014, Grant 2021JDJQ0036, and Grant SQ2020YFA070244. \textit{(Corresponding author: Jiajie Sun.)}}
\thanks{Haiqi Liu, Jiajie Sun, Xuqi Zhang and Xiaojing Shen are with School of Mathematics, Sichuan University, Chengdu, Sichuan 610064, China (e-mail: haiqiliu0330@163.com, sunjiajie369@126.com, zxqcc@stu.scu.edu.cn, shenxj@scu.edu.cn).}%
\thanks{Fanqin Meng is with Sichuan University of Science and Engineering Artificial Intelligence Key Laboratory, Yibin, Sichuan 644000, China (e-mail: mengfanqin2008@163.com).}%
\thanks{P. K. Varshney is with the Department of Electrical Engineering and Computer Science, Syracuse University, Syracuse, NY 13244 USA (e-mail: varshney@syr.edu).}
}

\maketitle

{\color{blue}\begin{abstract}
Multisensor track-to-track fusion for target tracking involves two primary operations: track association and estimation fusion. For estimation fusion, lossless measurement transformation of sensor measurements has been proposed for single target tracking. In this paper, we investigate track association which is a fundamental and important problem for multitarget tracking. First, since the optimal track association problem is a multi-dimensional assignment (MDA) problem, we demonstrate that MDA-based data association (with and without prior track information) using linear transformations of track measurements is lossless, and is equivalent to that using raw track measurements. Second, recent superior scalability and performance of belief propagation (BP) algorithms enable new real-time applications of multitarget tracking with resource-limited devices. Thus, we present a BP-based multisensor track association method with transformed measurements and show that it is equivalent to that with raw measurements. Third, considering communication constraints, it is more beneficial for local sensors to send in compressed data. Two analytical lossless transformations for track association are provided, and it is shown that their communication requirements from each sensor to the fusion center are less than those of fusion with raw track measurements. Numerical examples for tracking an unknown number of targets verify that track association with transformed track measurements has the same performance as that with raw measurements and requires fewer communication bandwidths.
\end{abstract}}

\begin{IEEEkeywords}
  Multitarget tracking,  track association, estimation fusion, measurement transformation.
\end{IEEEkeywords}

%
\IEEEpeerreviewmaketitle

\section{Introduction}
\IEEEPARstart{M}{ultisensor} multitarget tracking (MSMTT) is the problem of estimating the states of targets based on information provided by multiple sensors \cite{bar1990multitarget}. It originated in the military field \cite{Liggins2008_Handbook}, and now it is being applied to many nonmilitary fields. The applications of MSMTT include: surveillance \cite{blackman1999design}, autonomous driving \cite{urmson2008autonomous,patole2017automotive}, indoor
localization \cite{bartoletti2014sensor,shen2014estimating}, biomedical analytics \cite{mavska2014benchmark}, computer vision \cite{hoseinnezhad2012visual}, and robotics \cite{mullane2011random} etc.
There are three classical and popular frameworks for multitarget tracking \cite{Vo2015Multitarget}: joint
probabilistic data association (JPDA) \cite{BarShalom2009Theprobabilistic}, multiple hypotheses tracking (MHT) \cite{Reid1979An}, and random finite sets (RFS) \cite{mahler2007statistical}.

In this paper, we consider multisensor track-to-track fusion \cite{chong2000architectures}, where there is a fusion center. The sensor data are processed locally to
form sensor tracks, which are sent to the fusion center where they are fused to form system tracks. Track fusion is needed to associate the sensor tracks and generate an improved target state estimate. Therefore, multisensor track-to-track fusion for target tracking involves two main operations: estimation fusion and track association \cite{Liggins1997Distributed}.

Estimation fusion, or data fusion for estimation, is the problem of how to best utilize the information contained in multiple sets of data for the purpose of estimating a quantity--a parameter or process \cite{Li2003_Optimal}. Estimation fusion has been researched extensively and numerous results are available. Two general approaches have been used. One is the estimation approach that converts an estimation fusion problem to an estimation problem by treating available data from sensors as measurements \cite{Li2003_Optimal,bar1986effect,chang1997optimal,chen2003performance,chang2004performance,tian2009exact,chen2018new}. In \cite{bar1986effect,chang1997optimal,chen2003performance}, and the authors have proposed track-to-track fusion algorithms based on the maximum likelihood estimation (MLE)  and weighted least-squares (WLS) methods, respectively. In \cite{chang2004performance}, the authors proposed an optimal fusion algorithm based on the maximum a posteriori (MAP) formalism. In \cite{tian2009exact}, an estimation fusion algorithm was proposed, which is optimal in the sense of minimum mean-squared error (MMSE) for the Gaussian case. In \cite{Li2003_Optimal}, unified fusion rules were proposed in the sense of best linear unbiased estimate (BLUE) and WLS for all fusion architectures with arbitrary correlation. In \cite{chen2018new}, the authors proposed a  new approach to the nonlinear  fusion estimation problem. The other  estimation fusion approach is based on showing the equivalence between estimation fusion and centralized fusion \cite{Chong1979_Hierarchical,hashemipour1988decentralized,zhu2001optimality,zhu2012networked,sun2020distributed,2011Li_distributed}. The estimation fusion approaches are optimal in the sense that they are equivalent to the optimal centralized fusion. An optimal information filter fusion was proposed and discussed in \cite{Chong1979_Hierarchical,hashemipour1988decentralized}. The performance analysis for their fusion algorithm with feedback was given \cite{zhu2001optimality,zhu2012networked}. In \cite{sun2020distributed}, the author proposed  optimal linear fusion predictors and filters for systems with random parameter matrices and correlated noises. In \cite{2011Li_distributed}, by taking linear transformation of raw measurements of each sensor, two optimal fusion algorithms are proposed. Compared with existing fusion algorithms, their communication requirements from each sensor to the fusion center are equal to or less than those of the centralized and most existing fusion algorithms.

Track association refers to finding multiple tracks for the same target using different systems. Before the track state estimates can be fused, the sensor tracks have to be associated either with each other (sensor track to sensor track association) or with the system tracks (sensor track to system track association) \cite{chong2000architectures}. Track association consists of the two key steps: computing a table of association metrics and selecting the best association hypothesis, usually by some statistical algorithm or assignment algorithm. In \cite{singer1971computer}, the authors proposed the statistical algorithm by using the weighted distance test method \cite{kanyuck1970correlation}, which used the Chi-squared distribution to detect whether the two estimates belong to the same target. Bar-Shalom \cite{bar1995multitarget} provided a method to test for the weighted distance under relevant conditions that work by introducing two estimated covariance matrix cross terms. Later, in \cite{BarShalom1997Ageneralized}, the authors proposed a generalized S-D assignment algorithm for multisensor-multitarget state estimation. In \cite{Sathyan2011MDA}, the authors introduced a multi-dimensional assignment (MDA)-based data association approach with prior track information for passive multitarget tracking.
The optimal track association problem is an MDA problem. The MDA problem is NP-hard, and the optimal solution can only be obtained by a global search. The classical methods use the Lagrangian relaxation method \cite{BarShalom1992_Anewrelaxation,Poore97_LRA}, linear relaxation approach \cite{Storms2003An,Coraluppi2000_Allsource} and the sequential m-best algorithm to find the suboptimal solution of the problem, which achieve excellent results while limiting the computational cost \cite{popp2001m}. Recently, in \cite{williams2014approximate}, the authors proposed a novel scalable method for solving data association problems using belief propagation (BP) on a particular graphical model formulation. In \cite{meyer2017scalable}, the authors proposed a scalable algorithm for tracking an unknown number of targets using multiple sensors. In \cite{Win2018Message}, the authors summarized a recently proposed paradigm for scalable multitarget tracking based on the BP algorithm. Additionally, recent MSMTT algorithms based on RFS theory can be seen in \cite{fantacci2016scalable, Yi2017ditributedfusion,Vo2019Multisensor,gao2020multiobject,gostar2020centralized,van2021distributed} etc. 

To the best of our knowledge, there are many estimation fusion results for {\color{blue}single target tracking} in different scenarios (see, e.g., \cite{Chong1979_Hierarchical,hashemipour1988decentralized,zhu2001optimality,zhu2012networked,sun2020distributed}), which are optimal in the sense that they are equivalent to optimal centralized fusion with raw measurements. However, {\color{blue}the other main operation of multisensor track-to-track fusion, namely track association, which is a fundamental and important problem for multitarget tracking, the equivalence of measurement transformations has not been explored to the best of our knowledge. To solve this problem of track association, one usually needs to deal with difficult optimization problems, and it is unclear whether some measurement transformations can be selected that reduce communication bandwidths. In this paper, our goal is to analyze their equivalency in terms of track-to-track association likelihoods and reduce communication requirements by using suitable measurement transformations.
Our main contributions are as follows. 
\begin{itemize}
	\item {\it We demonstrate that the MDA-based track association method with linear transformations of track measurements is equivalent to that with raw track measurements.} The fundamental problem of multisensor track-to-track fusion for multitarget tracking is the data association problem of assigning the sensor track measurements to targets or clutter. If one knows which measurement originates from which target, then estimation techniques such as optimal information filter fusion or transformed measurement-based estimation \cite{2011Li_distributed} can be utilized to determine accurate state estimates. Since the optimal track association problem is an MDA problem, we demonstrate that MDA-based data association (with and without prior track information) using linear transformations of track measurements is lossless, i.e., it is equivalent to that using raw track measurements.
	\item {\it We derive a BP-based track association method via lossless linear transformations of track measurements.} Superior scalability and performance of BP algorithms enable new real-time applications of multitarget tracking with resource-limited devices \cite{Win2018Message}. This approach has advantages with respect to estimation accuracy, computational complexity, and implementation flexibility. Thus, we present a BP-based multisensor track association method using measurement transformations and show that they are equivalent to that using raw measurements. 
	\item {\it We provide two analytical lossless transformations for track association and analyze their communication requirements.} Considering communication constraints, it is more beneficial for local sensors to send compressed data with the goal of reducing the communication requirements. Two analytical lossless transformations for track association are provided, so that communication requirements from each sensor to the fusion center are less than that of fusion with raw track measurements. Therefore, communication-efficient transformations are suggested for different dynamic systems.
\end{itemize}
}

Numerical examples for tracking an unknown number of targets verify that the performance of fusion with transformed measurements is the same as that of fusion with raw measurements.

The rest of this paper is organized as follows. In Section \ref{sec_PF}, we formulate the problem of multisensor multitarget tracking and give preliminaries. In Sections \ref{sec_MDA} and \ref{sec_BF}, we derive the equivalence of measurement transformation and the use of raw measurements for
multisensor track association, where MDA and BP-based track association are analyzed respectively. Communication requirements are discussed in Section \ref{sec_CR}.  In Section \ref{sec_SR}, numerical examples are given. In Section \ref{sec_cl}, concluding remarks are provided.

\section{Problem Formulation}\label{sec_PF}
Let us consider sensor measurement transformations for multisensor track-to-track data association approaches:
1) MDA-based data association with or without prior track information \cite{Sathyan2011MDA,BarShalom1997Ageneralized}; 2)  a scalable BP-based data association \cite{meyer2017scalable,Win2018Message}.
We give some preliminaries as follows:

\subsection{Dynamic system}\label{Sec_2}
The linear dynamic model is given by
\begin{align}
\mathbf{x}_{k} = F_{k-1} \mathbf{x}_{k-1} + v_{k-1},
\end{align}
where $\mathbf{x}_{k} \in \mathbb{R}^{n}$ is the target state, $F_{k-1}\in \mathbb{R}^{n \times n}$ is the state transition matrix, and $v_{k-1}\in \mathbb{R}^{n}$ is the process noise, which is assumed Gaussian with zero mean and the covariance is $Q_{k-1} \in \mathbb{R}^{n \times n}$.

The measurement model is
\begin{align}\label{measure_model_1}
\mathbf{z}_{k} = H_{k}\mathbf{x}_{k} + w_{k},
\end{align}
where $H_k \in \mathbb{R}^{m\times n}$ is the measurement matrix, $\mathbf{z}_{k} \in \mathbb{R}^{m}$ and $w_{k}\in \mathbb{R}^{m}$ are the measurement vector and the measurement noise, respectively. Usually, the measurement noise is assumed Gaussian, where the covariance is $R_k \in \mathbb{R}^{m \times m}$.

\subsection{Sensor Measurement Transformation}
Let us consider a linear transformation of measurements for multisensor fusion. Compared with sending the raw measurement, each sensor sending the transformed data to the fusion center may potentially reduce communication requirements and  obtain lossless estimation \cite{2011Li_distributed}. Let $\breve{\mathbf{z}}_k := A_{k}\mathbf{z}_{k}$, 
where $A_{k}$ is a linear transformation matrix. By using (\ref{measure_model_1}), we have
\begin{align}\label{measure_type2}
\breve{\mathbf{z}}_k = \breve{H}_k\mathbf{x}_k + \breve{w}_k,
\end{align}
where
\begin{align}
&\breve{H}_k = A_{k}H_k, ~\breve{w}_{k}=A_{k}w_{k}, \\
& \breve{R}_{k}:=\mathnormal{Cov}(\breve{w}_k) = A_{k}R_k(A_{k})^{\text{T}}, \label{tran1_R_0}
\end{align}
and where $(\cdot)^{\text{T}}$ is the transpose of a matrix.


\subsection{Multisensor Multitarget Tracking}
In this subsection, we present the track association problem of multisensor multitarget tracking. {\color{blue}Let us consider a scenario, where $L$ heterogeneous sensors monitor a surveillance region of interest and send measurements at each time step to a fusion center. We assume that each sensor is equipped with a local computing unit that can compute local multitarget state estimates, and a transceiver that can transmit data to the fusion center. In this paper, we consider the case where each sensor tracks an unknown and time-varying number of targets by performing data association and track management, and transmits the data of updated tracks to the fusion center. The data of updated tracks sent by the local sensor can be raw track measurements or the estimates of targets, or the transformed track measurements. 

At time $k$, let $\mathbf{Z}_{k,l}= [(\mathbf{z}_{k,l}^{(1)})^{\text{T}},\cdots,(\mathbf{z}_{k,l}^{(M_{k,l})})^{\text{T}}]^{\text{T}}$ denote the track measurements sent by the $l$-th sensor, where $l=1,\cdots,L$ and $M_{k,l}$ is the number of measurements. On the other hand, let $\breve{\mathbf{Z}}_{k,l}= [(\breve{\mathbf{z}}_{k,l}^{(1)})^{\text{T}},\cdots,(\breve{\mathbf{z}}_{k,l}^{(M_{k,l})})^{\text{T}}]^{\text{T}}$ denote the transformed track measurements sent by the $l$-th sensor at time $k$. We define the data at the fusion center at time $k$ as $\mathbf{Z}_{k}=[\mathbf{Z}_{k,1}^{\text{T}},\cdots,\mathbf{Z}_{k,L}^{\text{T}}]^{\text{T}}$ or $\breve{\mathbf{Z}}_{k}=[\breve{\mathbf{Z}}_{k,1}^{\text{T}},\cdots,\breve{\mathbf{Z}}_{k,L}^{\text{T}}]^{\text{T}}$. Let $N_k$ denote the number of targets (possibly unknown) presented at time $k$. We use the target state to identify a target, i.e., the target $\tau$ at time $k$ is identified by its state $\mathbf{x}_{k}^{(\tau)} \in \mathbb{R}^{n}$. The stacked state at time $k$ is then denoted by $\mathbf{X}_k = [(\mathbf{x}_{k}^{(1)})^{\text{T}},\cdots,(\mathbf{x}_{k}^{(N_k)})^{\text{T}} ]^{\text{T}}$. 

Let the $(L+1)$-tuple $(\tau,i_1,\cdots,i_L)$ denote a track hypothesis formed by target $\tau$, and measurement $\mathbf{z}_{k,l}^{(i_l)}$ of the $l$-th sensor ($l=1,\cdots,L$), i.e., measurements $\mathbf{z}_{k,l}^{(i_l)}$ ($l=1,\cdots,L$) are originated from the same target $\tau$. Note that $\tau \in \{0,1,\cdots,N_{k}\}$ and $i_l \in \{0,1,\cdots,M_{k,l}\}$ in the $(L+1)$-tuple $(\tau,i_1,\cdots,i_L)$. If $\tau= 0$, the $(L+1)$-tuple means that measurements $\mathbf{z}_{k,l}^{(i_l)}$ ($l=1,\cdots,L$) are originated from clutter. If $\tau \neq 0$ and $i_l = 0$, the $(L+1)$-tuple means that target $\tau$ generates no measurements at the $l$-th sensor. Moreover, with the commonly used assumption that a target can generate at most one measurement, and a measurement can originate from at most one target, the track association problem can be formulated as an MDA problem \cite{bar1990multitarget,Liggins2008_Handbook,blackman1999design}. Before we formulate the MDA problem, the score function of the track hypothesis is given in the following subsection.}

In this paper, our goal is to analyze the equivalency  between track association with raw measurements and track association with transformed measurements in terms of the association score function. Moreover, an extension to the case of track association without prior track information, and a BP-based track association way to reduce the computational complexity are also discussed.

\section{Establishment of Equivalence for MDA Track Association}\label{sec_MDA}
In this section, we present the main results on equivalence between the  track association with raw measurements and track association with transformed measurements based on MDA with and without prior track information, respectively.

\vspace{-2mm}
\subsection{MDA Track Association with Prior Track Information }\label{sec_2_A}
At time $k$, assuming that the fusion center has maintained $N_k$ tracks and received data $\mathbf{Z}_k$ from the $L$ sensors. Let $\hat{\mathbf{x}}_{k-1|k-1}^{(\tau)}$ and $P_{k-1|k-1}^{(\tau)}$ denote the mean and the covariance of the state estimate of the track $\tau$ at time $k-1$, respectively, the prior track information at time $k$ can be represented by $\hat{\mathbf{x}}_{k|k-1}^{(\tau)}$ and $P_{k|k-1}^{(\tau)}$, where 
\begin{align}
	&\hat{\mathbf{x}}_{k|k-1}^{(\tau)} = F_{k-1} \hat{\mathbf{x}}_{k-1|k-1}^{(\tau)}, \label{MDA_pred_x} \\
	&P_{k|k-1}^{(\tau)} = F_{k-1} P_{k-1|k-1}^{(\tau)} F_{k-1}^{\text{T}} + Q_{k-1}. \label{MDA_pred_p}
\end{align}
For each track hypothesis $(\tau,i_{1},\cdots,i_{L})$, when $\tau \neq 0$, the score function $L_{(\tau,i_{1},\cdots,i_{L})}$ can be defined as a likelihood ratio~\cite{Sathyan2011MDA}, i.e., 
\begin{align}\label{score_function_prior}
	L_{(\tau,i_{1},\cdots,i_{L})} = \prod_{l=1}^L \frac{[1-P_d^{(l)}]^{1-u(i_l)}[P_d^{(l)} p(\mathbf{z}_{k,l}^{(i_l)}|\hat{\mathbf{x}}_{k|k-1}^{(\tau)})]^{u(i_l)}}{[\lambda_{f_l} p_{f_{l}}(\mathbf{z}_{k,l}^{(i_l)})]^{u(i_{l})}},
\end{align}
where $u(i_l) = 0$ if $i_l=0$, or $u(i_l) = 1$ otherwise. 
In the numerator of (\ref{score_function_prior}), $P_d^{(l)}$ is the probability of detection of the $l$-th sensor, and $p(\mathbf{z}_{k,l}^{(i_l)}|\hat{\mathbf{x}}_{k|k-1}^{(\tau)})$ is the likelihood that $\mathbf{z}_{k,l}^{(i_l)}$ originates from target $\tau$; in the denominator of (\ref{score_function_prior}), $\lambda_{f_l}$ is the mean number of clutter, and $p_{f_{l}}(\mathbf{z}_{k,l}^{(i_l)})$ is the {\it{probability density function}} (pdf) of the clutter measurement. {\color{blue}On the other hand, when $\tau= 0$, the track hypothesis $(0, i_1,\cdots,i_L)$ represents the fact that the measurements with indices $(i_1,\cdots,i_L)$ are clutter measurements, i.e., 
\begin{align}
	L_{(0,i_{1},\cdots,i_{L})} = \frac{\prod_{l=1}^L [\lambda_{f_l} p_{f_{l}}(\mathbf{z}_{k,l}^{(i_l)})]^{u(i_l)}}{\prod_{l=1}^L [\lambda_{f_l} p_{f_{l}}(\mathbf{z}_{k,l}^{(i_l)})]^{u(i_l)}}= 1.
\end{align}}

For fusion with raw measurements, the corresponding likelihood is
\begin{align}\label{likeli_centralized}
p(\mathbf{z}_{k,l}^{(i_l)} | \hat{\mathbf{x}}_{k|k-1}^{(\tau)}) =& \frac{(2\pi)^{-m/2}}{(|S_{k,l}^{(\tau)}|)^{1/2}} \exp \Bigl( -\frac{1}{2} (\mathbf{z}_{k,l}^{(i_l)}- \hat{\mathbf{z}}_{k|k-1,l}^{(\tau)})^{\text{T}} \nonumber \\
& \times  (S_{k,l}^{(\tau)})^{-1}(\mathbf{z}_{k,l}^{(i_l)}- \hat{\mathbf{z}}_{k|k-1,l}^{(\tau)}) \Bigr),
\end{align}
where $m$ is the dimension of the measurement, and the predicted measurement $\hat{\mathbf{z}}_{k|k-1,l}^{(\tau)}$ and the innovation covariance matrix $S_{k,l}^{(\tau)}$ are obtained as follows,
\begin{align}
&\hat{\mathbf{z}}_{k|k-1,l}^{(\tau)} = H_{k,l} \hat{\mathbf{x}}_{k|k-1}^{(\tau)}, \label{MDA_pred_z}\\
&S_{k,l}^{(\tau)} = H_{k,l} P_{k|k-1}^{(\tau)} H_{k,l}^{\text{T}} + R_{k,l}. \label{MDA_pred_s}
\end{align}
On the other hand, for fusion with transformed measurements, $\mathbf{z}_{k,l}^{(i_l)}$ in (\ref{score_function_prior}) is replaced by $\breve{\mathbf{z}}_{k,l}^{(i_l)}$, and a generalized likelihood \cite{Rao1973Linear} is applied:
\begin{align}\label{likeli_distributed}
p(\breve{\mathbf{z}}_{k,l}^{(i_l)} | \hat{\mathbf{x}}_{k|k-1}^{(\tau)}) =& \frac{(2\pi)^{-m/2}}{(\prod_{i=1}^{m}e_i)^{1/2} } \exp \Bigl(-\frac{1}{2} (\breve{\mathbf{z}}_{k,l}^{(i_l)}- \breve{\mathbf{z}}_{k|k-1,l}^{(\tau)})^{\text{T}} \nonumber\\
& \times  (\breve{S}_{k,l}^{(\tau)})^{\dagger} (\breve{\mathbf{z}}_{k,l}^{(i_l)}- \breve{\mathbf{z}}_{k|k-1,l}^{(\tau)}) \Bigr),%
\end{align}
where $\breve{\mathbf{z}}_{k|k-1,l}^{(\tau)} = A_{k,l}\hat{\mathbf{z}}_{k|k-1,l}^{(\tau)}$ and $\breve{S}_{k,l}^{(\tau)} =A_{k,l} S_{k,l}^{(\tau)} A_{k,l}^{\text{T}}$; $A_{k,l}$ represents the linear transformation of the $l$-th sensor at time $k$; $(\cdot)^{\dagger}$ is the Moore-Penrose pseudo inverse; $e_{i}$, $ i=1,\cdots,m$ are the nonzero eigenvalues of $\breve{S}_{k,l}^{(\tau)} $. 

By defining the cost of each track hypothesis $(\tau,i_1,\cdots,i_L)$: $C_{(\tau,i_1,\cdots,i_L)}=-\log L_{(\tau,i_1,\cdots,i_L)}$, 
the track-measurement association problem between the existing tracks and the measurements of the $L$ sensors is formulated as an MDA problem~\cite{Sathyan2011MDA}:
\begin{align}\label{LIP}
&\min_{\delta_{(\tau,i_1,\cdots,i_L)}}~\sum_{\tau}\sum_{\bm{i}}C_{(\tau,i_1,\cdots,i_L)}\delta_{(\tau,i_1,\cdots,i_L)} \notag
\\
&\text{s.t.} \nonumber \\
&\sum_{\bm{i}}\delta_{(\tau,i_1,\cdots,i_L)}=1,~\tau=1,\dots,N_{k}, \notag
\\
&\sum_{\tau}\sum_{\bm{i}\setminus \{i_l\}}\delta_{(\tau,i_1,\cdots,i_L)}=1,\notag
\\
&\text{for }~i_{l}=1,\dots,M_{k,l} ~\text{ and }~ l=1,\dots,L, \notag
\\
&\delta_{(\tau,i_1,\cdots,i_L)}\in \{0,1\}~ \text{ for all } ~(\tau,i_1,\cdots,i_L).
\end{align}
Note that $\delta_{(\tau,i_1,\cdots,i_L)} = 1$ means that the measurements with indices $(i_1,\cdots,i_L)$ originate from the track $\tau$. The constraints in (\ref{LIP}) mean that each measurement from the local sensors can only correspond to one track, and a track can only match with one measurement at the same sensor.

Here, we derive the relationship between data association with transformed measurements and that with raw measurements under some regularity conditions, which is summarized in the following Proposition \ref{equivalence_score}.

\begin{Proposition}\label{equivalence_score}
	Let $A_{k,l}$ be a full column rank matrix, the processed data $\breve{\mathbf{z}}_{k,l}^{(i_{l})}$ be a linear transformation of $\mathbf{z}_{k,l}^{(i_{l})}$, i.e., $\breve{\mathbf{z}}_{k,l}^{(i_{l})} = A_{k,l} \mathbf{z}_{k,l}^{(i_{l})}$, and the clutter is uniform in the region of interest. Then, the score function of the track association problem with raw measurements is equal to that of the problem with transformed measurements, i.e., $L_{(\tau,i_1,\cdots,i_L)}^c = L_{(\tau,i_1,\cdots,i_L)}^d$, where
	\begin{align}\label{score_centralized}
	L_{(\tau,i_1,\cdots,i_L)}^c = \prod_{l=1}^L \frac{[1-P_d^{(l)}]^{1-u(i_l)}[P_d^{(l)} p(\mathbf{z}_{k,l}^{(i_{l})}|\hat{\mathbf{x}}_{k|k-1}^{(\tau)})]^{u(i_l)}}{[\lambda_{f_l} p_{f_{l}}(\mathbf{z}_{k,l}^{(i_{l})})]^{u(i_{l})}},
	\end{align}
	and
	\begin{align}\label{score_distributed}
	L_{(\tau,i_1,\cdots,i_L)}^d = \prod_{l=1}^L \frac{[1-P_d^{(l)}]^{1-u(i_l)}[P_d^{(l)} p(\breve{\mathbf{z}}_{k,l}^{(i_{l})}|\hat{\mathbf{x}}_{k|k-1}^{(\tau)})]^{u(i_l)}}{[\lambda_{f_l} p_{f_{l}}(\breve{\mathbf{z}}_{k,l}^{(i_{l})})]^{u(i_{l})}}.
	\end{align}
	Moreover, the MDA  problem (\ref{LIP}) for track association with raw measurements is equivalent to that with transformed measurements.		
\end{Proposition}

The proof of Proposition \ref{equivalence_score} is in \cite[app \ref{proof_prop1}]{Liu2023Oncommunication}. The assumption that $A_{k,l}$ is full column rank is not too stringent. We will provide some transformation matrices with full column rank which can reduce communication requirements in Section \ref{sec_CR}.

\subsection{MDA Track Association without Prior Track Information} \label{sen_2_b}

In this subsection, we formulate MDA track association without prior track information. There are some scenarios where the prior track information at the fusion center is not available, such as sensor track to sensor track fusion (but not to system track fusion), track initialization \cite{BarShalom1997Ageneralized,chong2000architectures}, etc. In this case, the score function for each hypothesis $(i_1,\cdots,i_L)$ is defined as the following generalized likelihood ratio \cite{BarShalom1997Ageneralized}, i.e., 
\begin{align}\label{score_function}
	L_{(i_{1},\cdots,i_{L})} = \prod_{l=1}^L \frac{[1-P_d^{(l)}]^{1-u(i_l)}[P_d^{(l)} p(\mathbf{z}_{k,l}^{(i_l)}|\hat{\mathbf{x}}_{k,\text{ML}}^{(\tau)})]^{u(i_l)}}{[\lambda_{f_l} p_{f_{l}}(\mathbf{z}_{k,l}^{(i_l)})]^{u(i_{l})}},
\end{align}
where $\hat{\mathbf{x}}_{k,\text{ML}}^{(\tau)}$ is the MLE of the target state $\mathbf{x}_{k}^{(\tau)}$: 
\begin{align}\label{Ml_EST}
	\hat{\mathbf{x}}_{k,\text{ML}}^{(\tau)}=\argmax_{\mathbf{x}_{k}^{(\tau)}} \prod_{l \in \{l|u(i_{l})=1\}}p(\mathbf{z}_{k,l}^{(i_{l})}|\mathbf{x}_{k}^{(\tau)}).
\end{align} 

For fusion with raw measurements, the conditional pdf $p(\mathbf{z}_{k,l}^{(i_l)}|\hat{\mathbf{x}}_{k,\text{ML}}^{(\tau)})$ has the form as follows,
\begin{align}\label{likeli_meas_init}
&p(\mathbf{z}_{k,l}^{(i_{l})}|\hat{\mathbf{x}}_{k,\text{ML}}^{(\tau)}) = \frac{(2\pi)^{-m/2}}{(|R_{k,l}|)^{1/2}}\notag\\
& \times \exp \Bigl(-\frac{1}{2}(\mathbf{z}_{k,l}^{(i_{l})}- H_{k,l}\hat{\mathbf{x}}_{k,\text{ML}}^{(\tau)})^{\text{T}} R_{k,l}^{-1}(\mathbf{z}_{k,l}^{(i_{l})}- H_{k,l}\hat{\mathbf{x}}_{k,\text{ML}}^{(\tau)}) \Bigr).
\end{align}
On the other hand, for fusion with transformed measurements, the conditional pdf of the likelihood of transformed measurements is,
\begin{align}
&p(\breve{\mathbf{z}}_{k,l}^{(i_{l})}|\hat{\mathbf{x}}_{k,\text{ML}}^{(\tau)}) = \frac{(2\pi)^{-m/2}}{(\prod_{i=1}^{m}e_i)^{1/2} } \nonumber\\
& \times \exp \Bigl(-\frac{1}{2} [\breve{\mathbf{z}}_{k,l}^{(i_{l})}-f_l(\hat{\mathbf{x}}_{k,\text{ML}}^{(\tau)}) ]^{\text{T}} [g_l(R_{k,l}) ]^{\dagger} [\breve{\mathbf{z}}_{k,l}^{(i_{l})}-f_l(\hat{\mathbf{x}}_{k,\text{ML}}^{(\tau)}) ] \Bigr),%
\end{align}
where the linear transformation $f_l(\hat{\mathbf{x}}_{k,\text{ML}}^{(\tau)}) =A_{k,l}H_{k,l}\hat{\mathbf{x}}_{k,\text{ML}}^{(\tau)}$, and the transformed covariance $g_l(R_{k,l}) =A_{k,l}R_{k,l}^{-1} A_{k,l}^{\text{T}}$.

Define the cost of the candidate association $(i_1,\cdots,i_L)$ as $C_{(i_1,\cdots,i_L)} = -\log L_{(i_{1},\cdots,i_{L})}$, the problem of track association without prior track information can be formulated as an $L$-D assignment problem:
\begin{align}\label{LIP_noinit}
	&\min_{\delta_{(i_1,\cdots,i_L)}}~\sum_{\bm{i}}C_{(i_1,\cdots,i_L)}\delta_{(i_1,\cdots,i_L)} \notag
	\\
	&\text{s.t.}\notag
	\\
	&\sum_{\bm{i} \setminus \{i_l\}}\delta_{(i_1,\cdots,i_L)}=1,\notag
	\\
	&\text{for }~i_{l}=1,\dots,M_{k,l} ~\text{ and } l=1,\dots,L, \notag
	\\
	&\delta_{(i_1,\cdots,i_L)}\in \{0,1\}~ \text{ for all } (i_1,\cdots,i_L).
\end{align}
Note that $\delta_{(i_1,\cdots,i_L)} = 1$ means that the measurements with indices $(i_1,\cdots,i_L)$ originate from the same target, i.e., they can be used to initialized a new track. The constraints in (\ref{LIP_noinit}) mean that each measurement from the local sensors can only correspond to one target.
	
For MDA track association without prior track information, we have the following equivalency result.
\begin{Corollary}\label{corro_nopiror}
Under the conditions of Proposition \ref{equivalence_score}, the score function of fusion with raw measurements is equal to that of fusion with transformed measurements, i.e., $L_{(i_{1},\cdots,i_{L})}^{c}=L_{(i_{1},\cdots,i_{L})}^{d}$, 
where
\begin{align}\label{score_centralized_nopior}
	L_{(i_1,\cdots,i_L)}^c = \prod_{l=1}^L \frac{[1-P_d^{(l)}]^{1-u(i_l)}[P_d^{(l)} p(\mathbf{z}_{k,l}^{(i_{l})}|\hat{\mathbf{x}}_{k,\text{ML}}^{(\tau),c})]^{u(i_l)}}{[\lambda_{f_l} p_{f_{l}}(\mathbf{z}_{k,l}^{(i_{l})})]^{u(i_{l})}},
\end{align}
and
	\begin{align}\label{score_distributed_nopior}
	L_{(i_1,\cdots,i_L)}^d = \prod_{l=1}^L \frac{[1-P_d^{(l)}]^{1-u(i_l)}[P_d^{(l)} p(\breve{\mathbf{z}}_{k,l}^{(i_{l})}|\hat{\mathbf{x}}_{k,\text{ML}}^{(\tau),d})]^{u(i_l)}}{[\lambda_{f_l} p_{f_{l}}(\breve{\mathbf{z}}_{k,l}^{(i_{l})})]^{u(i_{l})}}.
	\end{align}
Moreover, the MDA problem (\ref{LIP_noinit})  for track association  with raw measurements is equivalent to that for track association  with transformed measurements.
\end{Corollary}

The proof of Corollary \ref{corro_nopiror} can be found in \cite[app \ref{proof_coro_1}]{Liu2023Oncommunication}.

\subsection{Summary of MDA Track Association}\label{sec_c}
Combining MDA Track association method with prior information with the estimation fusion method in \cite{2011Li_distributed}, the complete multisensor track-to-track fusion algorithm with transformed measurements is summarized in \cite[app \ref{app_algorithm_1}]{Liu2023Oncommunication}, which is equivalent to that with raw measurements.

The MDA track association with or without prior information requires the solution of an $(L+1)$-D or $L$-D assignment problem, respectively. The $L$-D assignment problem is an NP-hard problem for $L>2$. Although it is NP-hard, the original or primal $L$-D assignment problem can be relaxed, via successive constraint relaxation, to a two-dimensional (2-D) subproblem, which is optimally solvable at each iteration in polynomial time $\mathcal{O}(C N^{3})$, where $N$ is the number of tracks or sensor measurements, and $C$ is the range of the of the cost coefficient. Moreover, the worst case complexity of the relaxed $L$-D assignment algorithm in \cite{BarShalom1997Ageneralized} is  $\mathcal{O}((L-1)C N^{3})$.
Furthermore, to reduce computation complexity, we consider the scalable track association using the BP method \cite{williams2014approximate,Win2018Message}.

\section{Equivalence of BP-Based Track Association and Fusion}\label{sec_BF}
In this section, we analyze the equivalency between BP-based track association  with raw measurements and that with transformed measurements for tracking an unknown, time-varying number of targets \cite{williams2014approximate,Win2018Message}.

\subsection{The BP-based MSMTT algorithm}\label{facot_graph}
At time $k$, the fusion center receives data from $L$ sensors, which are sequentially processed for the $l$-th sensor, where $l=1,\cdots,L$. A target is either a newborn one or a target established in the past and survived to the present. The states of the potential targets for the $l$-th sensor are represented by
$\overline{\mathbf{X}}_{k,l} = [(\overline{\mathbf{x}}_{k,l}^{(1)})^{\text{T}},\cdots,(\overline{\mathbf{x}}_{k,l}^{(M_{k,l})})^{\text{T}}]^{\text{T}}$, where $M_{k,l}$ is the number of measurements. The states of the survived targets up to receiving the data of the $l$-th sensor at the fusion center are $\underline{\mathbf{X}}_{k,l} = [(\underline{\mathbf{x}}_{k,l}^{(1)})^{\text{T}},\cdots,(\underline{\mathbf{x}}_{k,l}^{(N_{k,l})})^{\text{T}}]^{\text{T}}$, where the number $N_{k,l}$ of the survived targets that are updated by using data of the $1$-st sensor to the $(l-1)$-th sensor. Meanwhile, the 0-1 variable $\overline{r}_{k,l}^{(i_{l})}=1$ represents that the measurement $i_{l}$ generated by a new target and $\underline{r}_{k,l}^{(\tau)}=1$ means that the survived target $\tau$ exists up to the $l$-th sensor at time $k$. We define that $\overline{r}_{k,l} = [\overline{r}_{k,l}^{(1)},\cdots,\overline{r}_{k,l}^{(M_{k,l})}]^{\text{T}}$ for the new targets, and $\underline{r}_{k,l} = [\underline{r}_{k,l}^{(1)},\cdots,\underline{r}_{k,l}^{(N_{k,l})}]^{\text{T}}$ for the survived targets. After $L$ iterations, the states of the survived targets at the fusion center are also denoted as $\tilde{f}(\mathbf{x}_{k}^{(\tau)},r_{k}^{(\tau)})$, $\tau= 1,\cdots,N_k$, where
\begin{equation}\label{bp_prior}
	\setlength{\nulldelimiterspace}{0pt}
	\tilde{f}(\mathbf{x}_{k}^{(\tau)},r_{k}^{(\tau)}) = \left\{
	\begin{IEEEeqnarraybox}[][c]{ls}
		\tilde{f}(\underline{\mathbf{x}}_{k,L}^{(\tau)},\underline{r}_{k,L}^{(\tau)}) ,\quad & $\tau \leq N_{k,L}$  \\
		\\
		\tilde{f}(\overline{\mathbf{x}}_{k,L}^{(i_L)},\overline{r}_{k,L}^{(i_L)}), & $i_{L}= \tau - N_{k,L} $,%
	\end{IEEEeqnarraybox}\right.
\end{equation}
and $N_k = N_{k,L}+M_{k,L}$.

Let $a_{k,l} = [a_{k,l}^{(1)},\cdots,a_{k,l}^{(N_{k,l})}]^{\text{T}}$ denote the (unknown) data association variable vector at time $k$, where $a_{k,l}^{(\tau)} = i_l \in \{ 1,\cdots,M_{k,l} \}$ if target $\tau$ generates a measurement $i_l$ at the $l$-th sensor and $a_{k,l}^{(\tau)} = 0 $ if target $\tau$ does not generate a measurement at the $l$-th sensor. On the other hand, an alternative association vector $b_{k,l} = [b_{k,l}^{(1)},\cdots,b_{k,l}^{(M_{k,l})}]^{\text{T}}$ is introduced for the $l$-th sensor, where $b_{k,l}^{(i_l)} = \tau \in \{ 1,\cdots,N_{k,l} \} $ if measurement $i_l$ originates from target $\tau$ and $b_{k,l}^{(i_l)} = 0 $ if measurement $i_l$ is a clutter measurement. 
The constraints for data association, i.e., at time $k$, one target can only generate at most one measurement at each sensor, and one measurement can only originate from one target or clutter, can be represented by an indicator function:
\begin{align}
	\psi(a_{k,l},b_{k,l}) = \prod_{\tau = 1}^{N_{k,l}} \prod_{i_l = 1}^{M_{k,l}} \psi(a_{k,l}^{(\tau)},b_{k,l}^{(i_l)}),
\end{align}
where $\psi(a_{k,l}^{(\tau)},b_{k,l}^{(i_l)}) = 0$ if $a_{k,l}^{(\tau)} = i_l$ and $b_{k,l}^{(i_l)}\neq \tau$ or $b_{k,l}^{(i_l)} = \tau$ and $a_{k,l}^{(\tau)}\neq i_l$, and $\psi(a_{k,l}^{(\tau)},b_{k,l}^{(i_l)}) = 1$ otherwise. 

At time $k$, when processing the $l$-th sensor, we assume that the beliefs $\tilde{f}_{l-1}(\underline{\mathbf{x}}_{k,l-1}^{(\tau)},\underline{r}_{k,l-1}^{(\tau)})$ and $\tilde{f}_{l-1}(\overline{\mathbf{x}}_{k,l-1}^{(i_{l-1})},\overline{r}_{k,l-1}^{(i_{l-1})})$ for both survived targets and new targets are calculated up to the $(l-1)$-th sensor. Main steps of the BP data association and fusion update algorithm for the measurements of the $l$-th sensor are as follows \cite{Win2018Message}:
\subsubsection{Initialization}
For $l=1$, the beliefs $\tilde{f}_0(\underline{\mathbf{x}}_{k,1}^{(\tau)},\underline{r}_{k,1}^{(\tau)})$ are initialized by $\alpha(\underline{\mathbf{x}}_{k}^{(\tau)},\underline{r}_{k}^{(\tau)})$, i.e., $\tilde{f}_0(\underline{\mathbf{x}}_{k,1}^{(\tau)},\underline{r}_{k,1}^{(\tau)}) = \alpha(\underline{\mathbf{x}}_{k}^{(\tau)},\underline{r}_{k}^{(\tau)})$, 
where
\begin{align}\label{bp_init_1}
	\nonumber
	\alpha(\underline{\mathbf{x}}_{k}^{(\tau)},\underline{r}_{k}^{(\tau)}) =\sum_{\underline{r}_{k}^{(\tau)}\in \{0,1\}} \int & f(\underline{\mathbf{x}}_{k}^{(\tau)},\underline{r}_{k}^{(\tau)}| \mathbf{x}_{k-1}^{(\tau)},r_{k-1}^{(\tau)})\\
	&\times \tilde{f}(\mathbf{x}_{k-1}^{(\tau)},r_{k-1}^{(\tau)}) d \mathbf{x}_{k-1}^{(\tau)}.
\end{align}
Here, $\alpha(\underline{\mathbf{x}}_{k}^{(\tau)},\underline{r}_{k}^{(\tau)})$ is obtained by performing prediction from time $k-1$ to time $k$, $f(\underline{\mathbf{x}}_{k}^{(\tau)},\underline{r}_{k}^{(\tau)}|\mathbf{x}_{k-1}^{(\tau)},r_{k-1}^{(\tau)})$ is the single-target augmented state-transition pdf \cite{Win2018Message}, and $\tilde{f}(\mathbf{x}_{k-1}^{(\tau)},r_{k-1}^{(\tau)})$ is the belief calculated at time $k-1$.

For $l > 1$, the beliefs $\tilde{f}_{l-1}(\underline{\mathbf{x}}_{k,l}^{(\tau)},\underline{r}_{k,l}^{(\tau)})$ are initialized by,
\begin{align}\label{bp_init_2}
	\setlength{\nulldelimiterspace}{0pt}
	\tilde{f}_{l-1} & (\underline{\mathbf{x}}_{k,l}^{(\tau)},\underline{r}_{k,l}^{(\tau)}) \nonumber\\
	&= \left\{
	\begin{IEEEeqnarraybox}[][c]{ls}
		\tilde{f}_{l-1}(\underline{\mathbf{x}}_{k,l-1}^{(\tau)},\underline{r}_{k,l-1}^{(\tau)}) ,\quad & $\tau \leq N_{k,l-1}$  \\
		\\
		\tilde{f}_{l-1}(\overline{\mathbf{x}}_{k,l-1}^{(i_{l-1})},\overline{r}_{k,l-1}^{(i_{l-1})}), & $i_{l-1}= \tau - N_{k,l-1} $,%
	\end{IEEEeqnarraybox}\right.
\end{align}
where $\tau = 1,\cdots,N_{k,l}$, and $N_{k,l}= N_{k,l-1} + M_{k,l-1}$.

\subsubsection{Measurement evaluation}
For the survived targets,
\begin{align}\label{bp_2_meas}
	\beta(a_{k,l}^{(\tau)}) = \sum_{\underline{r}_k^{\tau}\in \{0,1\}} \int & q(\underline{\mathbf{x}}_{k,l}^{(\tau)},\underline{r}_{k,l}^{(\tau)}, a_{k,l}^{(\tau)} ; \mathbf{Z}_{k,l}) \nonumber \\
	&\qquad \times \tilde{f}_{l-1}(\underline{\mathbf{x}}_{k,l}^{(\tau)},\underline{r}_{k,l}^{(\tau)}) d \underline{\mathbf{x}}_{k,l}^{(\tau)},
\end{align}
where $q(\underline{\mathbf{x}}_{k,l}^{(\tau)},\underline{r}_{k,l}^{(\tau)}, a_{k,l}^{(\tau)} ; \mathbf{Z}_{k,l})$ is defined as follows,
\begin{align}
	\setlength{\nulldelimiterspace}{0pt}
	q(&\underline{\mathbf{x}}_{k,l}^{(\tau)},1, a_{k,l}^{(\tau)} ; \mathbf{Z}_{k,l})
	\nonumber \\
	&=\left\{
	\begin{IEEEeqnarraybox}[][c]{ls}
	\frac{P_d^{(l)}(\underline{\mathbf{x}}_{k,l}^{(\tau)}) p(\mathbf{z}_{k,l}^{(i_{l})}|\underline{\mathbf{x}}_{k,l}^{(\tau)})}{\lambda_{f_l} p_{f_l}(\mathbf{z}_{k,l}^{(i_{l})})},\quad & if $a_{k,l}^{(\tau)}\in \{1,\cdots,M_{k,l}\}$\\
	1-P_{d}^{(l)}(\underline{\mathbf{x}}_{k,l}^{(\tau)}),\quad & if $a_{k,l}^{(\tau)}= 0$.
	\end{IEEEeqnarraybox}\right. \label{bp_factor_q_1}
	\\
	q(&\underline{\mathbf{x}}_{k,l}^{(\tau)},0, a_{k,l}^{(\tau)} ; \mathbf{Z}_{k,l})
		= \mathbf{1}(a_{k,l}^{(\tau)}). \label{bp_q_x_r0}
\end{align}
Here, $P_{d}^{(l)}(\underline{\mathbf{x}}_{k,l}^{(\tau)})$ is the detection probability that the target $\tau$ is detected by the $l$-th sensor. $\mathbf{1}(a_{k,l}^{(\tau)}) = 0$ when $a_{k,l}^{(\tau)}\in \{1,\cdots,M_{k,l}\}$, and $\mathbf{1}(a_{k,l}^{(\tau)}) = 1$ when $a_{k,l}^{(\tau)}=0$. {\color{blue}Note that in the settings of limited field-of-view sensors, we assume that the fusion center knows the field-of-view information. When the target $\underline{\mathbf{x}}_{k,l}^{(\tau)}$ is not within the observation area of the sensor, we modify the probability so that $P_{d}^{(l)}(\underline{\mathbf{x}}_{k,l}^{(\tau)}) = 0$. }

For the new targets,
\begin{align}\label{bp_mess_xi}
	\xi(b_{k}^{(i_l)}) = \sum_{\overline{r}_{k,l}^{(i_l)} \in \{0,1\}} \int v(\overline{\mathbf{x}}_{k,l}^{(i_{l})},\overline{r}_{k,l}^{(i_{l})},b_{k,l}^{(i_{l})};\mathbf{z}_{k,l}^{(i_{l})}) d \overline{\mathbf{x}}_{k,l}^{(i_{l})},
\end{align}
where $v(\overline{\mathbf{x}}_{k,l}^{(i_{l})},\overline{r}_{k,l}^{(i_{l})},b_{k,l}^{(i_{l})};\mathbf{z}_{k,l}^{(i_{l})})$ is defined as follows,
\begin{align}
	\setlength{\nulldelimiterspace}{0pt}
	v(&\overline{\mathbf{x}}_{k,l}^{(i_{l})},1,b_{k,l}^{(i_{l})};\mathbf{z}_{k,l}^{(i_{l})})
	\notag\\
	&=\left\{
	\begin{IEEEeqnarraybox}[][c]{ls}
	\frac{\lambda_{n_l} f_{n}(\overline{\mathbf{x}}_{k,l}^{(i_{l})})p(\mathbf{z}_{k,l}^{(i_{l})}|\overline{\mathbf{x}}_{k,l}^{(i_{l})})}{\lambda_{f_l} p_{f_l}(\mathbf{z}_{k,l}^{(i_{l})})},\quad & if $b_{k,l}^{(i_{l})}=0$\\
	0,\quad & if $b_{k,l}^{(i_{l})}\in \{1,\cdots,N_{k,l}\}$,
	\end{IEEEeqnarraybox}\right. \label{bp_factor_v_1}
	\\
	v(&\overline{\mathbf{x}}_{k,l}^{(i_{l})},0,b_{k,l}^{(i_{l})};\mathbf{z}_{k,l}^{(i_{l})})=f_{D}(\overline{\mathbf{x}}_{k,l}^{(i_{l})}). \label{bp_factor_v_2}
\end{align}
Here, $\lambda_{n_l}$ is the mean number of new targets, and $f_{D}(\overline{\mathbf{x}}_{k,l}^{(i_{l})})$ represents a dummy pdf, which means that the target $i_l$ does not exist.

\subsubsection{Iterative data association}
For the data of the $l$-th sensor, at each iteration $p\in \{1,\cdots,P\}$, the following recursions are excuted for all measurements $i_l\in I_l$:
\begin{align}\label{ite_1_v}
	\nu_{i_l \rightarrow \tau}^{(p)}(a_{k,l}^{(\tau)}) = \sum_{b_{k,l}^{(i_l)}=0}^{N_{k,l}} \xi(b_{k,l}^{(i_l)}) \psi(a_{k,l}^{(\tau)},b_{k,l}^{(i_l)}) \prod_{ \genfrac{}{}{0pt}{2}{\tau' = 1}{\tau'\neq \tau} }^{N_{k,l}} \varphi_{\tau' \rightarrow i_l}^{(p-1)}(b_{k,l}^{(i_l)}),
\end{align}
and (when $p\neq P$)
\begin{align}\label{ite_2_kesai}
	\varphi_{\tau \rightarrow i_l}^{(p)}(b_{k,l}^{(i_l)})= \sum_{a_{k,l}^{(\tau)}=0}^{M_{k,l}} \beta(a_{k,l}^{(\tau)}) \psi(a_{k,l}^{(\tau)},b_{k,l}^{(i_l)}) \prod_{ \genfrac{}{}{0pt}{2}{i'_l= 1}{i'_l \neq i_l} }^{M_{k,l}} \nu_{i'_l \rightarrow \tau}^{(p)}(a_{k,l}^{(\tau)}).
\end{align}
For the initialization, i.e., $p = 0$,
\begin{align}\label{ite_init}
	\zeta_{\tau \rightarrow i_l}^{(0)}(b_{k,l}^{(i_l)}) = \sum_{a_{k,l}^{(\tau)}=0}^{M_{k,l}} \beta(a_{k,l}^{(\tau)}) \psi(a_{k,l}^{(\tau)},b_{k,l}^{(i_l)}).
\end{align}
After the last iteration $p=P$ is executed, we multiply the messages $ \nu_{i_l \rightarrow \tau}^{(p)}(a_{k,l}^{(\tau)})$ for $i_l = 1,\cdots,M_{k,l}$,
\begin{align}\label{ite_end}
	\kappa(a_{k,l}^{(\tau)}) = \prod_{i_l=1}^{M_{k,l}} \nu_{i_l \rightarrow \tau}^{(P)}(a_{k,l}^{(\tau)}),
\end{align}
and multiply the messages $\varphi_{\tau \rightarrow i_l}^{(p)}(b_{k,l}^{(i_l)})$ for $\tau = 1,\cdots,N_{k,l}$,
\begin{align}\label{bp_prob_b}
	\iota (b_{k,l}^{(i_l)}) = \prod_{\tau=1}^{N_{k,l}} \varphi_{\tau \rightarrow i_l}^{(P)}(b_{k,l}^{(i_l)}).
\end{align}

\subsubsection{Measurement update}
For the survived targets,
\begin{align}\label{bp_meas_up}
	\gamma(\underline{\mathbf{x}}_{k,l}^{(\tau)},1) &= \sum_{a_{k,l}^{(\tau)} = 0}^{M_{k,l}} q(\underline{\mathbf{x}}_{k,l}^{(\tau)},1, a_{k,l}^{(\tau)} ; \mathbf{Z}_{k,l})  \kappa(a_{k,l}^{(\tau)}),\\
	\gamma(\underline{\mathbf{x}}_{k,l}^{(\tau)},0) &= \kappa(a_k^{(\tau)}=0).\label{bp_meas_up_2}
\end{align}
For the new targets,
\begin{align}
	\varsigma  (\overline{\mathbf{x}}_{k,l}^{(i_{l})},1) &= v(\overline{\mathbf{x}}_{k,l}^{(i_{l})},1,b_{k,l}^{(i_{l})} = 0;\mathbf{z}_{k,l}^{(i_{l})}) \iota (0), \label{bp_meas_up_3} \\
	\varsigma  (\overline{\mathbf{x}}_{k,l}^{(i_{l})},0) &= \sum_{b_{k,l}^{(i_l)}=0}^{N_{k,l}} \iota (b_{k,l}^{(i_l)}) f_D(\overline{\mathbf{x}}_{k,l}^{(i_{l})}). \label{bp_meas_up_4}
\end{align}

\subsubsection{Belief Calculation}
For the survived targets, the fusion beliefs are calculated by 
\begin{align}\label{bp_appx}
	\tilde{f}_{l}(\underline{\mathbf{x}}_{k,l}^{(\tau)},1) &= \frac{1}{\underline{C}_k^{\tau}} \tilde{f}_{l-1}(\underline{\mathbf{x}}_{k,l}^{(\tau)},1) \gamma^{(\tau)}(\underline{\mathbf{x}}_{k,l}^{(\tau)},1),\\
	\label{bp_appx_2}
	\tilde{f}_{l}(\underline{\mathbf{x}}_{k,l}^{(\tau)},0) &= \frac{1}{\underline{C}_k^{\tau}} \tilde{f}_{l-1}(\underline{\mathbf{x}}_{k,l}^{(\tau)},0) \gamma^{(\tau)}(\underline{\mathbf{x}}_{k,l}^{(\tau)},0),
\end{align}
where the constant $\underline{C}_k^{\tau} = \int \tilde{f}_{l-1}(\underline{\mathbf{x}}_{k,l}^{(\tau)},1) \gamma^{(\tau)}(\underline{\mathbf{x}}_{k,l}^{(\tau)},1) d \underline{\mathbf{x}}_{k,l}^{(\tau)} + \tilde{f}_{l-1}(\underline{\mathbf{x}}_{k,l}^{(\tau)},0) \gamma^{(\tau)}(\underline{\mathbf{x}}_{k,l}^{(\tau)},0)$. 
For the new targets, the beliefs are calculated by
\begin{align}
	\label{bp_belief_new_1}
	\tilde{f}_l (\overline{\mathbf{x}}_{k,l}^{(i_{l})},1) &= \frac{1}{\overline{C}_k^{i_l}} \varsigma  (\overline{\mathbf{x}}_{k,l}^{(i_{l})},1)\\
	\label{bp_belief_new_2}
	\tilde{f}_l (\overline{\mathbf{x}}_{k,l}^{(i_{l})},0) &= \frac{1}{\overline{C}_k^{i_l}} \varsigma  (\overline{\mathbf{x}}_{k,l}^{(i_{l})},0),
\end{align}
where $\overline{C}_k^{i_l} := \int \varsigma  (\overline{\mathbf{x}}_{k,l}^{(i_{l})},1) d \overline{\mathbf{x}}_{k,l}^{(i_{l})} + \varsigma  (\overline{\mathbf{x}}_{k,l}^{(i_{l})},0)$.

\subsubsection{Target Declaration, State Estimation, and Pruning}
For survived targets and new targets, we use their existence beliefs $\tilde{p}(\underline{r}_{k,l}^{(\tau)}=1)$ and $\tilde{p}(\overline{r}_{k,l}^{(i_l)}=1)$ to declare whether the targets exist. {\color{blue}Here, the existence beliefs are calculated as follows. 
\begin{align}
	\tilde{p}(\underline{r}_{k,l}^{\tau}= 1)= \int \tilde{f}_{l}(\underline{\mathbf{x}}_{k,l}^{\tau},1)  d \underline{\mathbf{x}}_{k,l}^{\tau},\\
	\tilde{p}(\overline{r}_{k,l}^{i_l}= 1)= \int \tilde{f}_{l}(\overline{\mathbf{x}}_{k,l}^{i_l},1)  d \overline{\mathbf{x}}_{k,l}^{i_l}.
\end{align}
}
Given an appropriate threshold $P_{th}$ \cite{Win2018Message}, the survived target or new target is declared to exist if $\tilde{p}(\underline{r}_{k,l}^{(\tau)}= 1) > P_{th}$ or $\tilde{p}(\overline{r}_{k,l}^{(i_l)}= 1) > P_{th}$.

State estimation is performed by
\begin{align}\label{bp_est_1}
	\hat{\underline{\mathbf{x}}}_{k,l}^{\tau,\text{MMSE}} = \int \underline{\mathbf{x}}_{k,l}^{(\tau)} \tilde{f}_{l}(\underline{\mathbf{x}}_{k,l}^{(\tau)},1)/ \tilde{p}(\underline{r}_{k,l}^{(\tau)} = 1) d \underline{\mathbf{x}}_{k,l}^{(\tau)},
\end{align}
for survived targets, and
\begin{align}\label{bp_est_2}
	\hat{\overline{\mathbf{x}}}_{k,l}^{i_l,\text{MMSE}} = \int \overline{\mathbf{x}}_{k,l}^{(i_l)} \tilde{f}_{l}(\overline{\mathbf{x}}_{k,l}^{(i_l)},1)/ \tilde{p}(\overline{r}_{k,l}^{(i_l)} = 1) d \overline{\mathbf{x}}_{k,l}^{(i_l)},
\end{align}
for new targets.

Finally, similar to the target declaration, given appropriate thresholds $P_{pr}$ and $N_{pr}$, survived and new targets are removed when their existence beliefs $\tilde{p}(\underline{r}_{k,l}^{(\tau)}=1)$ and $\tilde{p}(\overline{r}_{k,l}^{(i_l)}=1)$ are below~$P_{pr}$ or they lose measurements more than $N_{pr}$ scans.

\subsection{Equivalency}\label{equ_bp}
Here, we derive the relationship between BP track association with transformed measurements and that with raw measurements in the following Proposition \ref{bp_Prop1}. 

\begin{Proposition}\label{bp_Prop1}
	Under the conditions of Proposition \ref{equivalence_score}, in the measurement evaluation step, the calculation of the factor nodes $q$ and $v$ with raw measurements are equal to those with transformed measurements, respectively, i.e.,
	\begin{align}
		q(\underline{\mathbf{x}}_{k,l}^{(\tau)},\underline{r}_{k,l}^{(\tau)}, a_{k,l}^{(\tau)} ; \mathbf{Z}_{k,l})&=q(\underline{\mathbf{x}}_{k,l}^{(\tau)},\underline{r}_{k,l}^{(\tau)}, a_{k,l}^{(\tau)} ; \breve{\mathbf{Z}}_{k,l}), \\
		v(\overline{\mathbf{x}}_{k,l}^{(i_{l})},\overline{r}_{k,l}^{(i_{l})},b_{k,l}^{(i_{l})};\mathbf{z}_{k,l}^{(i_{l})})&=v(\overline{\mathbf{x}}_{k,l}^{(i_{l})},\overline{r}_{k,l}^{(i_{l})},b_{k,l}^{(i_{l})};\breve{\mathbf{z}}_{k,l}^{(i_{l})}), 
	\end{align}
	where $q(\cdot ; \breve{\mathbf{Z}}_{k,l})$ and $v(\cdot ; \breve{\mathbf{z}}_{k,l}^{(i_{l})})$ are defined by (\ref{bp_factor_q_1})--(\ref{bp_q_x_r0}) and (\ref{bp_factor_v_1})--(\ref{bp_factor_v_2}) by replacing $\mathbf{Z}_{k,l}$ and $\mathbf{z}_{k,l}^{(i_{l})}$ with transformed measurements $\breve{\mathbf{Z}}_{k,l}$ and $\breve{\mathbf{z}}_{k,l}^{(i_{l})}$, respectively. 
	Moreover, for survived targets $\tau= 1,\cdots,N_{k,l-1}$, the fusion beliefs $\tilde{f}_{l}^c(\underline{\mathbf{x}}_{k,l}^{(\tau)},\underline{r}_{k,l}^{(\tau)})$ with raw measurements are equal to the fusion beliefs $\tilde{f}_{l}^d(\underline{\mathbf{x}}_{k,l}^{(\tau)},\underline{r}_{k,l}^{(\tau)})$ with transformed measurements; for new targets $i_l= 1,\cdots,M_{k,l}$, the beliefs $\tilde{f}_{l}^c(\overline{\mathbf{x}}_{k,l}^{(i_l)},\overline{r}_{k,l}^{(i_l)})$ with raw measurements are equal to the beliefs $\tilde{f}_{l}^d(\overline{\mathbf{x}}_{k,l}^{(i_l)},\overline{r}_{k,l}^{(i_l)})$ with transformed measurements.
\end{Proposition}

The proof of Proposition \ref{bp_Prop1} is given in \cite[app \ref{appe_bp_lemma1}]{Liu2023Oncommunication}. Proposition \ref{bp_Prop1} shows that, under some regularity conditions, the BP track association and fusion with transformed measurements are equivalent to that with raw measurements. Moreover, the complete algorithm of BP track association and fusion with transformed measurements is summarized in \cite[app \ref{app_algorithm_2}]{Liu2023Oncommunication}.

\begin{Remark}
{\color{blue}The main advantage of the BP method is its scalability. For a fixed number of iterations $P$ of message passing, the computation complexity of calculating the marginal posterior pdfs of all the target states is only linear in the number of sensors $L$. The complexity of an iteration of the scalable scheme (\ref{ite_1_v})--(\ref{ite_init}) performed for the $l$-th sensor scales as $\mathcal{O}(N_{k,l} M_{k,l})$. If the number of measurements $M_{k,l}$ increases linearly with the number of targets $N_{k,l}$, then the overall complexity of the method scales linearly in the number of sensors and quadratically in the number of targets \cite{meyer2017scalable}. Moreover, if the maximum number of targets is $N_{\text{max}}$, then the worst case computational complexity is $\mathcal{O}(L (P N_{\text{max}}^2))$. } 
\end{Remark}

\section{Communication Requirements of MDA and BP-Based Track Association}\label{sec_CR}
In this section, the communication requirements of different types of transformation matrices $A_k$ are discussed, and a comparison of communication requirements between sending transformed measurements, raw measurements, and filter information is provided. 

\subsection{Two Types of Transformations}
Let us consider two linear transformations, which have lower communication requirements without loss of information, compared with centralized fusion with raw measurements and information filter fuison.

\begin{table*}[tbp]
	\captionsetup[subtable]{position=top,font={ rm,md,up },justification=raggedright, captionskip= 1pt,farskip= 2pt}%
	\centering
	\caption{Summary of communication requirements}\label{table_1}
	\resizebox{.99\textwidth}{!}{
		\begin{tabular}{l|c|c|c|c}
			\hline
			\multirow{2}{*}{Fusion type}
			&\multirow{2}{*}{Fusion with raw measurements}
			&\multirow{2}{*}{Information filter fusion}
			& \multicolumn{2}{c}{ Fusion with transformed measurements}
			\\ \cline{4-5}
			&\multicolumn{1}{c|}{} &\multicolumn{1}{c|}{}
			&Type 1 &Type 2    \\ \hline
			\multicolumn{1}{l|}{Communication variables}
			&  $\mathbf{z}_{k},H_{k}$, $R_{k}$
			&$\hat{\mathbf{x}}_{k|k},~P_{k|k},~$$ \hat{\mathbf{x}}_{k|k-1},~P_{k|k-1}$
			&~~~~$\breve{\mathbf{z}}_{k}^{(1)},\breve{H}_{k}^{(1)}$~~~~
			&$\breve{\mathbf{z}}_{k}^{(2)},\breve{R}_{k}^{(2)}$  \\ \hline
			\multirow{2}{3.5cm}{Communication requirements (8-byte for one dimension)}
			& \multirow{2}{*}{$8 (m+ mn + \frac{m(m+1)}{2}) N_{\text{max}}$} & \multirow{2}{*}{$8 (2n+ n(n+1))N_{\text{max}}$}  
			& \multirow{2}{*}{$8 (m+ mn)N_{\text{max}}$} & \multirow{2}{*}{$8 (m+ \frac{m(m+1)}{2})N_{\text{max}}$} \\ 
			& & & & \\ \hline
			\multirow{3}{3.5cm}{Communication requirements in Kilobytes (KB) for $N_{\text{max}}= 100$, $m=2$, $n=4$} & \multirow{3}{*}{10.16 KB} & \multirow{3}{*}{21.88 KB} & \multirow{3}{*}{7.81 KB} & \multirow{3}{*}{3.91 KB} \\ 
			& & & & \\
			& & & & \\ \hline
		\end{tabular}}
\end{table*}

\subsubsection{Type 1 Transformation}
The first type defined in \cite{2011Li_distributed} is as follows:
\begin{align}\label{measure_trans3}
&\breve{\mathbf{z}}_{k}^{(1)}:=C_{k}\mathbf{z}_{k},\\
&\breve{H}_{k}^{(1)}:=\left[ B_{k}^{\text{T}}R_{k}^{-1}B_{k}\right]^{\frac{1}{2}}D_{k},~\breve{\eta}_{k}^{(1)}:=C_{k}\eta_{k},\\
&\breve{R}_{k}^{(1)}:=\mathnormal{Cov}(\breve{\eta}_k^{(2)})=C_{k}\mathnormal{Cov}(\eta_{k}) C_{k}^{\text{T}}=\mathbf{I},\label{tran2_R}
\end{align}
where $C_{k}=\left[ B_{k}^{\text{T}}R_{k}^{-1}B_{k}\right]^{-\frac{1}{2}} B_{k}^{\text{T}}R_{k}^{-1}$ and $B_{k}$ satisfies $H_{k}=B_{k}D_{k}$ which is the full rank decomposition of $H_{k}$. The rank of $H_{k}$ satisfies, $rank(H_{k})=r_{H}\leq  min(n,m)$ and $\mathbf{I}$ is an identity matrix with dimension $r_{H}$. The transformation matrix for Type 1 is $A_k = C_k= \left[ B_{k}^{\text{T}}R_{k}^{-1}B_{k}\right]^{-\frac{1}{2}} B_{k}^{\text{T}}R_{k}^{-1}$. Note that, for the Type 1 transformation,  $\breve{R}_{k}^{(1)}$ equals an identity matrix $\mathbf{I}$. Thus, each local sensor only needs to send $\breve{\mathbf{z}}_{k}^{(1)}$ and $\breve{H}_{k}^{(1)}$ to the fusion center and the corresponding communication  requirement is $(r_{H}+r_{H}\times n)$ \cite{2011Li_distributed}. Specifically, if $H_k$ is full row matrix, then $C_{k}$ is full column rank matrix and the communication requirement is $m+mn$.

\subsubsection{Type 2 Transformation}The second type is based on a special case of measurement matrix $H_{k}=[E_{k},\mathbf{O}]$, which is defined as follows:
\begin{align}
&\breve{\mathbf{z}}_{k}^{(2)}:= E_{k}^{-1}\mathbf{z}_{k},\\
&\breve{H}_{k}^{(2)}:=\left[\mathbf{I},\mathbf{O}\right],\label{tran_R3} \ \breve{\eta}_{k}^{(2)}:=E_{k}^{-1}\eta_{k},\\
&\breve{R}_k^{(2)}:=\mathnormal{cov}(\breve{\eta}_{k}^{(2)})=E_{k}^{-1}\mathnormal{cov}(\eta_{k})(E_{k}^{-1})^{\text{T}}.
\end{align}
where $E_{k}$ and $\mathbf{O}$ are a full rank matrix and zero matrix, respectively. The transformation matrix for Type 2 is $A_k = E_{k}^{-1}$. Since $\breve{H}_{k}^{(2)}:=\left[\mathbf{I},\mathbf{O}\right]$ is time invariant,  each local sensor only needs send $\breve{\mathbf{z}}_{k}^{(2)}$ and $\breve{R}_k^{(2)}$ to the fusion center and the corresponding transformation requirement  is $m+m(m+1)/2$.

\subsection{Comparison of Communication Requirements between Transformed Measurements and Other Data}
For fusion with raw measurements, each sensor needs to transmit measurement $\mathbf{z}_{k}$, measurement matrix $H_{k}$ and the covariance of the measurement noise $R_{k}$ to the fusion center. Thus, the communication requirements for each sensor is $m+mn+\frac{m(m+1)}{2}$.

In multisensor information  filter fusion \cite{Chong1979_Hierarchical}, each sensor needs to send data $P_{k|k}$, $P_{k|k-1}$, $\hat{\mathbf{x}}_{k|k}$ and $\hat{\mathbf{x}}_{k|k-1}$ to the fusion center. Thus, the communication requirement for information matrix filtering is ($2n+n(n+1)$) \cite{2011Li_distributed}.

The communication requirements for lossless measurement transformations are summarized  in Table \ref{table_1}. Note that, for time-invariant systems, only the measurements are sent to the fusion center. It is not necessary  to send $H_{k}$ and $R_{k}$  to the fusion center at each time $k$. In this case, the communication requirements of fusion with raw measurements are the same as those of Type 1 and Type 2. The communication dimension equals $m$. However, for time-varying systems, Table \uppercase\expandafter{\romannumeral1} shows that the communication requirements of  fusion with lossless transformation Types 1--2 are less than those of fusion with raw measurements or the information filter fusion, where the transformation matrix $A_k \in \mathbb{R}^{n\times m}$ is full column rank. {\color{blue}Furthermore, if the local sensor sends transformed measurements and the corresponding transformation matrix $A_k$ to the fusion center, then the fusion center can reconstruct the raw measurements and raw measurement model. However, the fusion center is sometimes unaware about the information of the transformation matrix (i.e., the local sensor does not share the transformation matrix to the fusion center). The analysis for the track association and fusion algorithms with transformed measurements in Section \ref{sec_MDA} and Section \ref{sec_BF} shows that they are equivalent to the track association and fusion algorithms with raw measurements, which does not require knowledge of the transformation matrix. }

{\color{blue}
\begin{Remark}
	Suppose that $N_{\text{max}}$ is the maximum number of targets seen by the sensor network, i.e.,  $N_{\text{max}}= \max (N_{1}, \cdots, N_{L})$, where $L$ is the number of sensors, then the order of magnitude of data that a sensor needs to transmit to the fusion center is upper bounded by $N_{\text{max}}$. Suppose that each dimension is represented by an 8-byte floating-point value. Then the communication bandwidth requirements of the sensor are summarized in Row 3 of Table \ref{table_1}. Specifically, let $n= 4$, $m= 2$, and $N_{\text{max}}= 100$. The corresponding communication bandwidth requirements of the sensor in Kilobytes are given in Row 4 of Table \ref{table_1}. It also shows that the communication bandwidth requirements of fusion with lossless transformation Types 1--2 are less than those of fusion with raw measurements and information filter fusion, where the transformation matrix $A_k \in \mathbb{R}^{n \times m}$ is full column rank.
\end{Remark}}

{\color{blue}
\begin{Remark}
	The communication rate is one of the important factors affecting the performance of multisensor fusion. Under the assumption of full-rate communication, multisensor estimation fusion is usually equivalent to centralized measurement fusion \cite{Li2003_Optimal,chang1997optimal,Chong1979_Hierarchical,2011Li_distributed}. However, if full-rate communication is not available, the performance of estimation fusion is not optimal in general since information sent to the fusion center is reduced. A track-to-track fusion method at arbitrary communication rates can be seen proposed in \cite{Koch2008On,Koch2009Exact,Govaers2012An}. Our paper discusses the lossless track measurement transformations under full-rate communication. We show that the transformed track measurements reduce the communication requirements (see Table \ref{table_1}), and are equivalent to the raw track measurements. On the other hand, in the case of reduced-rate communication, track association with transformed track measurements is still equivalent to that with raw track measurements, since the proof of equivalence does not depend on the communication rate. But the performance is worse than that of the full-rate communication. 
\end{Remark}}


\section{Simulation Results}\label{sec_SR}
In this section, we consider two scenarios with different numbers of sensors and targets, where the first one is a simple case with two sensors and three targets, and the second one contains ten sensors and ten targets. They are used to verify the main equivalency results of the MDA track association and BP-based track association, respectively. The performance of the proposed algorithms is evaluated by OSPA distance (with cutoff $c=50$ m, order $p=2$) \cite{Schuhmacher2008Aconsistent} as well as OSPA$^{(2)}$ distance (with the same cutoff $c$ and order $p$, and window length $w=10$) \cite{Beard2017OSPA}, and the estimated number of targets. Similar scenarios can be seen in \cite{van2021distributed}.
\subsection{Simulation Setting}
Let us consider the scenario where the targets are moving in the 2-D plane. The state of each target is modeled as 2-D position and velocity, i.e., the state of target $\tau$ is denoted by $\mathbf{x}_{k}^{(\tau)} = [x_{k}^{(\tau),1}, x_{k}^{(\tau),2},\dot{x}_{k}^{(\tau),1},\dot{x}_{k}^{(\tau),2}]^{\text{T}}$. Each target follows a nearly constant velocity model: $\mathbf{x}_{k}^{(\tau)} = F \mathbf{x}_{k-1}^{(\tau)} + \Gamma v_{k-1}^{(\tau)}$. Here, $F= [1, \Delta T; 0, 1] \otimes \mathbf{I}_2$ and $\Gamma = [\Delta T^2/2; \Delta T] \otimes \mathbf{I}_2$, where $\otimes$ denotes for the Kronecker tensor product, $\mathbf{I}_2$ is the 2-D identity matrix, and $\Delta T$ is the sampling period;  
the process noise $v_{k-1}^{(\tau)} \sim \mathcal{N}(\mathbf{0},q^2 \mathbf{I}_2) $ is a zero-mean Gaussian process noise, where $q$ characterizes the average increment of target speed in $\Delta T$. 
The raw measurement $\mathbf{z}_{k}^{(i_l)}$ originates from target $\tau$ at the $l$-th sensor is modeled according to $\mathbf{z}_{k}^{(i_l)}=H_{k,l} \mathbf{x}_{k}^{(\tau)} + w_{k,l}^{(i_l)}$,
where $H_{k,l}= [\text{diag}(1+ \theta_{k,l}^{(1)},1+ \theta_{k,l}^{(2)}), \mathbf{O}]$,  
$\theta_{k,l}^{(1)}$ and $\theta_{k,l}^{(2)}$ are time-varying uncertain parameters which may be sensor bias estimates. Here, they are known and are uniformly generated from [-0.02, 0.02]. The measurement noise $w_{k,l}^{(i_l)} \sim \mathcal{N}(\mathbf{0}, R_{k,l})$ is a zero-mean Gaussian noise with covariance $R_{k,l} = \text{diag}(\sigma^2+ \vartheta_{k,l}^{(1)},\sigma^2+ \vartheta_{k,l}^{(2)})$, where $\sigma$ is the reference standard deviation, $\vartheta_{k,l}^{(1)}$ and $\vartheta_{k,l}^{(2)}$ are time-varying uncertain parameters which may be the covariance of the sensor bias estimates. $\vartheta_{k,l}^{(1)}$ and $\vartheta_{k,l}^{(2)}$ are known and are uniformly generated from  [0, 1]. Each sensor can only detect targets within its field-of-view (angle of $[-45^\circ,45^\circ]$, range of $1200$ m) with probability $P_d^{(l)}$, $l=1,\cdots,L$. The clutter pdf is assumed uniform in the field-of-view of each sensor, and the number of clutter is assumed Poisson distributed with a mean number of $\lambda_{f_l}$, i.e., the clutter rate is $\lambda_{f_l}$. For this measurement equation, Type 1 transformation is $A_{k,l}= C_{k,l}$, where $C_{k,l}= \text{diag}(\sigma^2 + \vartheta_{k,l}^{(1)}, \sigma^2 + \vartheta_{k,l}^{(2)})^{-\frac{1}{2}}$; Type 2 transformation is $A_{k,l} =E_{k,l}^{-1}$, where $E_{k,l}= \text{diag}(1+ \theta_{k,l}^{(1)},1+ \theta_{k,l}^{(2)})$. Since the communication requirement of Type~2 transformation is less than that of Type~1 transformation in the scenarios, we use Type~2 transformation to verify the equivalency results throughout this section.

At each local sensor, initialization and tracking of the local tracks are obtained by the global nearest neighbor tracker (other trackers can be used, such as JPDA and MHT, etc.). The track management settings for the local sensor are as follows: a track is confirmed if it has been associated with at least four measurements and deleted if it loses measurements over $N_{pr}$ consecutive scans, where $N_{pr}= 3$. At each scan, once the tracks at the local sensor are updated, the local sensor sends the confirmed track measurements to the fusion center. 

{\color{blue}
We describe the two scenarios and present the corresponding simulation results in the following two subsections. The parameters common to both scenarios are set as follows. The sampling period is $\Delta T = 1$~s; the standard deviation of the process noise is $q = 0.1$~m/s$^2$; the reference standard deviation of the measurement noise is $\sigma = 5$~m.
}

\subsection{Scenario 1: Two Nodes with Three Targets}
\begin{figure}[H]%
	\vspace{-5mm}
	\centering 
	\subfloat[][]{%
	\label{fig_scen_1_tracks_1}%
	\includegraphics[width= .22\textwidth]{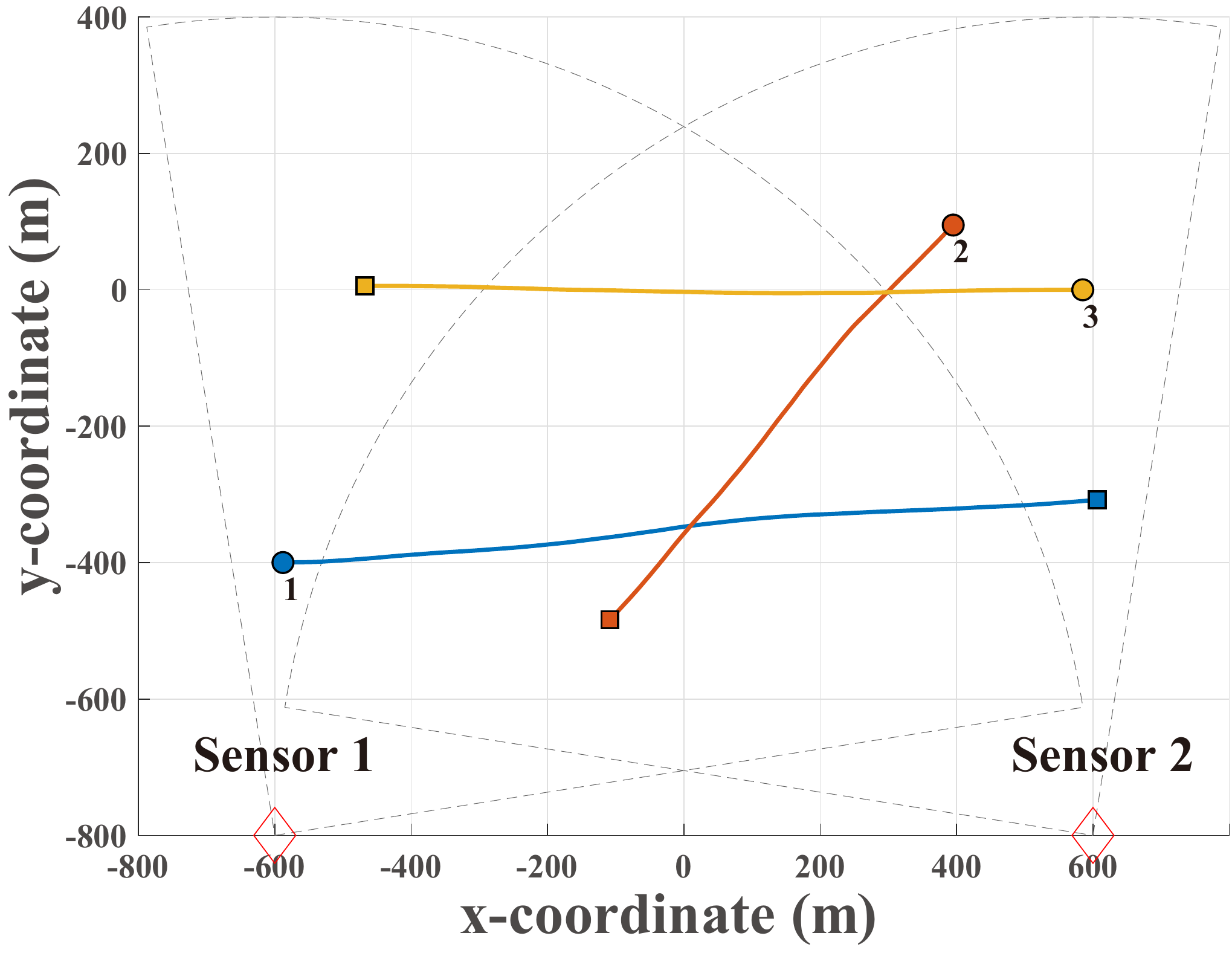}}%
	\subfloat[][]{%
	\label{fig_scen_1_tracks_4}%
	\includegraphics[width= .22\textwidth]{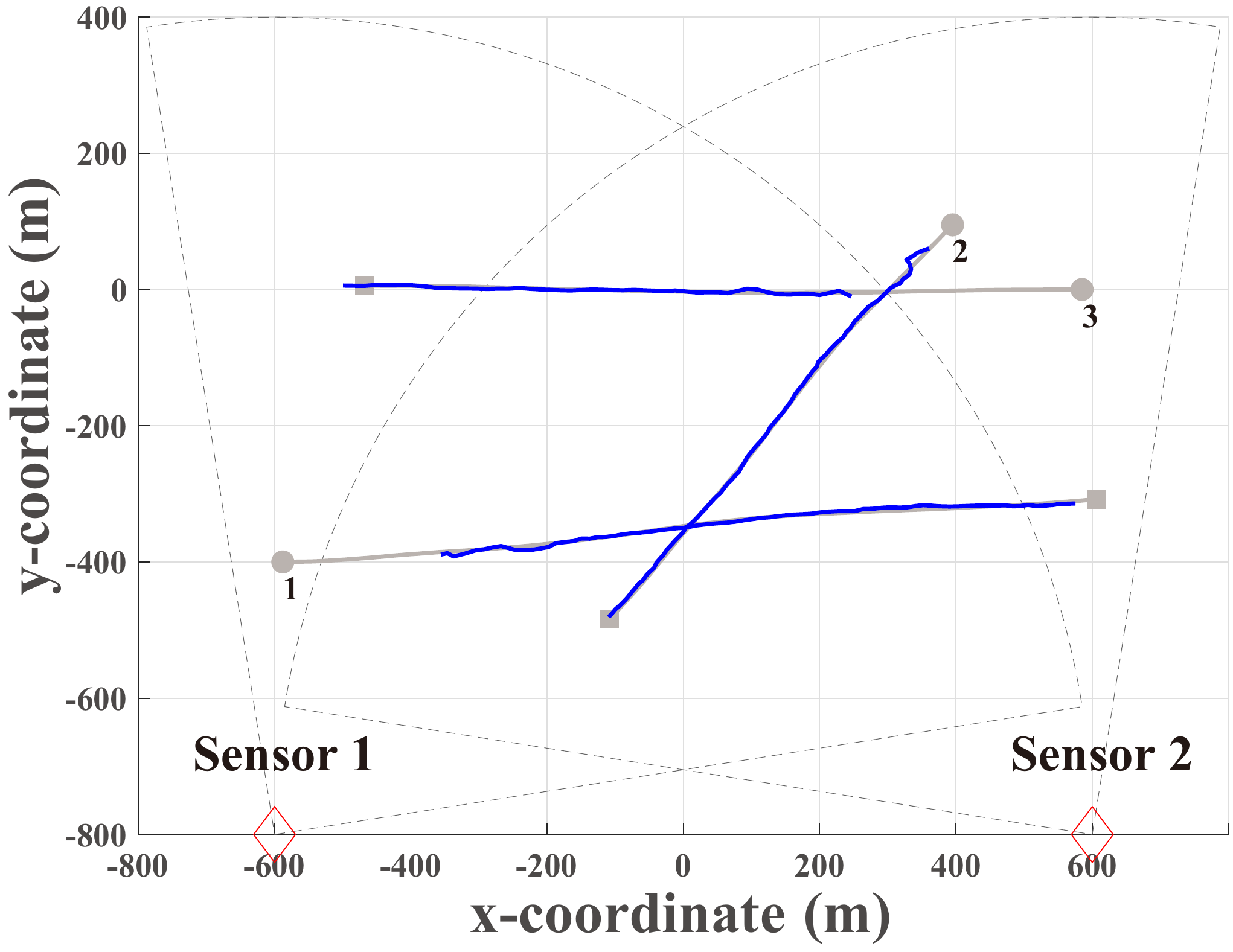}}%
	\\
	\subfloat[][]{%
	\label{fig_scen_1_tracks_5}%
	\includegraphics[width= .22\textwidth]{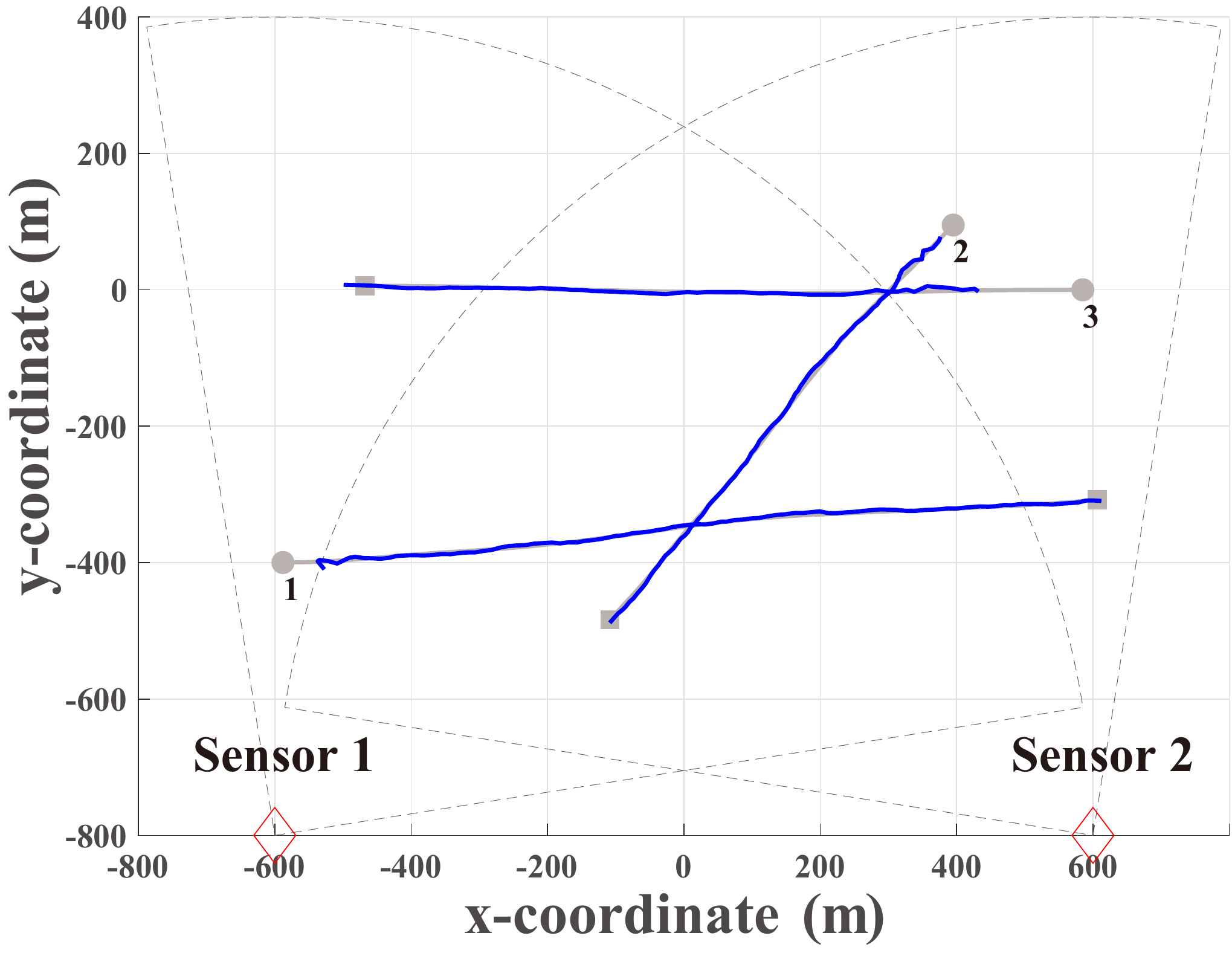}}%
	\subfloat[][]{%
	\label{fig_scen_1_tracks_8}%
	\includegraphics[width= .22\textwidth]{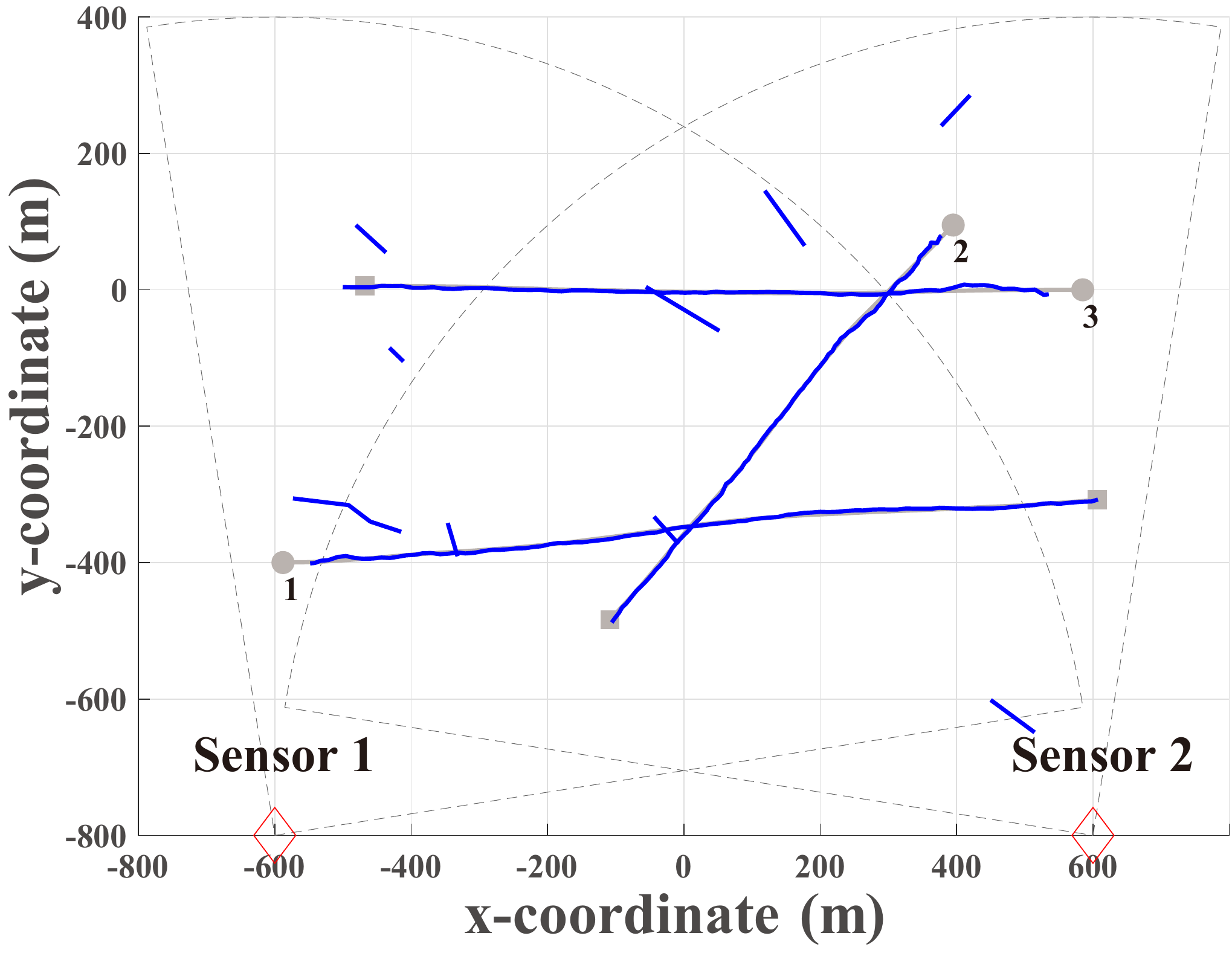}}%
	\caption[Scenario 1 results.]{Scenario 1 ground truth and estimated trajectories under different probabilities of detection and clutter rates: \subref{fig_scen_1_tracks_1} groud truth, \subref{fig_scen_1_tracks_4} estimated trajectories with $P_d^{(l)}= 0.9$, $\lambda_{f_l}= 10$, \subref{fig_scen_1_tracks_5} estimated trajectories (blue lines) with $P_d^{(l)}= 0.9$, $\lambda_{f_l}= 40$, \subref{fig_scen_1_tracks_8} estimated trajectories with $P_d^{(l)}= 0.7$, $\lambda_{f_l}= 10$. Starting and stopping positions are denoted by $\circ $ and $\square $, respectively. }%
	\label{fig_scen_1_tracks}%
	\vspace{-2mm}
\end{figure}
As shown in Fig. \ref{fig_scen_1_tracks}\subref{fig_scen_1_tracks_1}, we placed three targets moving in the 2-D plane $[-800 \text{ m}, 800 \text{ m}]\times[-800\text{ m},400\text{ m}]$ and used two sensors with limited field-of-view to monitor the targets. The whole period from the first target's birth to the last target's death is 100~s. Specifically, Targets 1 and 2 are born at time 1 s and die at time 100~s, and Target~3 is born at time 10~s and dies at time 80~s. The two sensors are located at $[-600,-800]^{\text{T}}$ m and $[600,-800]^{\text{T}}$ m, respectively. {\color{blue}To test the performance of the MDA track fusion algorithm, we evaluate it under different values of the clutter rate $\lambda_{f_l}= 10, 20, 30$, and $40$ when $P_d^{(l)}= 0.9$, and under different values of probability of detection $P_d^{(l)}= 0.7, 0.8, 0.9$, and $0.99$ when $\lambda_{f_l}= 10$.} We applied the LP relaxation-based algorithm \cite{Storms2003An} to solve the MDA track association problem. The track management settings for the fusion center are as follows: a track is confirmed if it is associated with at least two measurements and deleted if it loses measurements over $N_{pr}$ consecutive scans, where $N_{pr}= 3$.

{\color{blue}
Figs. \ref{fig_scen_1_tracks}\subref{fig_scen_1_tracks_4}--\ref{fig_scen_1_tracks}\subref{fig_scen_1_tracks_8} show the estimated trajectories under different probabilities of detection and clutter rates, the common parameter of Figs. \ref{fig_scen_1_tracks}\subref{fig_scen_1_tracks_4}--\ref{fig_scen_1_tracks}\subref{fig_scen_1_tracks_5} is the clutter rate $\lambda_{f_l}= 10$, and that of Figs. \ref{fig_scen_1_tracks}\subref{fig_scen_1_tracks_5}--\ref{fig_scen_1_tracks}\subref{fig_scen_1_tracks_8} is the probability of detection $P_d^{(l)}= 0.9$. Furthermore, as expected, Figs. \ref{fig_scen_1_tracks}\subref{fig_scen_1_tracks_4}--\ref{fig_scen_1_tracks}\subref{fig_scen_1_tracks_5} indicate that the lower the probability of detection, the slower the track initialization, and Figs. \ref{fig_scen_1_tracks}\subref{fig_scen_1_tracks_5}--\ref{fig_scen_1_tracks}\subref{fig_scen_1_tracks_8} show that the higher the clutter rate, the more the false tracks. 
}

\begin{figure}%
	\centering \subfloat[][]{%
	\label{fig_2_1}%
	\includegraphics[width= .48\textwidth]{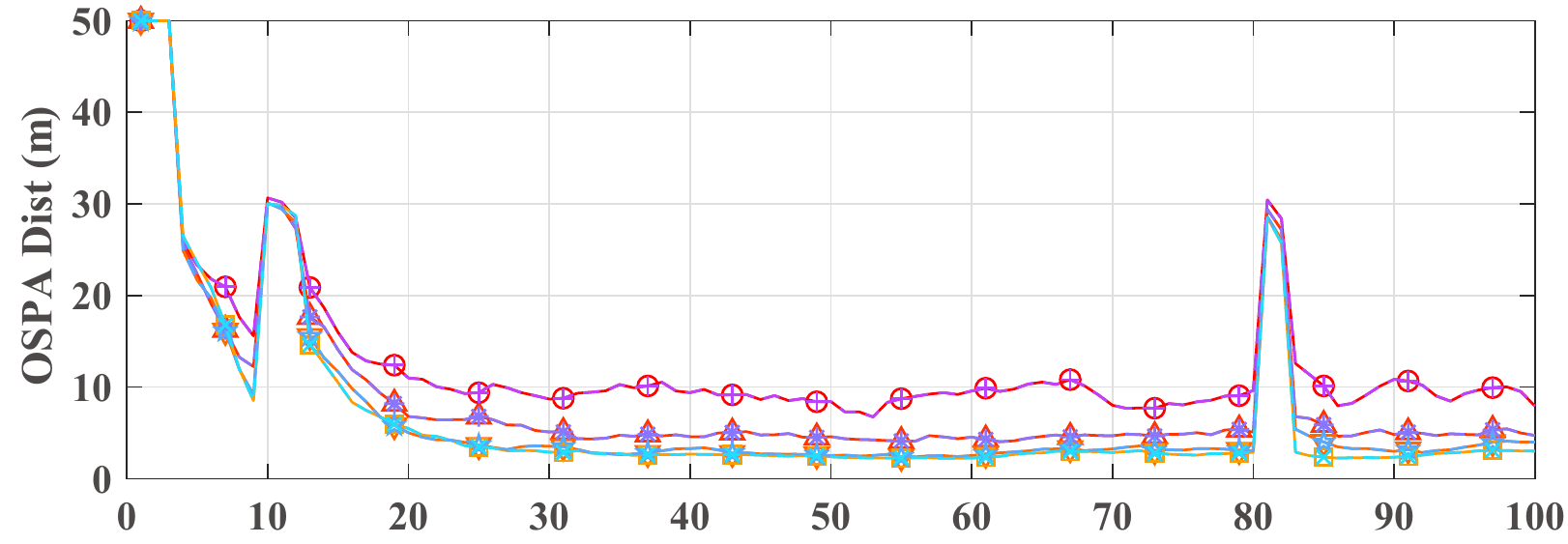}}%
	\hspace{8pt}%
	\subfloat[][]{%
	\label{fig_2_2}%
	\includegraphics[width= .48\textwidth]{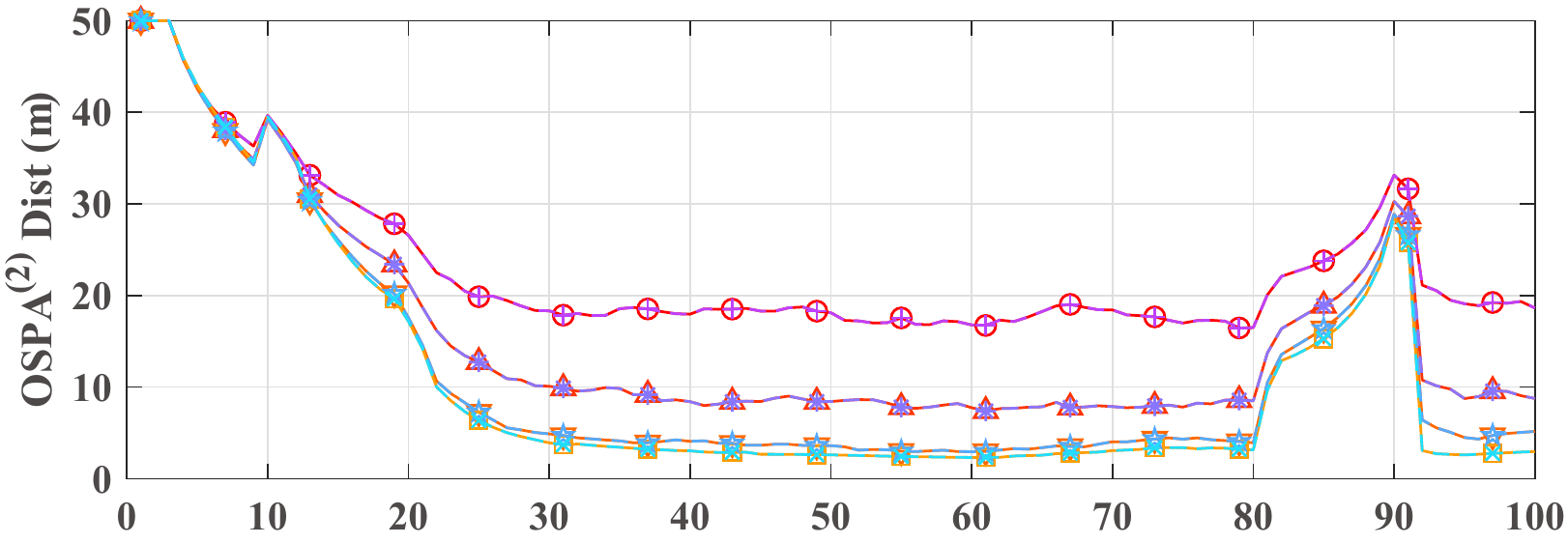}} 
	\hspace{8pt}%
	\subfloat[][]{%
	\label{fig_2_3}%
	\includegraphics[width= .48\textwidth]{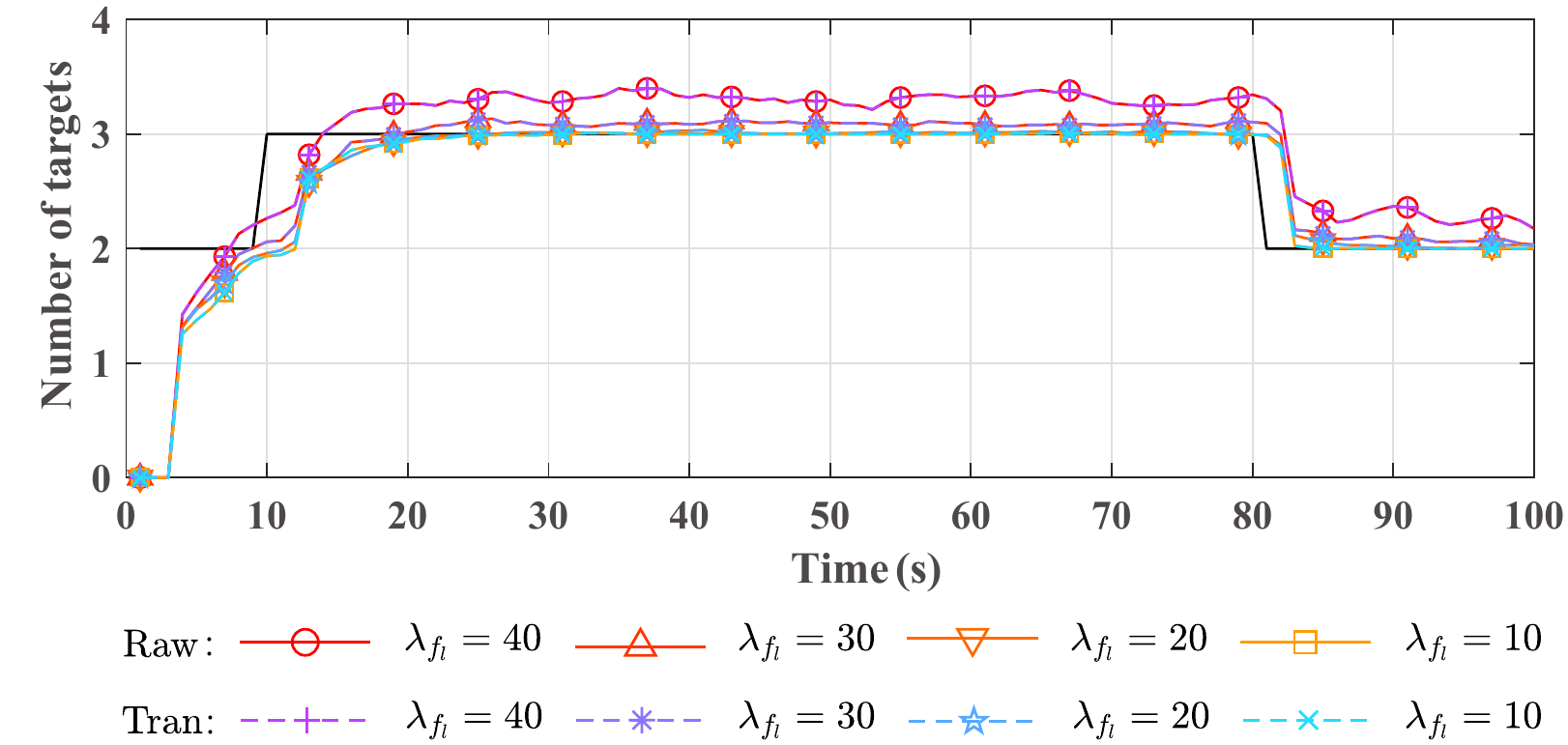}}%
	\caption[Scenario 1 results.]{Comparison results under different clutter rates: \subref{fig_2_1} OSPA distance, \subref{fig_2_2} OSPA$^{(2)}$ distance, \subref{fig_2_3} estimated number of targets (the black line represents the true number of targets). In the legends, ``Raw'' and ``Tran'' are the abbreviations of ``Raw measurements'' and ``Transformed measurements'', respectively.}%
	\label{fig_2_all}%
\end{figure}

{\color{blue}Figs. \ref{fig_2_all}--\ref{fig_3_all} illustrate the performance results of the MDA association and fusion algorithm with raw measurements and transformed measurements, where the OSPA distance, the OSPA$^{(2)}$ distance, and the estimated number of targets are plotted as a function of time steps, respectively. Figs. \ref{fig_2_all}--\ref{fig_3_all} confirm that the performance of the MDA association and fusion algorithm with raw measurements is equivalent to that with transformed measurements, which corroborates Proposition~\ref{equivalence_score}. Moreover, Figs. \ref{fig_2_all}\subref{fig_2_1}--\ref{fig_2_all}\subref{fig_2_3} show the performance results under $\lambda_{f_l}= 10,20,30$, and $40$ when $P_d^{(l)}= 0.9$, respectively. The curves of the OSPA distances shown in Fig. \ref{fig_2_all}\subref{fig_2_1} exhibit peaks at time $k = 10$ and $80$ s, respectively. The reason is that Target 3 is born at time $k=10$ s and dies at time $k=80$ s. Since a local track is confirmed when it has at least four measurements, the duration of the first two peaks of OSPA is about 4 s. Since a track is deleted if it loses measurements over three consecutive scans, the durations of last peak of OSPA are about 2 s. As expected, the OSPA distance, the OSPA$^{(2)}$ distance, and the estimated number of targets increase as the clutter rate $\lambda_{f_l}$ increases, which is consistent with the phenomenon in Figs. \ref{fig_scen_1_tracks}\subref{fig_scen_1_tracks_5}--\ref{fig_scen_1_tracks}\subref{fig_scen_1_tracks_8}. The reason is that the number of false tracks at the local sensors increases as $\lambda_{f_l}$ increases, and the number of false track measurements sent to the fusion center also increases, which results in an increase in the number of false tracks at the fusion center. Besides, a high clutter rate results in the curve of estimated number of targets in Fig. \ref{fig_2_all}\subref{fig_2_3} above that of the true number of targets.

\begin{figure}%
	\centering \subfloat[][]{%
	\label{fig_3_1}%
	\includegraphics[width= .48\textwidth]{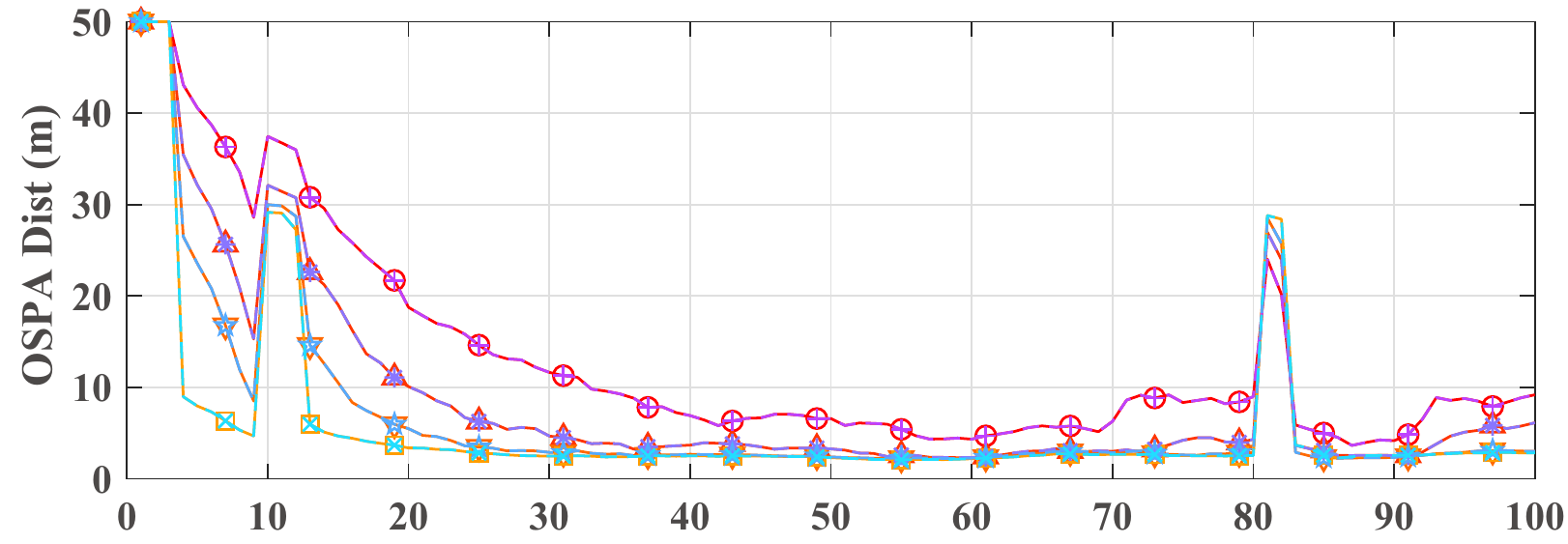}}%
	\hspace{8pt}%
	\subfloat[][]{%
	\label{fig_3_2}%
	\includegraphics[width= .48\textwidth]{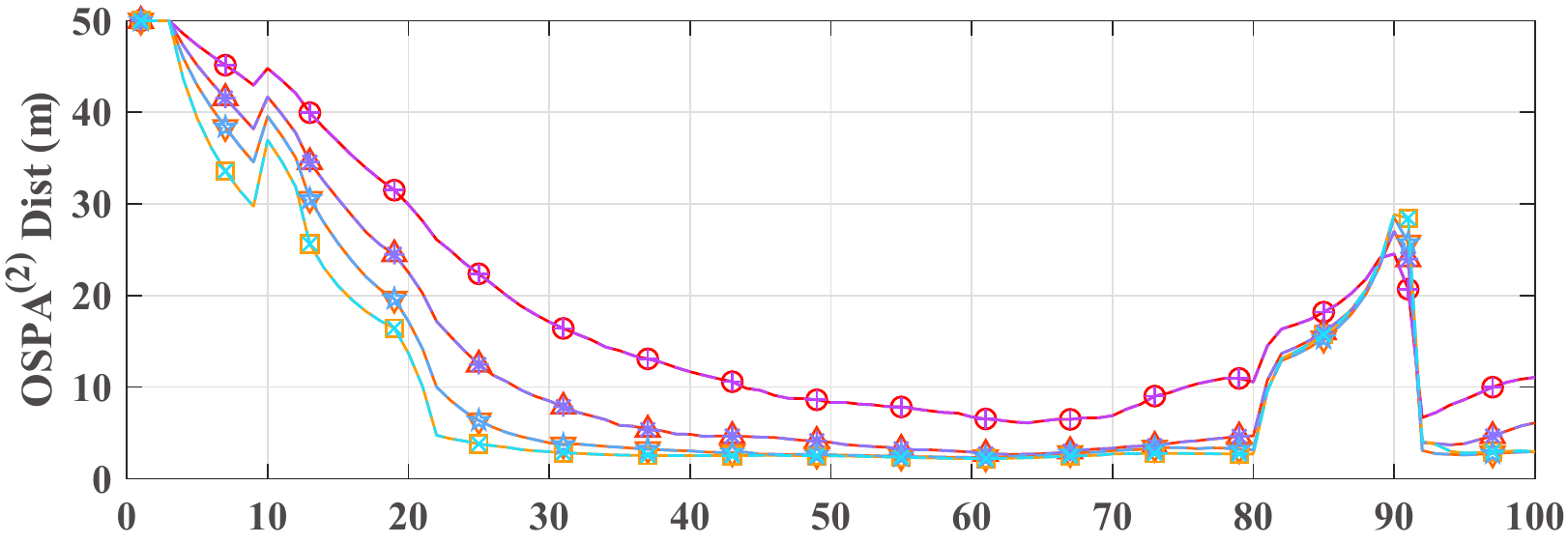}} 
	\hspace{8pt}%
	\subfloat[][]{%
	\label{fig_3_3}%
	\includegraphics[width= .48\textwidth]{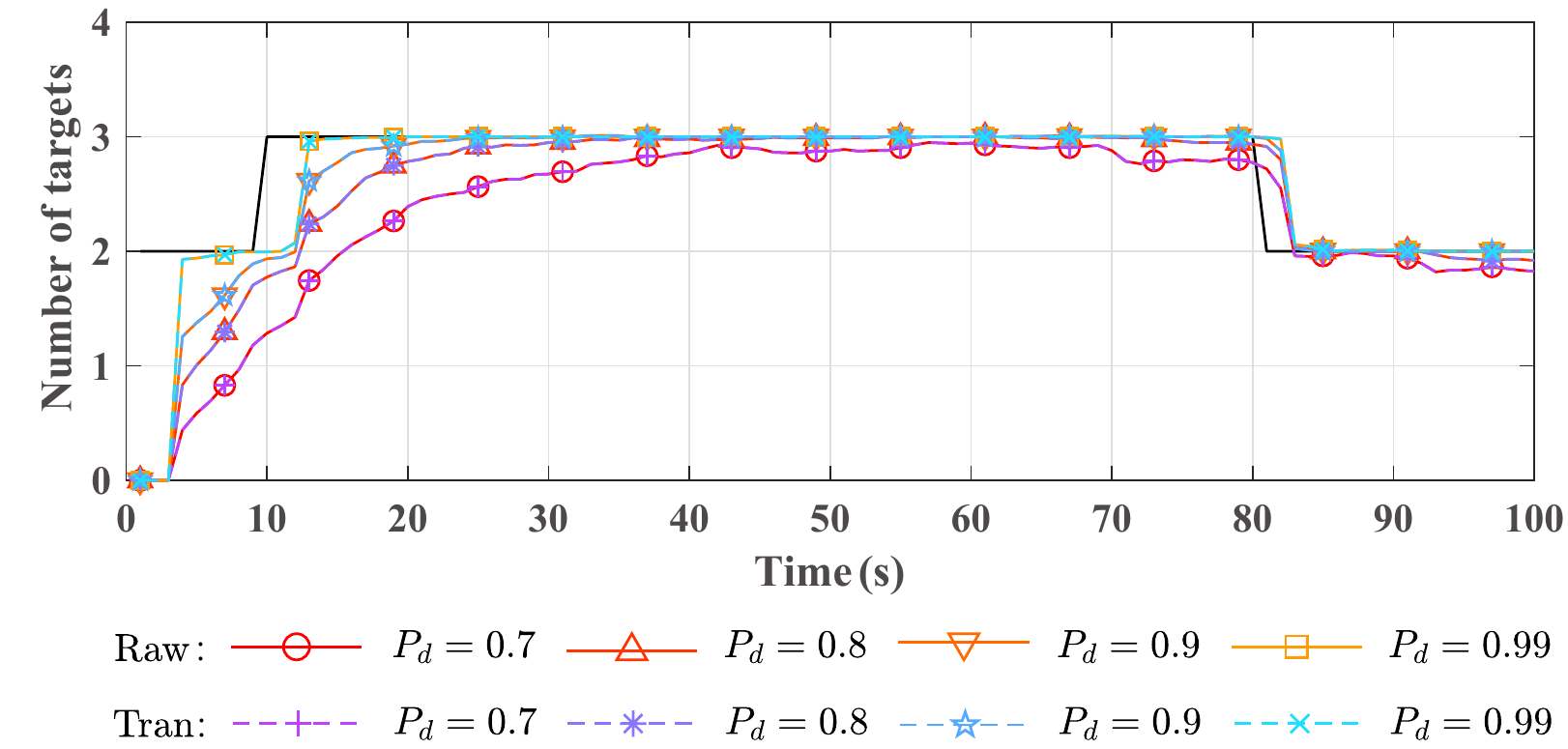}}%
	\caption[Scenario 1 results.]{Comparison results under different probabilities of detection: \subref{fig_3_1} OSPA distance, \subref{fig_3_2} OSPA$^{(2)}$ distance, \subref{fig_3_3} estimated number of targets. }%
	\label{fig_3_all}%
\end{figure}

Figs. \ref{fig_3_all}\subref{fig_3_1}--\ref{fig_3_all}\subref{fig_3_3} show the performance results under $P_d^{(l)}= 0.7,0.8,0.9$, and $0.99$ when $\lambda_{f_l}= 10$, respectively. For the same reason as before, the curves of the OSPA distances shown in Fig. \ref{fig_3_all}\subref{fig_3_1} exhibit two peaks. As expected, the OSPA and OSPA$^{(2)}$ distances decrease as the probability $P_d^{(l)}$ of detection increases, and the estimated number of targets increases as $P_d^{(l)}$ increases, which is consistent with the phenomenon in Figs. \ref{fig_scen_1_tracks}\subref{fig_scen_1_tracks_4}--\ref{fig_scen_1_tracks}\subref{fig_scen_1_tracks_5}. The reason is that a low $P_d^{(l)}$ results in a late track initialization at the local sensor, which is the reason that the curve of the estimated number of targets in Fig.~\ref{fig_3_all}\subref{fig_3_3} is below that of the true number of targets. 

Finally, Table \ref{table_3} summarizes the averaged communication requirements over 100 scans and 100 Monte Carlo runs at the fusion center. Under different values of $\lambda_{f_l}$ and $P_d^{(l)}$, the communication requirements for sending transformed measurements are less than those for sending raw measurements, which corroborates the communication requirement analysis in Section \ref{sec_CR}.

\begin{table}[H]
	\centering
	\caption{Communication Requirements in Bytes (B) of Scenario 1}\label{table_3}
	\subfloat[Different values of clutter rate $\lambda_{f_l}$]{
		\resizebox{.48\textwidth}{!}{
			\begin{tabular}{|l|c|c|c|c|c|}
				\hline
				\multicolumn{2}{|l}{\multirow{2}{*}{Fusion type}} & \multicolumn{4}{|c|}{Clutter rate $\lambda_{f_l}$} \\
				\cline{3-6}
				\multicolumn{2}{|c|}{} & 10 & 20 & 30 & 40\\
				\hline
				\multicolumn{2}{|l|}{Fusion with raw measurements} & 499.4 B & 506.3 B & 526.6 B & 579.9 B \\
				\hline 
				\multirow{2}{2.8cm}{Fusion with transformed measurements} &Type 1 & 307.4 B & 311.5 B & 324.2 B & 356.8 B\\ 
				\cline{2-6}
				& Type 2 & 192.1 B & 194.7 B & 202.6 B & 223.0 B\\ \hline
			\end{tabular}
		}
	}\\
	\subfloat[Different values of probability of detection $P_d^{(l)}$]{
		\resizebox{.48\textwidth}{!}{
			\begin{tabular}{|l|c|c|c|c|c|}
				\hline
				\multicolumn{2}{|l}{\multirow{2}{*}{Fusion type}} & \multicolumn{4}{|c|}{Probability of detection $P_d^{(l)}$} \\
				\cline{3-6}
				\multicolumn{2}{|c|}{} & 0.7 & 0.8 & 0.9 & 0.99\\
				\hline
				\multicolumn{2}{|l|}{Fusion with raw measurements} & 363.6 B & 444.2 B & 499.4 B & 533.1 B \\
				\hline 
				\multirow{2}{2.8cm}{Fusion with transformed measurements} & Type 1 & 223.7 B & 273.3 B & 307.4 B & 328.0 B\\ \cline{2-6}
				& Type 2 & 139.8 B & 170.8 B & 192.1 B & 205.0 B\\ \hline
			\end{tabular}
		}
	}
\end{table}

}

\subsection{Scenario 2: Ten Nodes with Ten Targets}
\begin{figure}[H]%
	\vspace{-3mm}
	\captionsetup[subfigure]{position=top,font={ rm,md,up 	},justification=raggedright, captionskip=-3pt,farskip= -3pt,singlelinecheck=false}
	\centering 
	\subfloat[][]{%
	\label{fig_scen_2_truth}%
	\includegraphics[width= .24\textwidth]{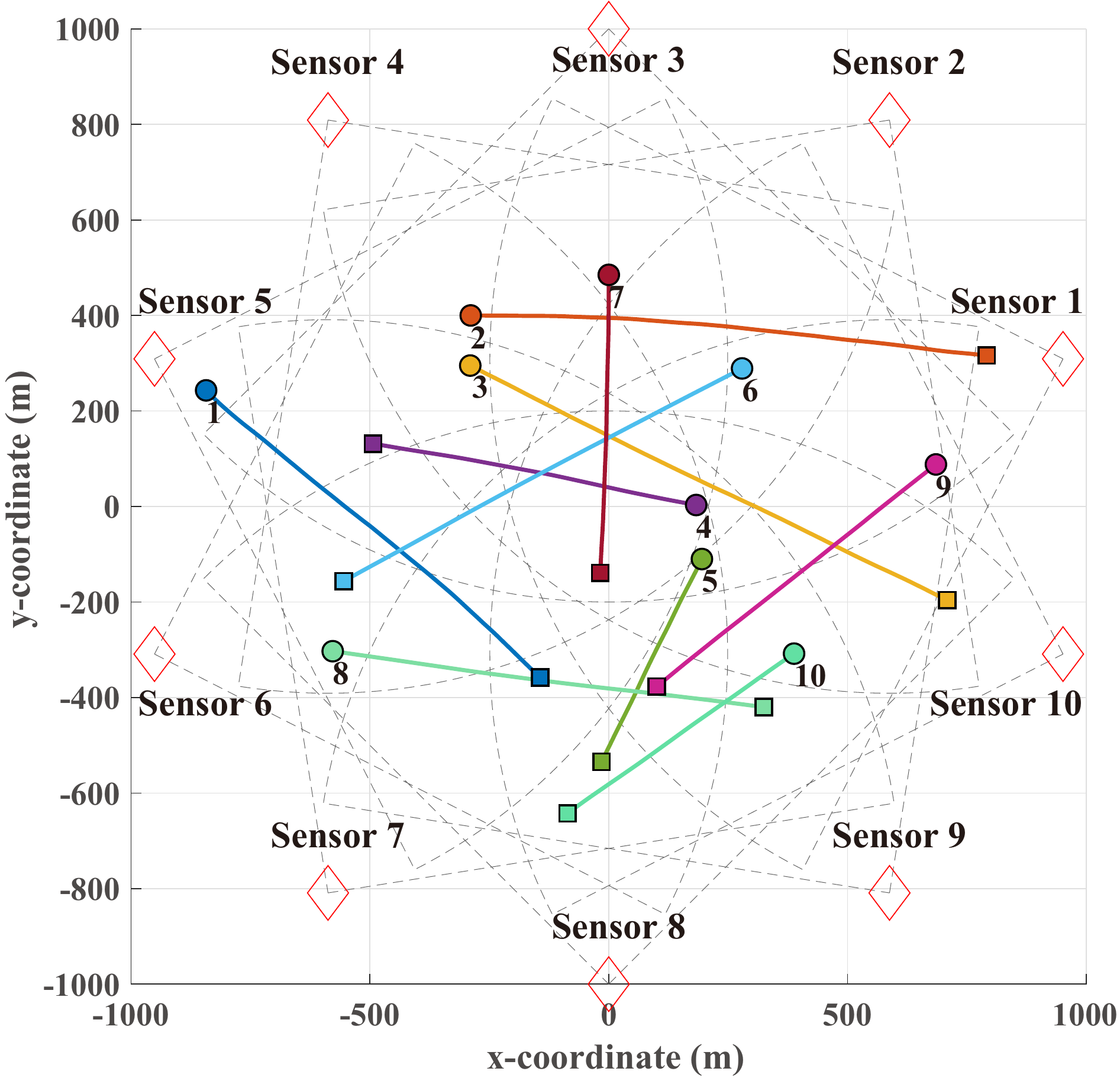}}%
	\subfloat[][]{%
	\label{fig_scen_2_tracks_1}%
	\includegraphics[width= .24\textwidth]{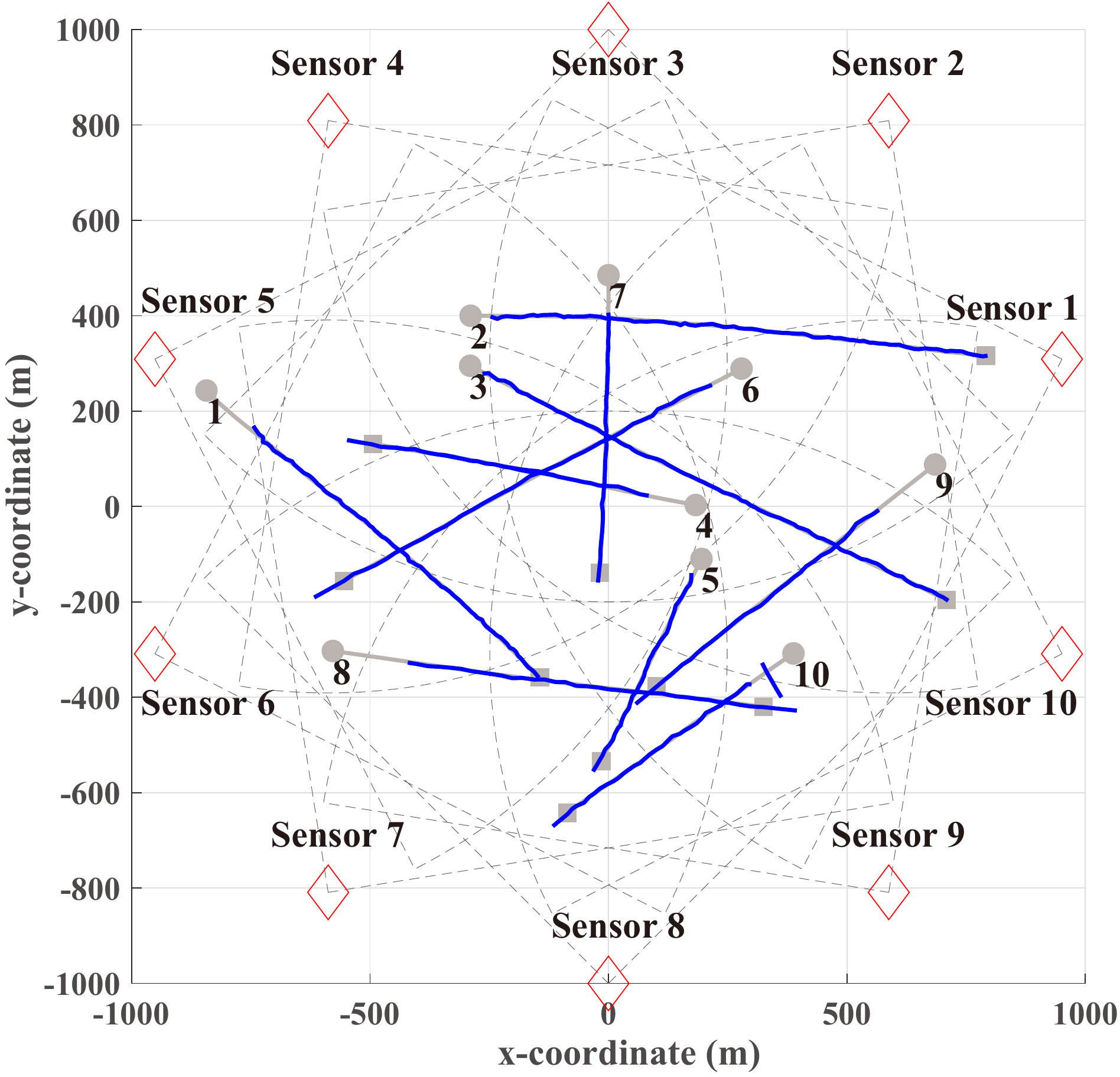}}%
	\\
	\subfloat[][]{%
	\label{fig_scen_2_tracks_4}%
	\includegraphics[width= .24\textwidth]{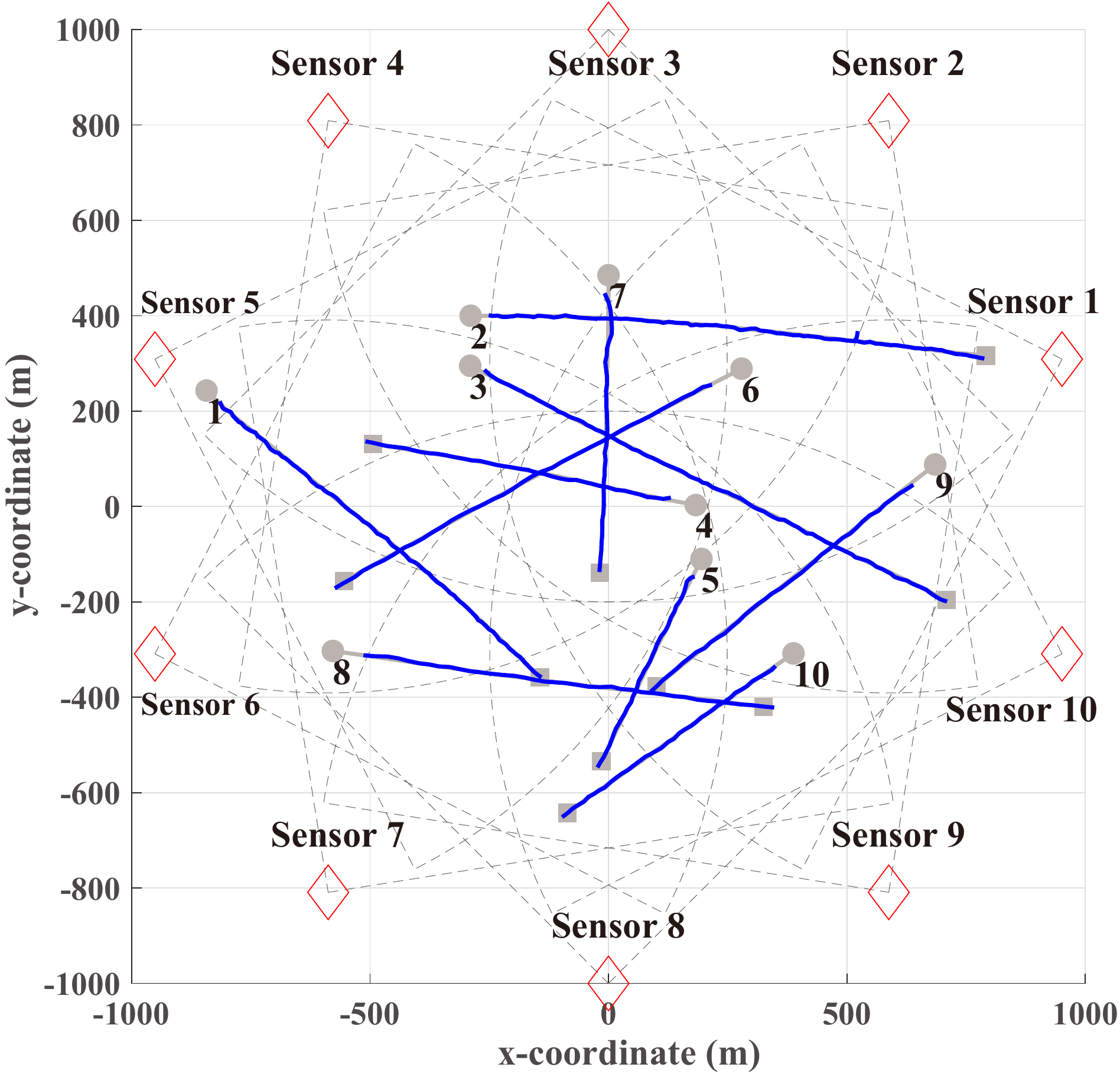}}%
	\subfloat[][]{%
	\label{fig_scen_2_tracks_5}%
	\includegraphics[width= .24\textwidth]{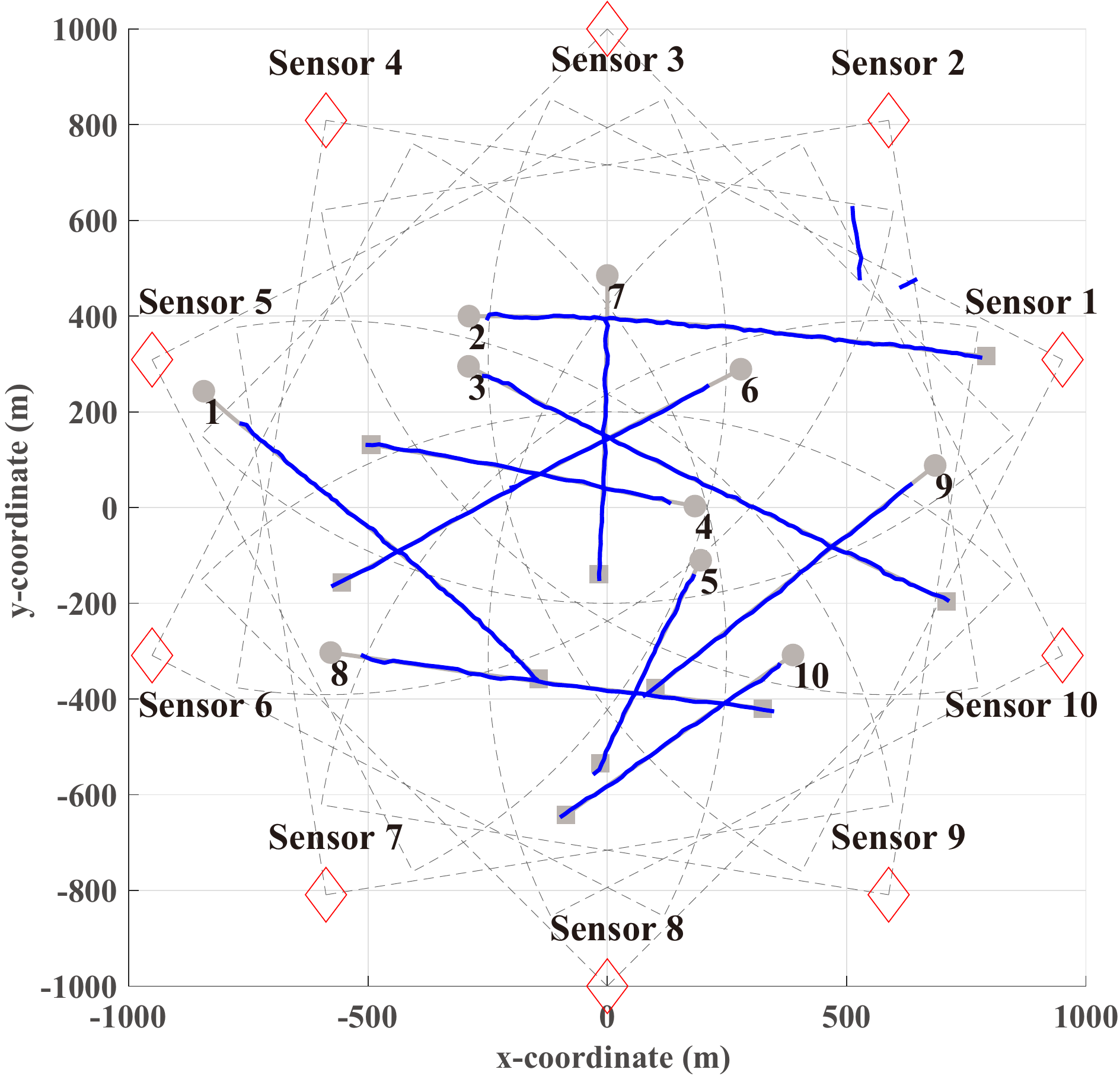}}%
	\caption[Scenario 2 results.]{Scenario 2 ground truth and estimated trajectories under different probabilities of detection and clutter rates: \subref{fig_scen_2_truth} groud truth, \subref{fig_scen_2_tracks_1} $P_d^{(l)}= 0.7$, $\lambda_{f_l}= 10$, \subref{fig_scen_2_tracks_4} $P_d^{(l)}= 0.9$, $\lambda_{f_l}= 10$, \subref{fig_scen_2_tracks_5} $P_d^{(l)}= 0.9$, $\lambda_{f_l}= 40$.}%
	\label{fig_scen_2_tracks}%
\end{figure}
In this scenario, we use ten sensors to track ten targets and validate that the BP-based track association and fusion algorithm with transformed measurements is equivalent to that with raw measurements. The ten targets move in the 2-D plane $[-1000 \text{ m}, 1000 \text{ m}]\times[-1000\text{ m},1000\text{ m}]$, where the whole period from the first target's birth to the last target's death is 100~s. Specifically, Targets 1--3 are born at time 1~s and die at time 100 s, Targets 4--6 are born at time 20~s and die at time 60~s, and Targets 7--10 are born at time 40~s and die at time 80~s. We set up ten sensors located on a circle with a radius of $1000$ m, where all the sensors are equidistant from each other. The sensors and the targets of Scenario~2 are shown in Fig.~\ref{fig_scen_2_tracks}\subref{fig_scen_2_truth}. {\color{blue}To test the performance of the BP track association and fusion algorithm, we evaluate it under different $\lambda_{f_l}= 10, 20, 30$, and $40$ when $P_d^{(l)}= 0.9$, and under different $P_d^{(l)}= 0.7, 0.8, 0.9$, and $0.99$ when $\lambda_{f_l}= 10$.} We used a particle-based implementation of the BP-based track association algorithm, where for each target, a set of 1000 particles is applied to approximate its belief. The number of iterations in the step of iterative data association is set to $P=10$. For target declaration, a target is declared if its existence belief $\tilde{p}(\underline{r}_{k,l}^{(\tau)}= 1) > P_{th}$ or $\tilde{p}(\overline{r}_{k,l}^{(i_l)}= 1) > P_{th}$, where $P_{th} = 0.7$. For target pruning, a target is removed if its existence belief $\tilde{p}(\underline{r}_{k,l}^{(\tau)}= 1) < P_{pr}$ or $\tilde{p}(\overline{r}_{k,l}^{(i_l)}= 1) < P_{pr}$, or if it loses measurements over $N_{pr}$ consecutive scans, where $P_{pr} = 1\text{e}^{-6}$ and $N_{pr}= 3$.


{\color{blue}
Figs. \ref{fig_scen_2_tracks}\subref{fig_scen_2_tracks_1}--\ref{fig_scen_2_tracks}\subref{fig_scen_2_tracks_5} shows the estimated trajectories under different probabilities of detection and clutter rates. As expected, Figs. \ref{fig_scen_2_tracks}\subref{fig_scen_2_tracks_1}--\ref{fig_scen_2_tracks}\subref{fig_scen_2_tracks_4} indicate that the lower the probability of detection, the slower the track initialization, and Figs. \ref{fig_scen_2_tracks}\subref{fig_scen_2_tracks_4}--\ref{fig_scen_2_tracks}\subref{fig_scen_2_tracks_5} show that the higher the clutter rate, the more the false tracks. Besides, when compared with Fig. \ref{fig_scen_1_tracks} in Scenario 1, Fig. \ref{fig_scen_2_tracks} shows fewer false tracks at high clutter rates and faster track initialization at low detection rates. The reason is that in this scenario, the fusion center utilizes complementary information from more sensors. 

Figs. \ref{fig_5_all}--\ref{fig_6_all} show that the performance results of the BP-based association and fusion algorithm with raw measurements and transformed measurements, where the OSPA distance, the OSPA$^{(2)}$ distance, and the estimated number of targets are plotted as a function of time steps, respectively. The results shown in Figs.~\ref{fig_5_all}--\ref{fig_6_all} confirm that the performance of the BP-based association and fusion algorithm with raw measurements is equivalent to that with transformed measurements, which corroborates Proposition \ref{bp_Prop1}. Besides, the curves of Figs. \ref{fig_5_all}--\ref{fig_6_all} are closer to each other than those of Figs. \ref{fig_2_all}--\ref{fig_3_all} in Scenario 1 since the fusion center in Scenario 2 utilizes more complementary information from more sensors. Moreover, Figs.~\ref{fig_5_all}\subref{fig_5_1}--\ref{fig_5_all}\subref{fig_5_3} show the performance results under $\lambda_{f_l}= 10,20,30$, and $40$ when $P_d^{(l)}= 0.9$, respectively. The curves of the OSPA distances shown in Fig. \ref{fig_5_all}\subref{fig_5_1} exhibit peaks at time $k = 1$, $20$, $40$, $60$, and $80$ s, due to the targets being born and dying. For the same reason as discussed in Scenario 1, the duration of the first three peaks (due to the targets being born) of OSPA is about 4 s. Since a track is deleted if its existence belief $\tilde{p}(\underline{r}_{k,l}^{(\tau)}= 1) < P_{pr}= 1\text{e}^{-6}$ or it loses measurements over $N_{pr}=3$ consecutive scans, it yields that the duration of last two peaks of OSPA is about 1--2 s. As expected, the OSPA distance, the OSPA$^{(2)}$ distance, and the estimated number of targets increase as the clutter rate $\lambda_{f_l}$ increases, which is consistent with the phenomenon in Figs. \ref{fig_scen_2_tracks}\subref{fig_scen_2_tracks_4}--\ref{fig_scen_2_tracks}\subref{fig_scen_2_tracks_5}. The reason is the same as that in Scenario 1.

\begin{figure}%
	\centering \subfloat[][]{%
	\label{fig_5_1}%
	\includegraphics[width=.48\textwidth]{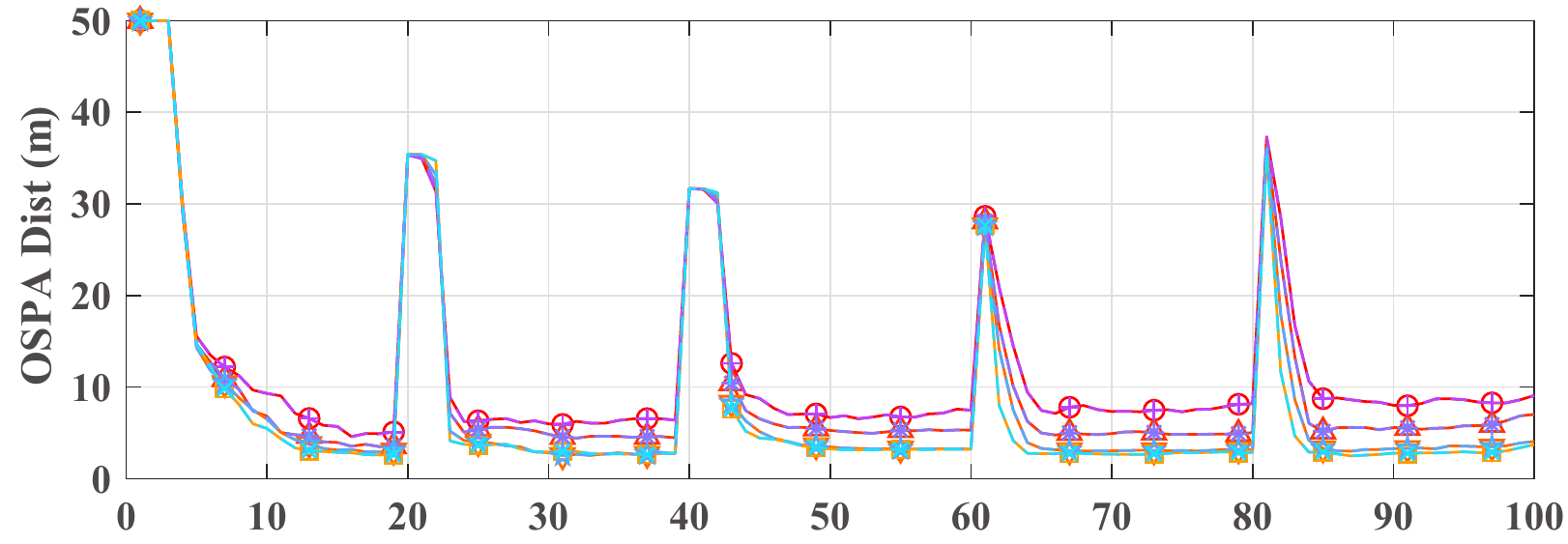}}%
	\hspace{8pt}%
	\subfloat[][]{%
	\label{fig_5_2}%
	\includegraphics[width=.48\textwidth]{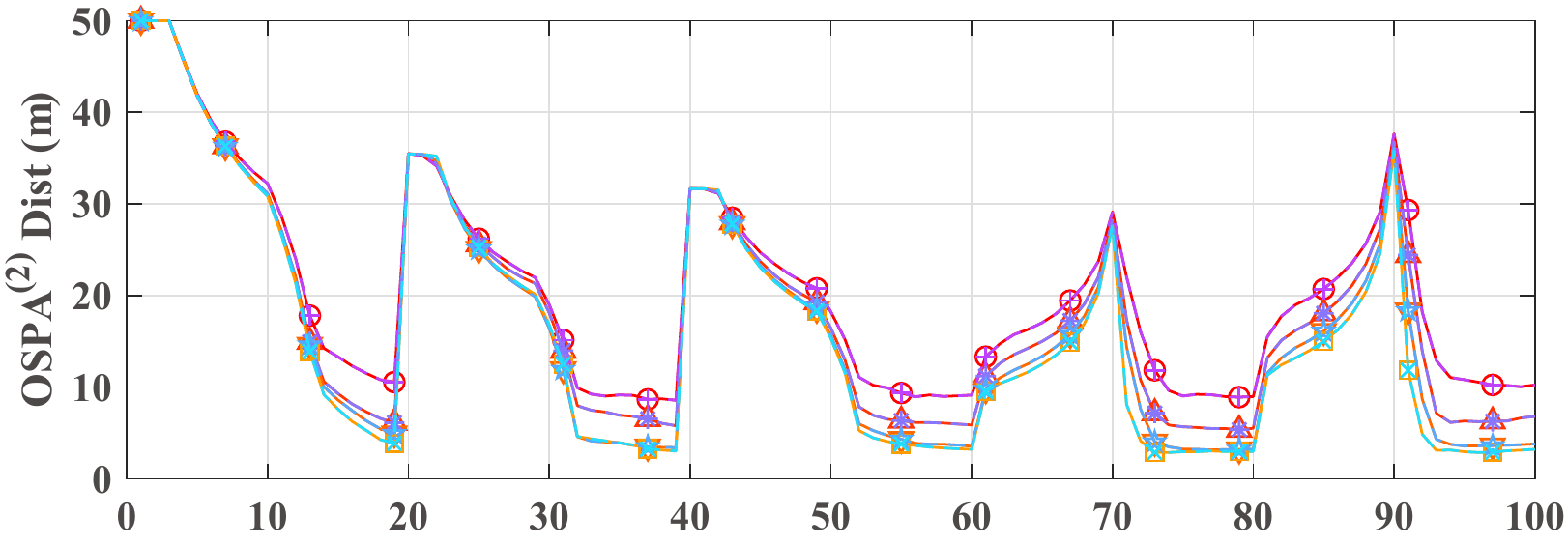}} 
	\hspace{8pt}%
	\subfloat[][]{%
	\label{fig_5_3}%
	\includegraphics[width=.48\textwidth]{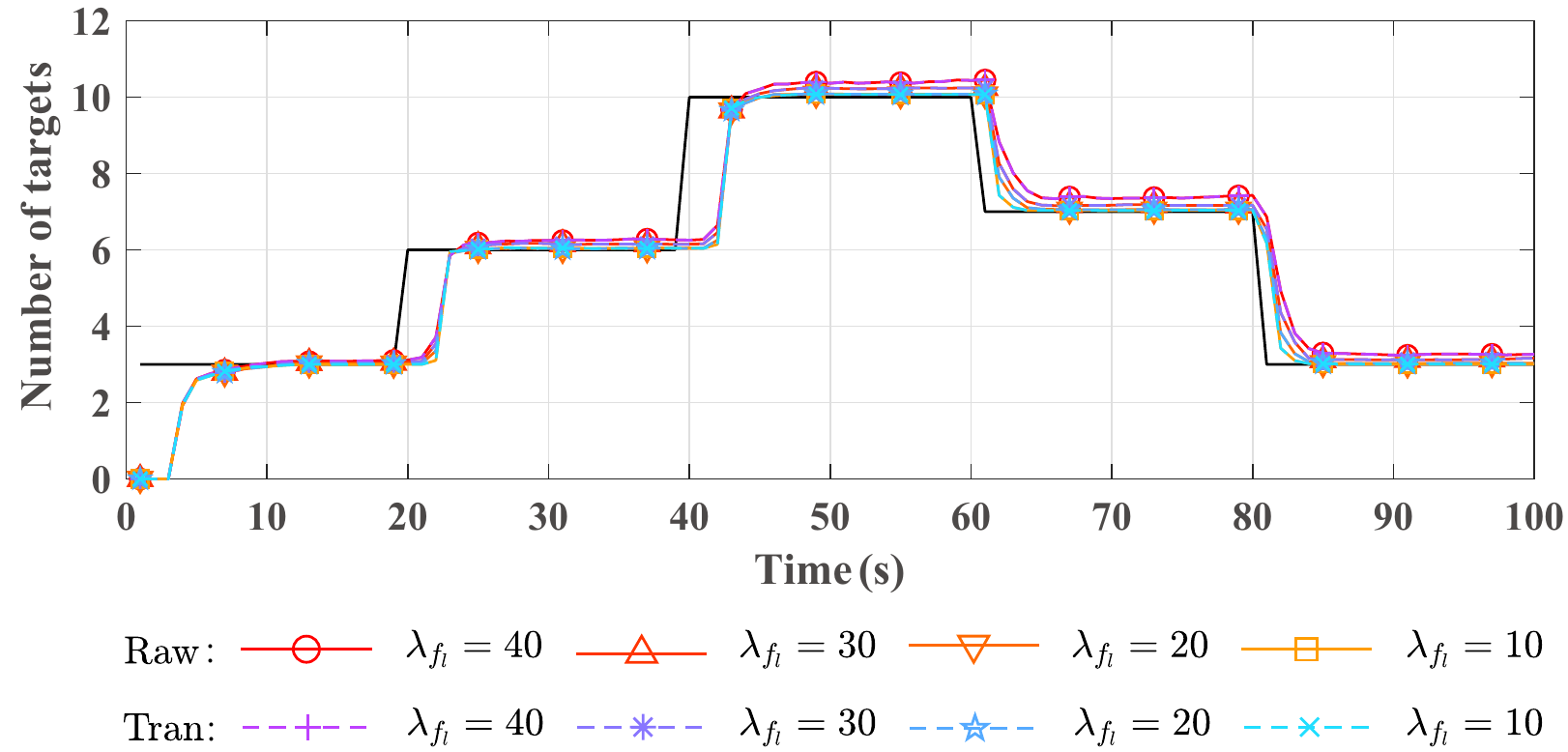}}%
	\caption[Scenario 2 results.]{Comparison results under different clutter rates: \subref{fig_5_1} OSPA distance, \subref{fig_5_2} OSPA$^{(2)}$ distance, \subref{fig_5_3} estimated number of targets.}%
	\label{fig_5_all}%
\end{figure}

\begin{figure}%
	\centering \subfloat[][]{%
	\label{fig_6_1}%
	\includegraphics[width=.48\textwidth]{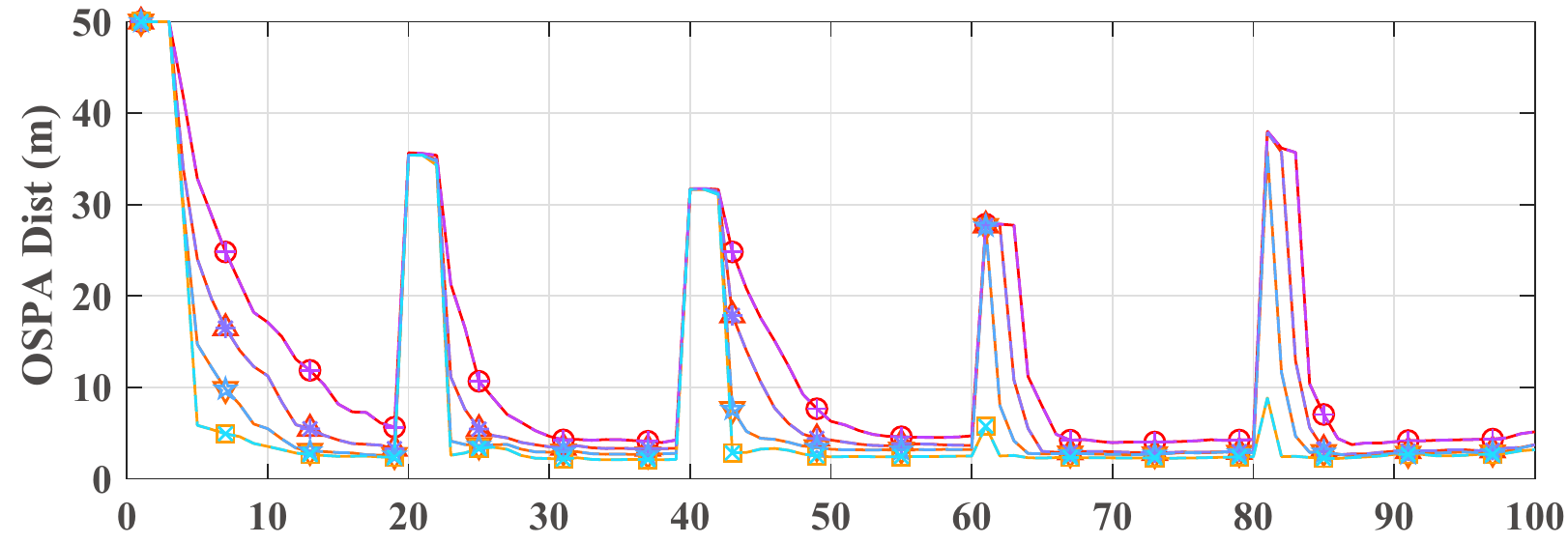}}%
	\hspace{8pt}%
	\subfloat[][]{%
	\label{fig_6_2}%
	\includegraphics[width=.48\textwidth]{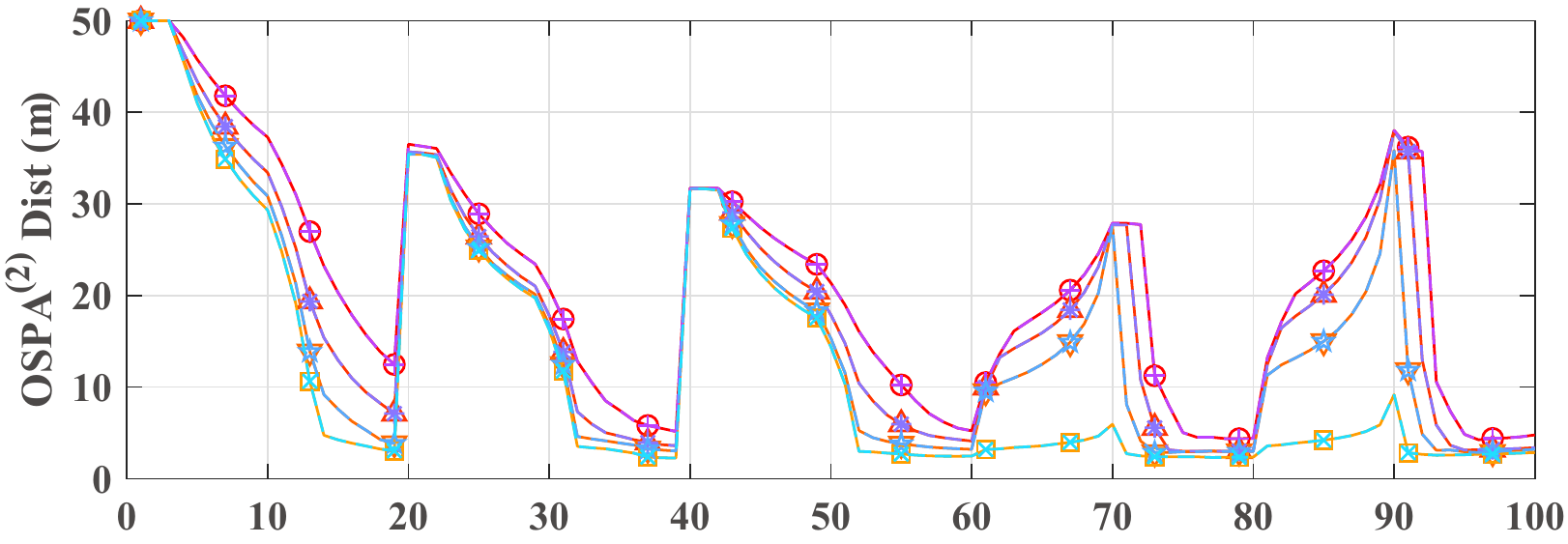}} 
	\hspace{8pt}%
	\subfloat[][]{%
	\label{fig_6_3}%
	\includegraphics[width=.48\textwidth]{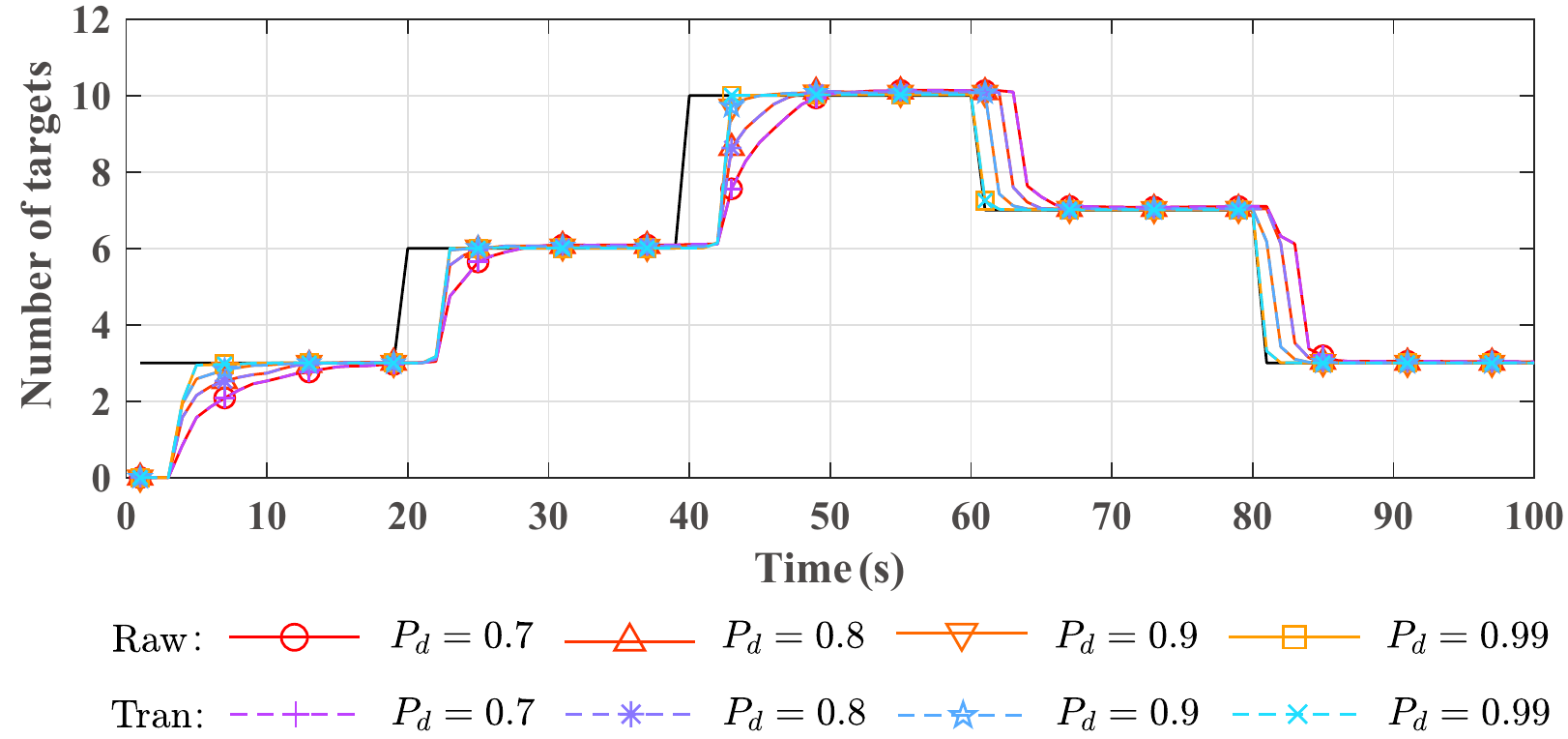}}%
	\caption[Scenario 2 results.]{Comparison results under different probabilities of detection: \subref{fig_5_1} OSPA distance, \subref{fig_5_2} OSPA$^{(2)}$ distance, \subref{fig_5_3} estimated number of targets.}%
	\label{fig_6_all}%
\end{figure}

Figs. \ref{fig_6_all}\subref{fig_6_1}--\ref{fig_6_all}\subref{fig_6_3} show the performance results under $P_d^{(l)}= 0.7,0.8,0.9$, and $0.99$ when $\lambda_{f_l}= 10$, respectively. Similarly, the curves of the OSPA distances shown in Fig. \ref{fig_6_all}\subref{fig_6_1} also exhibit five peaks. As expected, the OSPA and OSPA$^{(2)}$ distances decrease as the probability $P_d^{(l)}$ of detection increases, and the estimated number of targets increases as $P_d^{(l)}$ increases, which is consistent with the phenomenon in Figs. \ref{fig_scen_2_tracks}\subref{fig_scen_2_tracks_1}--\ref{fig_scen_2_tracks}\subref{fig_scen_2_tracks_4}. The reason is also the same as that in Scenario 1. 

Finally, Table \ref{table_4} summarizes the averaged communication requirements over 100 scans and 100 Monte Carlo runs at the fusion center. Under different values of $\lambda_{f_l}$ and $P_d^{(l)}$, the communication requirements for sending transformed measurements are less than those for sending raw measurements, which corroborates the analysis of communication requirement in Section \ref{sec_CR}. Compared to Table \ref{table_3} in Scenario 1, the communication requirements increase with the number of targets and the number of sensors. Moreover, the amount of communication bandwidths reduced by transmitting transformed measurements in Scenario 2 is more than that in Scenario 1.


\begin{table}
	\centering
	\caption{Communication Requirements in Kilobytes (KB) of Scenario~2}\label{table_4}
	\subfloat[Different values of clutter rate $\lambda_{f_l}$]{
		\resizebox{.48\textwidth}{!}{
			\begin{tabular}{|l|c|c|c|c|c|}
				\hline
				\multicolumn{2}{|l}{\multirow{2}{*}{Fusion type}} & \multicolumn{4}{|c|}{Clutter rate $\lambda_{f_l}$} \\
				\cline{3-6}
				\multicolumn{2}{|c|}{} & 10 & 20 & 30 & 40\\
				\hline
				\multicolumn{2}{|l|}{Fusion with raw measurements} & 4.64 KB & 4.70 KB  & 4.93 KB & 5.24 KB \\
				\hline 
				\multirow{2}{2.8cm}{Fusion with transformed measurements} &Type 1 & 2.87 KB & 2.90 KB & 3.02 KB & 3.23 KB\\ 
				\cline{2-6}
				& Type 2 & 1.79 KB & 1.81 KB & 1.89 KB & 2.02 KB\\ \hline
			\end{tabular}
		}
	}\\
	\subfloat[Different values of probability of detection $P_d^{(l)}$]{
		\resizebox{.48\textwidth}{!}{
			\begin{tabular}{|l|c|c|c|c|c|}
				\hline
				\multicolumn{2}{|l}{\multirow{2}{*}{Fusion type}} & \multicolumn{4}{|c|}{Probability of detection $P_d^{(l)}$} \\
				\cline{3-6}
				\multicolumn{2}{|c|}{} & 0.7 & 0.8 & 0.9 & 0.99\\
				\hline
				\multicolumn{2}{|l|}{Fusion with raw measurements} & 3.57 KB & 4.22 KB & 4.64 KB & 4.93 KB \\
				\hline 
				\multirow{2}{2.8cm}{Fusion with transformed measurements} & Type 1 & 2.19 KB & 2.59 KB & 2.87 KB & 3.04 KB\\ \cline{2-6}
				& Type 2 & 1.37 KB & 1.62 KB & 1.79 KB & 1.90 KB\\ \hline
			\end{tabular}
		}
	}
\end{table}

}

\section{Conclusion}\label{sec_cl}
{\color{blue}In this paper, for the fundamental problem of multisensor track-to-track fusion for multitarget tracking, we demonstrated the MDA-based data association (with and without prior track information) using linear transformations of track measurements is lossless, and is equivalent in terms of performance to that based on raw track measurements. Next, we presented a BP-based multisensor track association method based on measurement transformations and showed that it is equivalent to that with raw measurements. Finally, considering communication efficiency, two analytical lossless transformations for track association were provided, and communication requirements from each sensor to the fusion center were shown to be less than those of fusion with raw track measurements. Numerical examples for tracking an unknown number of targets using limited field-of-view sensors verified that the performance of fusion with transformed measurements is the same as that of fusion with raw measurements. Future works may include analyzing set-type track association methods \cite{van2021distributed}, distributed consensus fusion systems with or without feedback \cite{zhu2001optimality,zhu2012networked}, and non-linear dynamic systems \cite{Xiao2022Multisensor,Tan2023Nolinear}.}



%

\section*{Acknowledgment}
The authors would like to thank Yunmin Zhu for helpful suggestions.

\ifCLASSOPTIONcaptionsoff
  \newpage
\fi



\bibliographystyle{IEEEtran}
\bibliography{IEEEabrv,reference1(5)}

\newpage

\appendices

\section{Proof of Proposition \ref{equivalence_score}}\label{proof_prop1}
In this section, we prove Proposition \ref{equivalence_score}, which demonstrates that the MDA-based track association with prior track information with transformed measurements is equivalent to that with raw track measurements. Before we begin, we provide the preliminary results in Lemmas \ref{Lemma_1} and \ref{Lemma_2}.  
\begin{Lemma}\label{Lemma_1}
	Let $A_{k,l}$ be a full column rank matrix, and the processed data $\breve{\mathbf{z}}_{k,l}^{(i_{l})}$ be a linear transformation of $\mathbf{z}_{k,l}^{(i_{l})}$, i.e.,
	$\breve{\mathbf{z}}_{k,l}^{(i_{l})} = A_{k,l} \mathbf{z}_{k,l}^{(i_{l})}.$
	Then, the ratio of the likelihood $p(\mathbf{z}_{k,l}^{(i_{l})}|\hat{\mathbf{x}}_{k|k-1}^{(\tau)})$ for fusion with raw measurements (\ref{likeli_centralized}) and the likelihood $p(\breve{\mathbf{z}}_{k,l}^{(i_{l})}|\hat{\mathbf{x}}_{k|k-1}^{(\tau)})$ for fusion with transformed measurements (\ref{likeli_distributed}) is given by
	\begin{align}
		\frac{p(\mathbf{z}_{k,l}^{(i_{l})}|\hat{\mathbf{x}}_{k|k-1}^{(\tau)})}{p(\breve{\mathbf{z}}_{k,l}^{(i_{l})}|\hat{\mathbf{x}}_{k|k-1}^{(\tau)})} = \frac{(\prod_{i=1}^m e_i)^{1/2}}{(|S_{k,l}^{(\tau)}|)^{1/2}},
	\end{align}
	where $S_{k,l}^{(\tau)}$ is the innovation matrix and $e_i$ is the nonzero eigenvalue of $A_{k,l} S_{k,l}^{(\tau)} A_{k,l}^{\text{T}}$.
	\end{Lemma}
	\begin{IEEEproof}
		Without loss of generality, we omit the time index $k$, the sensor index $l$, and the target index $\tau$ to simplify writing. The likelihood function with raw measurements (\ref{likeli_centralized}) is
	\begin{equation}\label{likelihoodcenter}
		p\left(\mathbf{z} \mid \mathbf{x}\right)=\frac{(2\pi)^{-m/2}}{(|S|)^{1/2}}  e^{-\frac{1}{2}\left(\mathbf{z}-\hat{\mathbf{z}}\right)^{\text{T}}S^{-1}\left(\mathbf{z}-\hat{\mathbf{z}}\right)},
	\end{equation}
	and the likelihood function with transformed measurements (\ref{likeli_distributed}) is
	\begin{equation}\label{likelihooddistri}
		p\left(\breve{\mathbf{z}} \mid \mathbf{x}\right)=\frac{(2\pi)^{-m/2}}{(\prod_{i=1}^{m}e_i)^{1/2} } e^{-\frac{1}{2} (\mathbf{z}-\hat{\mathbf{z}})^{\text{T}} A^{\text{T}} (AS A^{\text{T}})^{\dagger}A (\mathbf{z}-\hat{\mathbf{z}}) }.
	\end{equation}
	The ratio of the above two pdfs $p\left(\mathbf{z} \mid \mathbf{x}\right)$ and $p\left(\breve{\mathbf{z}} \mid \mathbf{x}\right)$ is
	\begin{equation}\label{pf_likeli_ratio}
		\frac{p\left(\mathbf{z} \mid \mathbf{x}\right)}{p\left(\breve{\mathbf{z}} \mid \mathbf{x}\right)}=\frac{\sqrt{\prod_{i=1}^{m}e_{i}}}{\sqrt{|S|}}e^{-\frac{1}{2} (\mathbf{z}-\hat{\mathbf{z}})^{\text{T}}[S^{-1}- A^{\text{T}} (AS A^{\text{T}})^{\dagger}A] (\mathbf{z}-\hat{\mathbf{z}}) }.
	\end{equation}
	If the equation 
	\begin{align}\label{pf_eq}
		A^{\text{T}}(AS A^{\text{T}})^{\dagger}A=S^{-1}
	\end{align}
	holds, then Lemma \ref{Lemma_1} is proved. 

	Next, we give the proof of Equation (\ref{pf_eq}). The range spaces of matrices $A^{\text{T}}AS A^{\text{T}}$ and $SA^{\text{T}}$ can be written as follows:
	\begin{align}
		&\mathbf{W}_{1}=\{y|y= A^{\text{T}}ASA^{\text{T}}x, x \in \mathbb{R}^{n}\},\\
		\label{A16}
		&\mathbf{W}_{2}=\{y^{*}|y^{*}=S A^{\text{T}}x, x \in \mathbb{R}^{n}\},
	\end{align}
	and $\mathbf{W}_{1},\mathbf{W}_{2} \subseteq  \mathbb{R}^{m}$.
	Since $A$ and $S$ are full column rank matrix and fully rank matrix, respectively, the dimension of $\mathbf{W}_{2}$ is $m$ and $\mathbf{W}_{2} = \mathbb{R}^{m}$.
	Through (\ref{A16}), the range space $\mathbf{W}_{1}$ can be represented as follows:
	\begin{align*}
		\mathbf{W}_{1}=\{y|y= A^{\text{T}}Ay^{*}, y^{*} \in \mathbf{W}_{2}\},
	\end{align*}
	Since $A^{\text{T}} A$ is a fully rank matrix, the projection between $y$ and $y^{*}$ is a bijection, i.e., $\mathbf{W}_{1}=\mathbf{W}_{2}=\mathbb{R}^{m}$. Thus, we have
	\begin{align}\label{A(16)}
		\mathbb{R}(A^{\text{T}}AS A^{\text{T}})=\mathbb{R}(S A^{\text{T}}).
	\end{align}
	Similarly, we have the following relationships:
	\begin{align}\label{A(17)}
		&\mathbb{R}(S A^{\text{T}}AS A^{\text{T}})=\mathbb{R}(A^{\text{T}}),\\
		\label{mpinverse_3}
		&\mathbb{R}( S^{\text{T}}S A^{\text{T}})=\mathbb{R}( A^{\text{T}}),\\
		\label{mpinverse_4}
		&\mathbb{R}( A^{\text{T}}A S^{\text{T}})=\mathbb{R}( S^{\text{T}}).
	\end{align}
	Through (\ref{A(16)})--(\ref{A(17)}) and the reverse order laws for Moore-Penrose inverse, the following equation is established:
	\begin{align}\label{inverse_1}
		A^{\text{T}}(AS A^{\text{T}})^{\dagger}A= A^{\text{T}}(S A^{\text{T}})^{\dagger}.
	\end{align}
	Similarly, through (\ref{mpinverse_3})--(\ref{mpinverse_4}) and the reverse order laws for Moore-Penrose inverse, the equation
	\begin{align}\label{inverse_2}
		A^{\text{T}}(S A^{\text{T}})^{\dagger}=S^{-1},
	\end{align}
	holds. By (\ref{inverse_1}) and (\ref{inverse_2}), we have
	\begin{align}\label{dagger_lemma}
		A^{\text{T}}(AS A^{\text{T}})^{\dagger}A=S^{-1},
	\end{align}
	thus, the ratio of the two pdfs is
	\begin{align}
		\frac{p\left(\mathbf{z}\mid \mathbf{x}\right)}{p\left(\breve{\mathbf{z}} \mid \mathbf{x}\right)}=\frac{\sqrt{\prod_{i=1}^{m}e_{i}}}{\sqrt{|S|}},
	\end{align}
	and Lemma \ref{Lemma_1} is proved.
\end{IEEEproof}

\begin{Lemma}\label{Lemma_2}
	If $A_{k,l}$ is a full column rank matrix and the clutter is uniform in the region of interest, then the ratio of the clutter pdf $p_{f_{l}}(\mathbf{z}_{k,l}^{(i_{l})})$ of fusion with raw measurements and the clutter pdf $p_{f_{l}}(\breve{\mathbf{z}}_{k,l}^{(i_{l})})$ of fusion with transformed measurements is
	\begin{align}
		\frac{p_{f_{l}}(\mathbf{z}_{k,l}^{(i_{l})})}{p_{f_{l}}(\breve{\mathbf{z}}_{k,l}^{(i_{l})})} = \sqrt{|A_{k,l}^{\text{T}}A_{k,l}|}.
		\end{align}
	\end{Lemma}
	\begin{IEEEproof}
		Without loss of generality, we omit the time index $k$, the sensor index $l$, and the target index $\tau$ to simplify writing. Let $\mathbf{z}\in \mathbb{R}^m$ and $S_{\mathbf{z}}\subseteq \mathbb{R}^m$ denote the measurement and surveillance region, respectively. If the clutter is assumed uniform, the pdf of a clutter is
	\begin{align}
		p_{f_{l}}\left(\mathbf{z}\mid \gamma_{0}\right)=\frac{\lambda_{f_l}}{V_{\mathbf{z}}},
	\end{align}
	where $V_{\mathbf{z}}$ is the volume of the surveillance region $S_{\mathbf{z}}$, i.e., $V_{\mathbf{z}} = |S_{\mathbf{z}}|$.
	Since $\breve{\mathbf{z}}=A\mathbf{z}$, where $A$ is an $m_1 \times m$ matrix and $m_1 > m$, we have the transformed vector $\breve{\mathbf{z}} \in \mathbb{R}^{m_1}$ and the transformed region $S_{\breve{\mathbf{z}}}\subseteq \mathbb{R}^{m_1}$. Let $V_{\breve{\mathbf{z}}}$ denote the volume of $S_{\breve{\mathbf{z}}}$, the pdf of $\breve{\mathbf{z}}$ is
	\begin{align}
		p_{f_{l}}\left(\breve{\mathbf{z}}\mid \gamma_{0}\right)=\frac{\lambda_{f_l}}{V_{\breve{\mathbf{z}}}}.
	\end{align}
	Then, the ratio of $p_{f_{l}}(\mathbf{z}\mid \gamma_{0})$ and $p_{f_{l}}(\breve{\mathbf{z}}\mid \gamma_{0})$ is
	\begin{align}\label{eq_lem_1} 
		\frac{p_{f_{l}}\left(\mathbf{z}\mid \gamma_{0}\right)}{p_{f_{l}}\left(\breve{\mathbf{z}}\mid \gamma_{0}\right)} = \frac{V_{\breve{\mathbf{z}}}}{V_{\mathbf{z}}}.
	\end{align}
		Since $\breve{\mathbf{z}}=A\mathbf{z}$, where $A$ is an $m_1 \times m$ matrix ($m_1 > m$), by singular value decomposition (SVD) of $A$, we obtain
		\begin{align}\label{svd}
		\breve{\mathbf{z}}=U_{1}\Sigma_{A} V_{1}^{\text{T}}\mathbf{z},
		\end{align}
	where $A = U_{1}\Sigma_{A} V_{1}^{\text{T}}$, and $U_{1}$ and $ V_{1}^{\text{T}}$ are unitary matrices and $\Sigma_{A}$ is an $m_1 \times m$ matrix which nonzero elements are the singular values of $A$. Left multiply $ U_1^{\text{T}}$ by the both sides of (\ref{svd}), i.e.,
	\begin{align}\label{svd_1}
		U_1^{\text{T}}\breve{\mathbf{z}}&= U_1^{\text{T}}U_{1}\Sigma_{A} V_{1}^{\text{T}}\mathbf{z}=\Sigma_{A} V_{1}^{\text{T}}\mathbf{z}.
	\end{align}
	From (\ref{svd_1}), the first $m$ components of $(U_1)^{\text{T}}\breve{\mathbf{z}}$ are represented as
	\begin{align}
		( U_1^{\text{T}}\breve{\mathbf{z}})_{(i)} = \sigma_i ( V_{1}^{\text{T}}\mathbf{z})_{(i)},\quad i= 1,\cdots,m,
	\end{align}
	where $\sigma_i$ is the $i$-th singular value. The last $(m_1- m)$ components of $ U_1^{\text{T}}\breve{\mathbf{z}}$ are all zeros.
	Since $ U_1^{\text{T}}$ and $ V_{1}^{\text{T}}$ are the unitary matrices, which do not change the length of $\breve{\mathbf{z}}$ and $\mathbf{z}$, respectively, the ratio of the volumes $V_{\breve{\mathbf{z}}}$ and $V_{\mathbf{z}}$ is
	\begin{align}\label{eq_lem_2}
		\frac{V_{\breve{\mathbf{z}}}}{V_{\mathbf{z}}} = \prod_{i=1}^m \sigma_i.
	\end{align}
	Note that the singular values of $A$ are the square roots of the eigenvalues of $A^{\text{T}}A$, i.e.,
	\begin{align}\label{eq_lem_3}
		\prod_{i=1}^m \sigma_i = \sqrt{| A^{\text{T}} A|}.
	\end{align}
	Thus, based on (\ref{eq_lem_1}), (\ref{eq_lem_2}) and (\ref{eq_lem_3}), we conclude that
	\begin{align}
		\frac{p_{f_{l}}\left(\mathbf{z}\mid \gamma_{0}^{k}\right)}{p_{f_{l}}\left(\breve{\mathbf{z}}\mid \gamma_{0}^{k}\right)} = \sqrt{| A^{\text{T}}A|},
	\end{align}
	and Lemma \ref{Lemma_2} is proved.
\end{IEEEproof}

Lemma \ref{Lemma_1} provides the likelihood ratio for a measurement originating from a target, between raw measurements and transformed measurements. Lemma \ref{Lemma_2} provides the likelihood ratio for a clutter measurement, also between raw measurements and transformed measurements. In the following, we prove Proposition \ref{equivalence_score} using the results of Lemmas \ref{Lemma_1} and \ref{Lemma_2}. 

Through (\ref{score_centralized})--(\ref{score_distributed}), the ratio of the two score functions is rewritten as follows,
\begin{align}\label{A30}
&\frac{L_{(\tau,i_1,\cdots,i_L)}^{c}}{L_{(\tau,i_1,\cdots,i_L)}^{d}}=\prod_{l \in \{l|u(i_{l})=1\}}
\left[\frac{p(\mathbf{z}_{k,l}^{(i_{l})}|\hat{\mathbf{x}}_{k|k-1}^{(\tau)})}{p(\breve{\mathbf{z}}_{k,l}^{(i_{l})}|\hat{\mathbf{x}}_{k|k-1}^{(\tau)})} \frac{p_{f_{l}}(\breve{\mathbf{z}}_{k,l}^{(i_{l})}|\gamma^k_0)}{p_{f_{l}}(\mathbf{z}_{k,l}^{(i_{l})}|\gamma^k_0)}\right].
\end{align}
By Lemma {\ref{Lemma_1}} and Lemma {\ref{Lemma_2}}, the ratio (\ref{A30}) can be simplified as follows
\begin{align}\label{proof_ratio}
	\frac{L_{(\tau,i_1,\cdots,i_L)}^{c}}{L_{(\tau,i_1,\cdots,i_L)}^{d}}=\prod_{l \in \{l|u(i_{l})=1\}}\frac{(\prod_{i=1}^m e_{i,l})^{1/2}}{(|S_{k,l}^{(\tau)}| )^{1/2}(|A_{k,l}^{\text{T}} A_{k,l}|)^{1/2}},
\end{align}
where $e_{i,l}$ is the $i$-th eigenvalue of $A_{k,l} S_{k,l}^{(\tau)} A_{k,l}^{\text{T}}$. Next, we prove that $(|S_{k,l}^{(\tau)}| )^{1/2}(| A_{k,l}^{\text{T}}A_{k,l}|)^{1/2} = (\prod_{i=1}^m e_{i,l})^{1/2}$, then the conclusion of Proposition $\ref{equivalence_score}$  holds. Without loss of generality, we omit the time index $k$, the sensor index $l$, and the target index $\tau$ to simplify writing.
The SVD of $AS A^{\text{T}}$ and $A$ are obtained as follows,
\begin{align}
\label{A29}
AS A^{\text{T}}=U\Sigma U^{\text{T}},\quad A=U_{1} \Sigma_{A} V_{1}^{\text{T}},
\end{align}
where $U,U_{1},V_{1}$ are identity matrices, and
\begin{align}
  &\Sigma=\left[\begin{array}{ccccc}
    \varLambda & \mathbf{0}  \\
    \mathbf{0} & \mathbf{0}  \\
    \end{array}\right], \quad
    \varLambda = \text{diag}(e_{1},\cdots,e_{m}),\\
  &\Sigma_{A} = \left[\begin{array}{ccccc}
    \Sigma_{A}^{(1)}  \\
    \mathbf{0}   \\
    \end{array}\right], \quad
  \Sigma_{A}^{(1)} = \text{diag}(\sigma_1,\cdots,\sigma_m),
\end{align}
where $\sigma_i$ is the $i$-th singular value of $A$, and $e_i$ is the $i$-th eigenvalue of $A S A^{\text{T}}$.
Through (\ref{A29}), we obtain
\begin{IEEEeqnarray}{rCl}
	U \Sigma U^{\text{T}} = AS A^{\text{T}} = U_{1}\Sigma_{A} V_{1}^{\text{T}} S V_{1} \Sigma_{A}^{\text{T}} U_{1}^{\text{T}}.
\end{IEEEeqnarray}
Thus, we have $\Sigma\sim \Sigma_{A} V_{1}^{\text{T}} S V_{1} \Sigma_{A}^{\text{T}}$, 
where
\begin{align}
  \Sigma_{A} V_{1}^{\text{T}} S V_{1} \Sigma_{A}^{\text{T}} = \left[\begin{array}{cc}
    \Sigma_{A}^{(1)} V_{1}^{\text{T}} S V_{1} (\Sigma_{A}^{(1)})^{\text{T}} & \mathbf{0}\\
    \mathbf{0} & \mathbf{0}
  \end{array}\right].
\end{align}
Due to the similarity of $\Sigma$ and $\Sigma_{A} V_{1}^{\text{T}}S V_{1} \Sigma_{A}^{\text{T}}$, the eigenvalues of $\varLambda$  and $\Sigma_{A}^{(1)} V_{1}^{\text{T}}S V_{1} (\Sigma_{A}^{(1)})^{\text{T}}$ are the same. Based on the relationship between eigenvalue and determinant, we obtain $|\varLambda|=|\Sigma_{A}^{(1)} V_{1}^{\text{T}} S V_{1} (\Sigma_{A}^{(1)})^{\text{T}}|$, 
i.e.,
\begin{align}\label{A24}
\prod_{i=1}^m e_{i}=|S| |(\Sigma_A^{(1)})^{\text{T}}\Sigma_A^{(1)}| = |S | | A^{\text{T}} A |,
\end{align}
where the last equation follows from (\ref{eq_lem_3}).
Thus, by (\ref{proof_ratio}) and (\ref{A24}), the score function with raw measurements is equal to that with transformed measurements, i.e.,
\begin{align}
	\frac{L_{(\tau,i_1,\cdots,i_L)}^{c}}{L_{(\tau,i_1,\cdots,i_L)}^{d}}=1,
\end{align}
and Proposition {\ref{equivalence_score}} is proved.

{\section{Proof of Corollary \ref{corro_nopiror}}\label{proof_coro_1}
In this section, we prove Proposition \ref{corro_nopiror}, which states that the MDA-based track association without prior track information with transformed measurements is equivalent to that with raw track measurements. We begin by presenting the preliminary result in the following Lemma \ref{Lemma_3}. 
\begin{Lemma}\label{Lemma_3}
	Let $A_{k,l}$ be a full column rank matrix, and the processed data $\breve{\mathbf{z}}_{k,l}^{(i_{l})}$ be a linear transformation of $\mathbf{z}_{k,l}^{(i_{l})}$. The MLE solution based on raw measurements is the same as that based on linearly transformed measurements, i.e., $\hat{\mathbf{x}}_{k,\text{ML}}^{(\tau),c}=\hat{\mathbf{x}}_{k,\text{ML}}^{(\tau),d}$, 
	where
	\begin{align}\label{cen_ML}
	\hat{\mathbf{x}}_{k,\text{ML}}^{(\tau),c}=\argmax_{\mathbf{x}_{k}^{(\tau)}} \prod_{l \in \{l|u(i_{l})=1\}}p(\mathbf{z}_{k,l}^{(i_{l})}|\mathbf{x}_{k}^{(\tau)}),
		\end{align}
	and
	\begin{align}\label{dis_ML}
	\hat{\mathbf{x}}_{k,\text{ML}}^{(\tau),d}=\argmax_{\mathbf{x}_{k}^{(\tau)}} \prod_{l \in \{l|u(i_{l})=1\}}p(\breve{\mathbf{z}}_{k,l}^{(i_{l})}|\mathbf{x}_{k}^{(\tau)}).
	\end{align}
\end{Lemma}
\begin{IEEEproof}
	For a given track hypothesis $(i_{1},\cdots,i_{L})$ and $\mathbf{x}_{k}^{(\tau)}$, the problem (\ref{cen_ML}) is equivalent to the following problem,
	\begin{align}\label{equ_1_change}
	\hat{\mathbf{x}}_{k,\text{ML}}^{(\tau),c}=\argmax_{\mathbf{x}_{k}^{(\tau)}} \prod_{l \in \{l|u(i_{l})=1\}}p(\mathbf{z}_{k,l}^{(i_{l})}|\mathbf{x}_{k}^{(\tau)}).
	\end{align}
	By taking the logarithm of the objective function in problem (\ref{equ_1_change}), the original problem is equivalent to the problem
	\begin{align}
	\hat{\mathbf{x}}_{k,\text{ML}}^{(\tau),c}=\argmax_{\mathbf{x}_{k}^{(\tau)}}\sum_{l \in \{l|u(i_{l})=1\}}f_{k,l}^c(\mathbf{x}_{k}^{(\tau)}),
	\end{align}
	where
	\begin{align}
	f_{k,l}^c(\mathbf{x}_k^{\tau})=-\frac{1}{2}(\mathbf{z}_{k,l}^{(i_{l})}- H_{k,l}\mathbf{x}_{k}^{(\tau)})^{\text{T}} R_{k,l}^{-1}(\mathbf{z}_{k,l}^{(i_{l})}- H_{k,l}\mathbf{x}_{k}^{(\tau)}).
	\end{align}
	Similarly, the problem (\ref{dis_ML}) is equivalent to the following problem,
	\begin{align}
	\hat{\mathbf{x}}_{k,\text{ML}}^{(\tau),d}=\argmax_{\mathbf{x}_{k}^{(\tau)}}\sum_{l \in \{l|u(i_{l})=1\}}f_{k,l}^d(\mathbf{x}_{k}^{(\tau)}),
	\end{align}
	where
	\begin{align}
	\nonumber	f_{k,l}^d(\mathbf{x}_{k}^{(\tau)})&=-\frac{1}{2}(\mathbf{\breve{y}}_{k,l}^{(i_l)}- \breve{H}_{k,l}\mathbf{x}_{k}^{(\tau)})^{\text{T}} \check{R}_{k,l}^{\dagger}(\mathbf{\breve{y}}_{k,l}^{(i_l)}- \breve{H}_{k,l}\mathbf{x}_{k}^{(\tau)})\\
	\nonumber
	&=-\frac{1}{2}(\mathbf{z}_{k,l}^{(i_{l})}- H_{k,l}\mathbf{x}_{k}^{(\tau)})^{\text{T}} A_{k,l}^{\text{T}}(A_{k,l}R_{k,l} A_{k,l}^{\text{T}})^{\dagger}\\
	&\quad \times A_{k,l}(\mathbf{z}_{k,l}^{(i_{l})}- H_{k,l}\mathbf{x}_{k}^{(\tau)}).
	\end{align}
	By (\ref{dagger_lemma}), we have $f_{k,l}^c(\mathbf{x}_{k}^{(\tau)})=f_{k,l}^d(\mathbf{x}_{k}^{(\tau)})$. Moreover, the solution of problem (\ref{cen_ML}) is same as that of problem (\ref{dis_ML}). Lemma \ref{Lemma_3} is proved.
\end{IEEEproof}

Lemma \ref{Lemma_3} shows that the MLE solution based on transformed measurements is equivalent to that based on raw measurements. Based on the results of Lemma \ref{Lemma_3} and the proof of Proposition \ref{equivalence_score}, Corollary \ref{corro_nopiror} is proved by replacing $S$ with $R$ in Appendix \ref{proof_prop1}.

\section{The MDA Association and Fusion Algorithm}
\label{app_algorithm_1}
\begin{algorithm}
	\caption{MDA Association and Fusion}\label{alg:MDA}
	\SetKwBlock{uBegin}{}{}
	\SetKwData{Left}{left}\SetKwData{This}{this}\SetKwData{Up}{up} 
	\SetKwFunction{Union}{Union}\SetKwFunction{FindCompress}{FindCompress} 
	\SetKwInOut{Input}{input}\SetKwInOut{Output}{output}

	\Input{$\{\hat{\mathbf{x}}_{k-1|k-1}^{(\tau)}, P_{k-1|k-1}^{(\tau)} \}_{\tau=1}^{N_{k-1}}$, $\breve{\mathbf{Z}}_{k,l}$, $\{H_{k,l}\}_{l=1}^L$, or $\{R_{k,l}\}_{l=1}^L$;}
	\Output{$\{\hat{\mathbf{x}}_{k|k}^{(\tau)}, P_{k|k}^{(\tau)}\}_{\tau=1}^{N_k}$;}
	\emph{Prediction: }
	\uBegin{
		\emph{Calculate $\{\hat{\mathbf{x}}_{k|k-1}^{(\tau)}, P_{k|k-1}^{(\tau)} \}_{\tau=1}^{N_{k-1}}$ using (\ref{MDA_pred_x})--(\ref{MDA_pred_p})}\;
	}
	\emph{Track maintenance: }
	\uBegin{
		\emph{Establish the MDA problem (\ref{LIP})}\;
		\emph{Solve the MDA problem (\ref{LIP})}\;
		\emph{Track update $\{\hat{\mathbf{x}}_{k|k}^{(\tau)}, P_{k|k}^{(\tau)}\}_{\tau=1}^{N_{k-1}}$: Each track $\tau$ is updated by $\{\breve{\mathbf{z}}_{k,l}^{(i_l)}\}_{l=1}^{L}$ for $\delta_{(\tau,i_1,\cdots,i_L)}=1$ using the method in \cite{2011Li_distributed}}\;
		\emph{Delete $\{\breve{\mathbf{z}}_{k,l}^{(i_l)}\}_{l=1}^{L}$ in $\breve{\mathbf{Z}}_{k}$ for $\delta_{(\tau,i_1,\cdots,i_L)}=1$}\;
	}
	\emph{Newborn track initialization: }
	\uBegin{
		\emph{Establish the MDA problem (\ref{LIP_noinit})}\;
		\emph{Solve the MDA problem (\ref{LIP_noinit})}\;
		\emph{Initialize tracks $\{\hat{\mathbf{x}}_{k|k}^{(\tau')}, P_{k|k}^{(\tau')} \}_{\tau'=N_{k-1}+ 1}^{N_{k-1}+N'_k}$: Each newborn track $\tau'$ is initialized by $\{\breve{\mathbf{z}}_{k,l}^{(i_l)}\}_{l=1}^{L}$ for $\delta_{(i_1,\cdots,i_L)}=1$}\;
	}
	\emph{Output: }
	\uBegin{
		\emph{$N_k = N_{k-1} + N'_k$}\;
		\emph{$\{\hat{\mathbf{x}}_{k|k}^{(\tau)}\}_{\tau=1}^{N_{k}} = \left\{\{\hat{\mathbf{x}}_{k|k}^{(\tau)}\}_{\tau=1}^{N_{k-1}}, \{\hat{\mathbf{x}}_{k|k}^{(\tau')}\}_{\tau'=N_{k-1}+ 1}^{N_{k-1}+ N'_k}\right\}$}\;
		\emph{$\{P_{k|k}^{(\tau)}\}_{\tau=1}^{N_{k}} = \left\{\{P_{k|k}^{(\tau)}\}_{\tau=1}^{N_{k-1}}, \{P_{k|k}^{(\tau')}\}_{\tau'=N_{k-1}+ 1}^{N_{k-1}+ N'_k}\right\}$}\;
		\emph{Set $k=k+1$}.
	}
\end{algorithm}

\section{Proof of Proposition \ref{bp_Prop1}}\label{appe_bp_lemma1}
In this section, we show that the BP track association method with raw measurements is equivalent to that with transformed measurements. We first prove that $q(\cdot ; \mathbf{Z}_{k,l}) = q(\cdot ; \breve{\mathbf{Z}}_{k,l})$. When $\underline{r}_{k,l}^{(\tau)} = 0$, from (\ref{bp_q_x_r0}), $q(\underline{\mathbf{x}}_{k,l}^{(\tau)}, 0, a_{k,l}^{(\tau)} ; \mathbf{Z}_{k,l}) = q(\underline{\mathbf{x}}_{k,l}^{(\tau)}, 0, a_{k,l}^{(\tau)} ; \breve{\mathbf{Z}} _{k,l})$. When $\underline{r}_{k,l}^{(\tau)} = 1$,
from (\ref{bp_factor_q_1}), the equation holds when $a_{k,l}^{(\tau)} = 0$. Thus, we just need to consider the case of $\underline{r}_{k,l}^{(\tau)} = 1$ and $a_{k,l}^{(\tau)} \neq 0$, where the ratio of $q(\cdot ; \mathbf{Z}_{k,l})$ and $q(\cdot ; \breve{\mathbf{Z}}_{k,l})$ is 
\begin{align}
	\frac{q(\underline{\mathbf{x}}_{k,l}^{(\tau)}, 1, a_{k,l}^{(\tau)} ; \mathbf{Z}_{k,l})}{q(\underline{\mathbf{x}}_{k,l}^{(\tau)}, 1, a_{k,l}^{(\tau)} ; \breve{\mathbf{Z}} _{k,l})}
	= \frac{p(\mathbf{z}_{k,l}^{(a_{k,l}^{(\tau)})}|\underline{\mathbf{x}}_{k,l}^{(\tau)})}{p(\breve{\mathbf{z}}_{k,l}^{(a_{k,l}^{(\tau)})}|\underline{\mathbf{x}}_{k,l}^{(\tau)})} \frac{p_{f_{l}}(\breve{\mathbf{z}}_{k,l}^{(a_{k,l}^{(\tau)})})}{p_{f_{l}}(\mathbf{z}_{k,l}^{(a_{k,l}^{(\tau)})})}.
\end{align}
By using Lemma \ref{Lemma_1} and Lemma \ref{Lemma_2}, we have
\begin{align}
	\frac{q(\underline{\mathbf{x}}_{k,l}^{(\tau)}, 1, a_{k,l}^{(\tau)} ; \mathbf{Z}_{k,l})}{q(\underline{\mathbf{x}}_{k,l}^{(\tau)}, 1, a_{k,l}^{(\tau)} ; \breve{\mathbf{Z}} _{k,l})}  = \frac{\sqrt{\prod_{i=1}^{m}e_{i}}}{\sqrt{|R|}\sqrt{| A^{\text{T}}A|}}  = 1,
\end{align}
where $e_i$, $i=1,\cdots,m$ are the nonzero eigenvalues of $AR A^{\text{T}}$, and the last equation holds due to (\ref{A24}) in Appendix~\ref{proof_prop1} by replacing $S$ with $R$. Then, the following equation holds, 
\begin{align}\label{pf_bp_q}
	q(\underline{\mathbf{x}}_{k,l}^{(\tau)}, \underline{r}_{k,l}^{(\tau)}, a_{k,l}^{(\tau)} ; \mathbf{Z}_{k,l}) = q(\underline{\mathbf{x}}_{k,l}^{(\tau)}, \underline{r}_{k,l}^{(\tau)}, a_{k,l}^{(\tau)} ; \breve{\mathbf{Z}} _{k,l}).
\end{align}
Let $\beta^c(a_{k,l}^{(\tau)})$ denote the message (\ref{bp_2_meas}) with raw measurements and $\beta^d(a_{k,l}^{(\tau)})$ denote the message (\ref{bp_2_meas}) with transformed measurements.
We have
\begin{align}
	\nonumber
	\beta^c&(a_{k,l}^{(\tau)}) - \beta^d(a_{k,l}^{(\tau)}) \\
	\nonumber
	=& \sum_{r_{k,l}^{\tau} \in \{0,1\}} \int (q(\underline{\mathbf{x}}_{k,l}^{(\tau)}, \underline{r}_{k,l}^{(\tau)}, a_{k,l}^{(\tau)} ; \mathbf{Z}_{k,l})  \\
	\nonumber
	&\qquad\qquad - q(\underline{\mathbf{x}}_{k,l}^{(\tau)}, \underline{r}_{k,l}^{(\tau)}, a_{k,l}^{(\tau)} ; \breve{\mathbf{Z}} _{k,l}))  \tilde{f}_{l-1}(\underline{\mathbf{x}}_{k,l}^{(\tau)},\underline{r}_{k,l}^{(\tau)}) d \underline{\mathbf{x}}_{k,l}^{(\tau)}\\
	=& 0,
\end{align}
i.e., $\beta^c(a_{k,l}^{(\tau)}) = \beta^d(a_{k,l}^{(\tau)})$. Similarly, we can prove that
\begin{align}
	v(\overline{\mathbf{x}}_{k,l}^{(i_{l})},\overline{r}_{k,l}^{(i_{l})},b_{k,l}^{(i_{l})};\mathbf{z}_{k,l}^{(i_{l})}) = v(\overline{\mathbf{x}}_{k,l}^{(i_{l})},\overline{r}_{k,l}^{(i_{l})},b_{k,l}^{(i_{l})};\breve{\mathbf{z}}_{k,l}^{(i_{l})}),
\end{align}
and the message (\ref{bp_mess_xi}) with raw measurements is equal to the message (\ref{bp_mess_xi}) with transformed measurements, i.e., $\xi^{c} (b_{k,l}^{(i_l)}) = \xi^{d} (b_{k,l}^{(i_l)})$. From (\ref{ite_1_v})--(\ref{ite_init}), we know that the inputs of the step of iterative data association are $\beta(a_{k,l}^{(\tau)})$ and $\xi(b_{k,l}^{(i_l)})$. Since the inputs $\beta^c(a_{k,l}^{(\tau)})$ and $\xi^{c} (b_{k,l}^{(i_l)})$ are equal to the inputs $\beta^d(a_{k,l}^{(\tau)})$ and $\xi^{d} (b_{k,l}^{(i_l)})$, the outputs of iterative data association
\begin{align}
	\label{pf_bp_kappa}
	\kappa^{c}(a_{k,l}^{(\tau)})&=\kappa^{d}(a_{k,l}^{(\tau)}),\\
	\label{pf_bp_iota}
	\iota^c (b_{k,l}^{(i_l)})&=\iota^d (b_{k,l}^{(i_l)}),
\end{align}
where $\{\kappa^{c}(a_{k,l}^{(\tau)}),\iota^c (b_{k,l}^{(i_l)})\}$ and $\{\kappa^{d}(a_{k,l}^{(\tau)}),\iota^d (b_{k,l}^{(i_l)})\}$ are the outputs of the step of iterative data association with raw measurements and transformed measurements, respectively. 

Next, we prove that in the step of measurement update, the messages $\gamma^{c}(\underline{\mathbf{x}}_{k,l}^{(\tau)},\underline{r}_{k,l}^{(\tau)})$ and $\varsigma^{c} (\overline{\mathbf{x}}_{k,l}^{(i_{l})},\overline{r}_{k,l}^{(i_{l})}) $ with raw measurements are equal to the messages $\gamma^{d}(\underline{\mathbf{x}}_{k,l}^{(\tau)},\underline{r}_{k,l}^{(\tau)})$ and $\varsigma^{d} (\overline{\mathbf{x}}_{k,l}^{(i_{l})},\overline{r}_{k,l}^{(i_{l})})$ with transformed measurements, respectively,  i.e.,
\begin{align}
\gamma^{c}(\underline{\mathbf{x}}_{k,l}^{(\tau)},\underline{r}_{k,l}^{(\tau)})&=\gamma^{d}(\underline{\mathbf{x}}_{k,l}^{(\tau)},\underline{r}_{k,l}^{(\tau)}), \\
\varsigma^{c}  (\overline{\mathbf{x}}_{k,l}^{(i_{l})},\overline{r}_{k,l}^{(i_{l})})&=\varsigma^{d}  (\overline{\mathbf{x}}_{k,l}^{(i_{l})},\overline{r}_{k,l}^{(i_{l})}).
\end{align}
For survived targets, the following equations hold from Equations (\ref{pf_bp_q}) and (\ref{pf_bp_kappa}):
\begin{align}
	q(\underline{\mathbf{x}}_{k,l}^{(\tau)}, \underline{r}_{k,l}^{(\tau)}, a_{k,l}^{(\tau)} ; \mathbf{Z}_{k,l}) &= q(\underline{\mathbf{x}}_{k,l}^{(\tau)}, \underline{r}_{k,l}^{(\tau)}, a_{k,l}^{(\tau)} ; \breve{\mathbf{Z}} _{k,l}),\\
	\kappa^{c}(a_{k,l}^{(\tau)})&=\kappa^{d}(a_{k,l}^{(\tau)}).
\end{align}
From (\ref{bp_meas_up}) and (\ref{bp_meas_up_2}), we conclude that $\gamma^{c}(\underline{\mathbf{x}}_{k,l}^{(\tau)},\underline{r}_{k,l}^{(\tau)})=\gamma^{d}(\underline{\mathbf{x}}_{k,l}^{(\tau)},\underline{r}_{k,l}^{(\tau)})$.
Similarly, for new targets, 
\begin{align}
	v(\overline{\mathbf{x}}_{k,l}^{(i_{l})},\overline{r}_{k,l}^{(i_{l})},b_{k,l}^{(i_{l})};\mathbf{z}_{k,l}^{(i_{l})}) &= v(\overline{\mathbf{x}}_{k,l}^{(i_{l})},\overline{r}_{k,l}^{(i_{l})},b_{k,l}^{(i_{l})};\breve{\mathbf{z}}_{k,l}^{(i_{l})}),\\
	\iota^{c} (b_{k,l}^{(i_l)})&=\iota^{d} (b_{k,l}^{(i_l)}).
\end{align}
From (\ref{bp_meas_up_3}) and (\ref{bp_meas_up_4}), we have $\varsigma^{c}  (\overline{\mathbf{x}}_{k,l}^{(i_{l})},\overline{r}_{k,l}^{(i_{l})})=\varsigma^{d}  (\overline{\mathbf{x}}_{k,l}^{(i_{l})},\overline{r}_{k,l}^{(i_{l})})$.

Finally, the beliefs of survived targets $\tilde{f}_l^c(\underline{\mathbf{x}}_{k,l}^{(\tau)},\underline{r}_{k,l}^{(\tau)}) = \tilde{f}_l^d(\underline{\mathbf{x}}_{k,l}^{(\tau)},\underline{r}_{k,l}^{(\tau)})$ hold from (\ref{bp_appx}) and (\ref{bp_appx_2}), and the beliefs of new targets $\tilde{f}_l^c(\overline{\mathbf{x}}_{k,l}^{(i_l)},\overline{r}_{k,l}^{(i_{l})}) = \tilde{f}_l^d(\overline{\mathbf{x}}_{k,l}^{(i_l)},\overline{r}_{k,l}^{(i_{l})})$ hold from (\ref{bp_belief_new_1}) and (\ref{bp_belief_new_2}). Proposition \ref{bp_Prop1} is proved.

\section{The BP-Based Association and Fusion Algorithm}\label{app_algorithm_2}
\begin{algorithm}
	\caption{BP-Based Association and Fusion}\label{alg:BP}
	\SetKwBlock{uBegin}{}{}
	\SetKwData{Left}{left}\SetKwData{This}{this}\SetKwData{Up}{up} 
	\SetKwFunction{Union}{Union}\SetKwFunction{FindCompress}{FindCompress} 
	\SetKwInOut{Input}{input}\SetKwInOut{Output}{output}

	\Input{$\{\tilde{f}(\mathbf{x}_{k-1}^{(\tau)},r_{k-1}^{(\tau)})\}_{\tau=1}^{N_{k-1}}$, $\breve{\mathbf{Z}}_k$, $\{H_{k,l}\}_{l=1}^L$, or $\{R_{k,l}\}_{l=1}^L$;}
	\Output{$\{\tilde{f}(\mathbf{x}_{k}^{(\tau)},r_{k}^{(\tau)})\}_{\tau=1}^{N_k} $;}
	\emph{Prediction: }
	\uBegin{
		\emph{$\tilde{f}_0(\underline{\mathbf{x}}_{k,1}^{(\tau)},\underline{r}_{k,1}^{(\tau)}) = \alpha(\underline{\mathbf{x}}_{k}^{(\tau)},\underline{r}_{k}^{(\tau)})$ via (\ref{bp_init_1})}\;
		\emph{$N_{k,1} = N_{k-1}$}\;
	}
	\emph{Sequential processing: }\\
	\For{$l=1,\cdots,L$}{
		\emph{Measurement evaluation: (\ref{bp_2_meas})}\;
		\emph{Iterative data association: (\ref{ite_1_v})--(\ref{bp_prob_b})}\;
		\emph{Measurement update: (\ref{bp_meas_up})}\;
		\emph{Belief calculation for survived targets: $\tilde{f}_{l}(\underline{\mathbf{x}}_{k,l}^{(\tau)},\underline{r}_{k,l}^{(\tau)})$ using (\ref{bp_appx})--(\ref{bp_appx_2})}\;
		\emph{Belief calculation for new targets: $\tilde{f}_{l}(\overline{\mathbf{x}}_{k,l}^{(i_l)},\overline{r}_{k,l}^{(i_l)})$ using (\ref{bp_belief_new_1})--(\ref{bp_belief_new_2})}\;
		\emph{State Estimation: (\ref{bp_est_1})--(\ref{bp_est_2})}\;
		\emph{$\tilde{f}_{l}(\underline{\mathbf{x}}_{k,l+1}^{\tau},\underline{r}_{k,l+1}^{\tau}) = \tilde{f}_{l}(\underline{\mathbf{x}}_{k,l}^{(\tau)},\underline{r}_{k,l}^{(\tau)})$, for $\tau \leq N_{k,l}$}\;
		\emph{$\tilde{f}_{l}(\underline{\mathbf{x}}_{k,l+1}^{\tau},\underline{r}_{k,l+1}^{\tau}) = \tilde{f}_{l}(\overline{\mathbf{x}}_{k,l}^{\tau - N_{k,l}},\overline{r}_{k,l}^{\tau - N_{k,l}})$,
		for $N_{k,l} < \tau \leq N_{k,l} + M_{k,l}$}\;
	}
	\emph{Output: }
	\uBegin{
		\emph{$\tilde{f}(\mathbf{x}_{k}^{(\tau)},r_{k}^{(\tau)}) = \tilde{f}_{L}(\underline{\mathbf{x}}_{k,L}^{(\tau)},\underline{r}_{k,L}^{(\tau)})$, for $\tau \leq N_{k,L}$}\;
		\emph{$\tilde{f}(\mathbf{x}_{k}^{(\tau)},r_{k}^{(\tau)}) = \tilde{f}_{L}(\overline{\mathbf{x}}_{k,L}^{(\tau - N_{k,L})},\overline{r}_{k,L}^{(\tau - N_{k,L})})$}\;
		\emph{$N_{k,L} < \tau \leq N_{k,L} + M_{k,L}$}\;
		\emph{$N_{k} = N_{k,L} + M_{k,L}$}\;
		\emph{Set $k=k+1$}\;
	}
\end{algorithm}

\end{document}